\documentclass[11pt,a4paper]{article}
\pdfoutput=1
\usepackage{jheppub}
\usepackage{amsmath,amsfonts,amssymb,bbm, mathtools, mathrsfs}
\usepackage{hyperref, csquotes}
\usepackage{graphicx, color}
\usepackage{tikz,pgf,tikz-cd,float}
\usepackage{physics,tensor}
\usepackage[nottoc,numbib]{tocbibind}
\include{fdiagrams}

\newcommand{\one}{\mathbbm 1}
\newcommand{\R}{\mathbb R}
\newcommand{\C}{\mathbb C}

\newcommand{\cas}{\textrm{Cas}}     
\newcommand{\SUT}{\mathrm{SU}(2)}
\newcommand{\SUO}{\mathrm{SU}(1,1)}
\newcommand{\ISO}{\mathrm{ISO}(2)}
\newcommand{\SL}{\text{SL$(2,\C)$}}

\newcommand{\spl}{\mathfrak{sl}\left(2,\C\right)}
\renewcommand{\TH}{\text{H}^3}
\newcommand{\OH}{\mathrm{H}^{1,2}}
\newcommand{\defeq}{\vcentcolon=}

\begin{document}
\title{The Complete Barrett-Crane Model and its Causal Structure}

\author{Alexander F. Jercher,}
\author{Daniele Oriti,}
\author{Andreas G. A. Pithis}
\emailAdd{alexander.jercher@campus.lmu.de, daniele.oriti@physik.lmu.de, andreas.pithis@physik.lmu.de}

\affiliation{Arnold Sommerfeld Center for Theoretical Physics,\\ Ludwig-Maximilians-Universit\"at München \\ Theresienstrasse 37, 80333 M\"unchen, Germany, EU}
\affiliation{Munich Center for Quantum Science and Technology (MCQST), Schellingstr. 4, 80799 M\"unchen, Germany, EU}
\date{\today}

\begin{abstract}
{
The causal structure is a quintessential element of continuum spacetime physics and needs to be properly encoded in a theory of Lorentzian quantum gravity. Established spin foam (and tensorial group field theory (TGFT)) models mostly work with relatively special classes of Lorentzian triangulations (e.g. built from spacelike tetrahedra only), obscuring the explicit implementation of the local causal structure at the microscopic level. We overcome this limitation and construct a full-fledged model for Lorentzian quantum geometry the building blocks of which include spacelike, lightlike and timelike tetrahedra. We realize this within the context of the Barrett-Crane TGFT model. Following an explicit characterization of the amplitudes via methods of integral geometry, and the ensuing clear identification of local causal structure, we analyze the model's amplitudes with respect to its (space)time-orientation properties and provide also a more detailed comparison with the framework of causal dynamical triangulations (CDT). 
}
\end{abstract}

\maketitle

\section{Introduction}\label{sec:Introduction}

Among the most important conceptual and physical insights of special and general relativity is that space and time do not have an independent existence but are instead part of a single entity, spacetime. Spacetime carries a Lorentzian signature and therefore bears a causal structure. In fact, most of the geometric information of spacetime is encoded in the causal relations between events in spacetime. This is proven by Malament's theorem~\cite{Malament}, which states that a spacetime metric is obtained by causal relations up to a conformal factor. To make precise what we mean by causal structure, we introduce the terms of bare causality and time-orientation, following the arguments of~\cite{Livine:2002rh,Bianchi:2021ric}, and disentangle these two notions. In a continuum spacetime setting, bare causality refers to the possibility of dividing the tangent space at each spacetime point into three classes, being timelike, lightlike or spacelike. Extending this property globally, bare causality allows to determine whether two spacetime points have a timelike, lightlike or spacelike separation. Time-orientation on the other hand means that at every spacetime point, timelike vectors are distinguished to be either future pointing or past pointing, where future and past are defined by singling out an arbitrary timelike vector.\footnote{By continuity, this notion can be extended to lightlike vectors~\cite{HawkingBook}.} Globally (when a global extension can be defined) this property amounts to the possibility of arbitrarily determining an arrow of time. Bare causality and time-orientation together make up what is usually referred to as causal structure or causality.  

Taking the significance of causality in classical continuum gravity seriously, one naturally expects from a theory of quantum gravity (QG) that it addresses the role of causality, either by encoding it directly into the quantum theory or by providing good reasons why and how the causal structure should arise only in an appropriate classical and/or continuum limit. Among background-independent quantum gravity approaches, causal dynamical triangulations (CDT)~\cite{Loll:2019rdj,Ambjorn:2012jv} and causal set theory~\cite{Surya:2019ndm} explicitly include causal structure in their microscopic configurations\footnote{With respect to the above designation, CDT considers bare causality by including timelike and spacelike edges and time-orientation by distinguishing the $(4,1)$- and $(3,2)$-simplices from the \lq time-reversed\rq $(1,4)$- and $(2,3)$-simplices, respectively. In causal set theory, the starting point of quantization is a causal set (i.e. a poset with ordering relations interpreted as causal relations), which takes both aspects, i.e. bare causality and time-orientation, directly into account~\cite{Surya:2019ndm}.}, and make it a key ingredient of the quantum theory. Tensorial group field theories (TGFTs)~\cite{Oriti:2011jm,Krajewski:2011zzu,Carrozza:2013oiy,Oriti:2014uga} and spin foam models~\cite{Perez:2012wv}, the framework we choose to work with in this article, and closely related formalisms such as loop quantum gravity (LQG)~\cite{Ashtekar:2004eh,Rovelli:2011eq}, mostly focus on geometrical aspects of spacetime, encoded via holonomies and tetrads/fluxes. The main objective of this article is to develop a TGFT and spin foam model that transparently encodes all of the bare causal structure in the microscopic quantum degrees of freedom.

Tensorial group field theories~\cite{Oriti:2011jm,Krajewski:2011zzu,Carrozza:2013oiy} are combinatorially non-local quantum and statistical field theories defined on a group manifold\footnote{This group manifold is {\it not} interpreted as a spacetime on which physical fields are defined, but plays a different, auxiliary role of configuration space for the fundamental degrees of freedom {\it building up} spacetime.}. From a combinatorial perspective, they can be seen as the generalization of matrix models for two-dimensional quantum gravity~\cite{DiFrancesco:1993cyw} to higher dimensions, in the same way as tensor models~\cite{GurauBook}. The group theoretic structure on the other hand enriches these with quantum geometric degrees of freedom and turns them into proper field theories. Combining these two aspects in one formalism, TGFTs provide a link between several quantum gravity approaches, specifically LQG, spin foam models and simplicial gravity path integrals~\cite{Reisenberger:1997sk,Freidel:1998pt,Baratin:2011hp,Finocchiaro:2018hks}. As we show in the following, further connections to CDT and causal set theory can be envisaged and potentially exploited.

A specific GFT model (we indicate as \lq GFT\rq~models those TGFT models with a quantum geometric interpretation, inspired also by the other related quantum gravity formalisms) is characterized by a choice of group, an action and model-dependent constraints on the group field. The Feynman amplitudes of such quantum geometric models will take the form of spin foam models. The Barrett-Crane (BC) model~\cite{Barrett:1999qw,Perez:2000ec,Baratin:2011tx,Jercher:2021bie} and the EPRL model~\cite{Oriti:2016qtz} are the most prominent GFT (and spin foam) models, realizing a constrained BF-quantization of first-order Palatini and first-order Palatini-Holst gravity, respectively. Although imposing the simplicity constraint differently, in their most studied formulation the two models have in common that their elementary building blocks are spacelike tetrahedra only, implying that in perturbative expansion, only Lorentzian triangulations consisting of spacelike components are generated. In spite of working with the local symmetry group of Lorentzian gravity, $\SL$, the causal aspects of discrete and continuum geometry, in particular bare causality, are obscured by such restriction. This is because timelike and lightlike geometric objects need to be reconstructed from spacelike configurations. The other ensuing restriction is that the boundary states of the theory are necessarily spacelike. This is sufficient to describe important sectors of gravitational dynamics, for example the cosmological evolution of spacelike hypersurfaces~\cite{Oriti:2016qtz,Jercher:2021bie,Marchetti:2020umh}, but it leaves out equally interesting ones, e.g. Anti-de Sitter (AdS) space and much of black hole physics, since these spacetimes contain timelike and lightlike boundaries, respectively. 
Taking yet another perspective, the inclusion of additional configurations corresponding to timelike and lightlike tetrahedra leads to an extension of the TGFT models to new kinetic and interaction terms (and respective edge and vertex amplitudes). A possible consequence could be that the resulting model lies in a different universality class than that containing only the spacelike tetrahedra, therefore showing a different macroscopic continuum behavior. As an example, CDT and its Euclidean counterpart of dynamical triangulations (DT)~\cite{Loll:1998aj,Ambjorn:1991pq} show a different phase diagram because the two models are based on different building blocks, and a similar behavior can be determined for single- and multi-matrix models~\cite{Ambjorn:2001br,Eichhorn:2020sla,Benedetti:2008hc,Castro:2020dzt}. 

In the existing literature, several extensions of the aforementioned models have been proposed to include also timelike or lightlike tetrahedra. Already Barrett and Crane anticipated in~\cite{Barrett:1999qw} that there are in principle more building blocks that can be included than just spacelike tetrahedra. Shortly after,  Perez and Rovelli extended the analysis to a GFT model which describes timelike tetrahedra only, thus including spacelike as well as timelike faces~\cite{Perez:2000ep}. However this model is still rather restrictive in that it only includes timelike boundaries. From the perspective of canonical LQG, only little attention has been paid towards the inclusion of other than spacelike boundaries, due to the standard limitations of the canonical analysis. Timelike tetrahedra and faces are instead studied in~\cite{Alexandrov:2005ar} in the formalism of so-called \enquote{covariant LQG}~\cite{Alexandrov:2002br}, employing projected spin networks~\cite{Dupuis:2010jn}. Lightlike configurations have been even less explored in LQG and related approaches, see~\cite{Speziale:2013ifa,Neiman:2012fu}. In spin foam models, important progress to enlarge the set of causal configurations is represented by the Conrady-Hnybida (CH) extension~\cite{Conrady:2010kc,Conrady:2010vx} of the EPRL model (see~\cite{Simao:2021qno,Liu:2018gfc,Han:2021bln} for its asymptotic analysis). This extension includes spacelike \textit{and} timelike but not lightlike tetrahedra.

Going beyond the implementation of bare causal structures, work addressing the time-orientation aspect of causality in GFTs and spin foam models has been put forward in~\cite{Livine:2002rh} for the Barrett-Crane model which only incorporates spacelike tetrahedra. An analysis of the kernels which define the vertex amplitude revealed that the model does not encode any time-orientation and it possesses amplitudes that are invariant under time reversal~\cite{Livine:2002rh}. To turn the model into a time-oriented form, an explicit breaking of the time reversal invariance at the level of spin foam amplitudes was suggested therein. For the EPRL spin foam model, a related analysis has been performed in recent work~\cite{Bianchi:2021ric} with the same conclusions. 
This symmetry of the amplitudes may be a necessary feature from a canonical quantization point of view. A proper interpretation comes also from a covariant path integral perspective~\cite{Livine:2002rh, Halliwell:1990qr, Oriti:2004yu, Teitelboim:1981ua}, even though the role of a time-oriented version of the same quantum dynamics may well play an important role in quantum gravity and deserves to be studied with more attention. Moreover, such time-reversal symmetry may not produce any decisive consequence for what concerns the macroscopic arrow of time, which may well have a statistical, thermodynamic or otherwise collective and emergent origin, in quantum gravity. For that reason, we focus in this work on the implementation of bare causality in GFT and postpone an extension aiming at time-orientability to future research.

In order to have a complete and manifest implementation of bare causality in spin foam (TGFT) dynamics, we look for a model that properly includes all types of building blocks and their possible combinations. In the following, we show that the Barrett-Crane GFT model in its extended formulation can be formulated in a straightforward generalization to include timelike, lightlike and spacelike tetrahedra. 

\

Before proceeding with our analysis, let us digress briefly. The Barrett-Crane model, in both its GFT and spin foam formulation, is a rather straightforward constrained BF-quantization of first-order Palatini gravity. Several criticisms have been raised, however, against it, based on: the appearance of degenerate geometries (in the sense of vanishing $4$-volume) in the asymptotics~\cite{Barrett:2002ur,Freidel:2002mj}, the \enquote{wrong} boundary states~\cite{Alesci:2007tx}, missing secondary constraints~\cite{Baratin:2011tx,Baratin:2010wi}, imposing simplicity constraints \enquote{too strongly}~\cite{Engle:2007uq} and the non-covariant imposition of geometricity constraints~\cite{Baratin:2011tx}. While these criticisms are serious and have to be looked at carefully, we consider them inconclusive and the Barrett-Crane model still a potentially viable formulation of quantum gravity in the spin foam and TGFT context. The asymptotic analysis (e.g. in~\cite{Barrett:2002ur,Freidel:2002mj}) is mostly restricted to a single $4$-simplex and has not been extended to larger simplicial complexes, and thus its true implications are unclear. The findings of~\cite{Alesci:2007tx} (and similar works) simply reflect the mismatch between LQG states and the boundary states of the BC model. Since these boundary states result from a quantization of Palatini-Host gravity versus a quantization of Palatini gravity without additional topological contribution (and thus no Barbero-Immirzi parameter), respectively, the mismatch should not be surprising or upsetting, and has no bearing, per se, on the validity of the BC model. Next, the analysis of~\cite{Dittrich:2021kzs} in the context of \enquote{effective spin foams}~\cite{Asante:2020iwm,Asante:2020qpa,Asante:2021zzh} suggests that explicit secondary constraints may not be needed, in a spin foam context, to obtain the correct semiclassical and continuum behavior. Last, the Barrett-Crane model arises when imposing the simplicity constraints strongly (via a projector), as it should be, given that, in absence of the Barbero-Immirzi parameter, they are first class, and covariant, as shown in~\cite{Baratin:2011tx}. We conclude that the Barrett-Crane model is still a viable quantum gravity model, worth of any further attention, especially concerning its effective continuum gravitational description. The analysis of~\cite{Jercher:2021bie} support this viewpoint: the key results of GFT condensate cosmology~\cite{Gielen:2013kla,Gielen:2013naa,Oriti:2016qtz,Gielen:2016dss,Oriti:2016acw,Pithis:2019tvp,Jercher:2021bie}, such as an emergent Friedmann dynamics and a quantum bounce, can be recovered in the BC GFT model as well. 
In fact, this model offers an advantage that is particularly important for the inclusion of all bare causal configurations: Since the constraints of the Barrett-Crane GFT model, in a suitably extended formulation that includes explicitly normal vectors to the tetrahedra~\cite{Baratin:2011tx,Jercher:2021bie}, are covariantly imposed via a projector, the model is unique (at least from this point of view, in contrast with other existing models, in particular those including the Barbero-Immirzi parameter). Crucially, this extended formalism is also convenient for implementing full bare causality at the microscopic quantum level, and therefore we adopt it also in this article.

\

We pursue the following objectives. The main goal is the implementation of all possible bare causal configurations in the Barrett-Crane GFT model via a generalization to all possible normal vector signatures. Having obtained the corresponding complete model, the next objective is to study a selected number of properties of the model. First, we seek a clear quantum geometric interpretation of the configurations that enter its amplitudes. Second, we scrutinize the issue of time-orientation and clarify why the model is in fact orientation-symmetric. Last, we offer a detailed comparison to the CDT approach, which is made possible by the enlarged configuration space of the model. 
 
We begin the main body of this article by setting up the model in Section~\ref{subsec:Definition of the Complete Model} and give a geometric interpretation thereafter in Section~\ref{subsec:Bivector Variables and Geometric Interpretation}. Important for most explicit GFT computations, we derive the spin representation of the complete BC model in Section~\ref{subsec:Spin Representation of the Group Field and its Action} which allows us to present the spin form formulation in Section~\ref{subsec:Spin Foam Model Formulation}. As a last point of Section~\ref{sec:Extended Barrett-Crane Group Field Theory Model}, we provide a colored version of the model, necessary to generate topologically non-singular simplicial complexes and, more generally, to control the topology of the same complexes. In Sections~\ref{subsec:Kernels of Non-Mixed Type} and~\ref{subsec:Kernels of Mixed Type}, we explicitly compute the kernels that enter the vertex in the spin representation, using methods of integral geometry introduced in~\cite{VilenkinBook}. These are subsequently analyzed with regard to space, time and spacetime-orientation in Section~\ref{subsec:Spacetime Orientation}. Section~\ref{sec:Putting the Complete BC Model into Context of Other Models} provides a comparison between the model constructed here and other GFT models as well as the construction of a CDT-like GFT model, understood as a causal tensor model. We close in Section~\ref{sec:Discussion and Conclusion} with a discussion of the obtained results and outline future research directions that we consider as promising. Appendices~\ref{sec:Aspects of SL2C and its Representation Theory} and~\ref{sec:Integral Geometry} introduce the key notions of $\SL$-representation theory and integral geometry, crucial for the computations of Sections~\ref{subsec:Spin Representation of the Group Field and its Action} and~\ref{sec:Explicit Expressions for the Vertex Amplitudes}. Appendix~\ref{sec:Classical Computations of Simplicity Constraints and Bivector Signatures} contains a derivation of the classical results that enter Table~\ref{tab:bivectors} in Section~\ref{subsec:Bivector Variables and Geometric Interpretation}. 

\section{Extended Barrett-Crane group field theory model}\label{sec:Extended Barrett-Crane Group Field Theory Model}

An extended formulation of the Lorentzian Barrett-Crane GFT model, based on the Euclidean theory developed in~\cite{Baratin:2011tx}, has been defined in~\cite{Jercher:2021bie}. The extension consisted of adding an additional variable in the upper sheet of the $3$-hyperboloid $\TH_+$, interpreted as a timelike normal vector of tetrahedra. This served as an auxiliary variable (it drops from the dynamical amplitudes, but not from boundary states) allowing for a covariant and commuting imposition of simplicity and closure constraints.

\begin{figure}[htp]
\centering
\includegraphics[width=.3\textwidth]{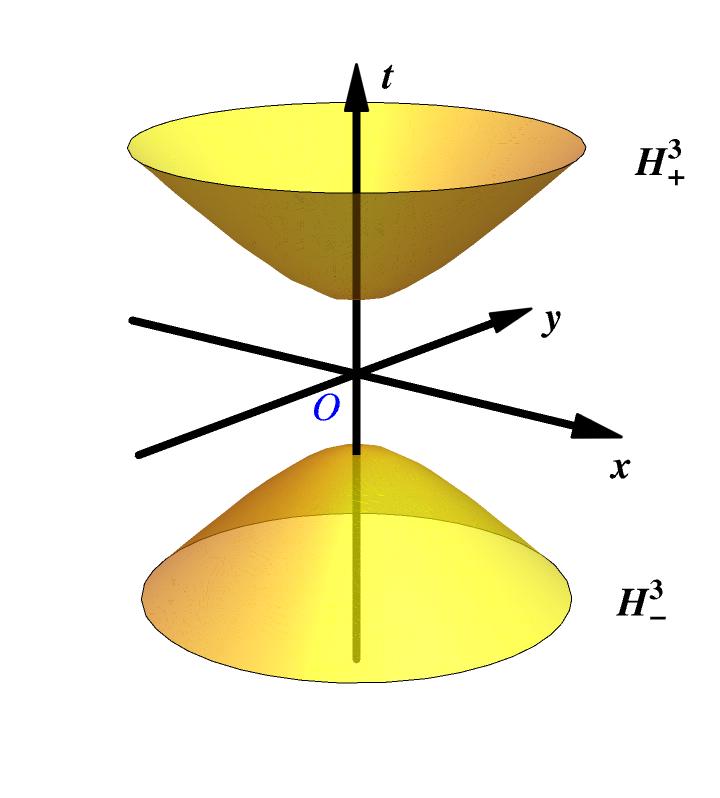}\hfill
\includegraphics[width=.3\textwidth]{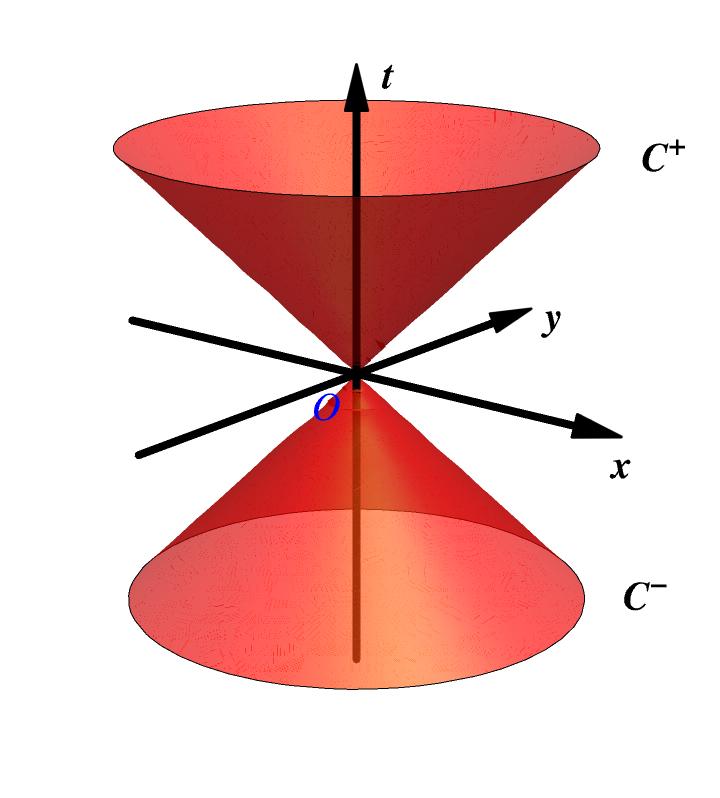}\hfill
\raisebox{+0.4cm}{\includegraphics[width=.3\textwidth]{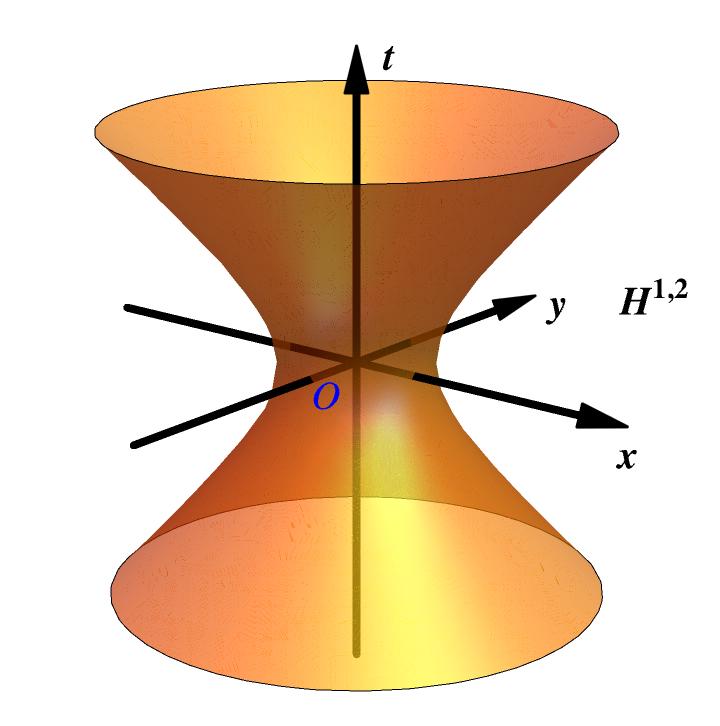}}\hfill
\caption{Distinguished hypersurfaces in Minkowski space from the perspective of a chosen observer at $O$. From left to right: The two-sheeted hyperboloid (defined via $y^{\mu}y_{\mu}=1$, with $y\in \mathbb{R}^{1,3}$), the lightcone (defined via $y^{\mu}y_{\mu}=0$) and the one-sheeted hyperboloid (defined via $y^{\mu}y_{\mu}=-1$). Note that the z-axis is suppressed to give three-dimensional pictures and that the so-called skirt radius of the hyperboloids is set to unity in this work.}
\label{fig:hypersurfaces}
\end{figure}

The Feynman diagrams of this model are dual to simplicial compelxes made only of spacelike tetrahedra. A natural way to remove this restriction is to consider additional normal vectors with lightlike and spacelike signature\footnote{We use the signature convention $(+,-,-,-)$, so that the square of timelike, lightlike and spacelike Minkowski vectors is positive, zero and negative, respectively.}, respectively. This is indicated in Table~\ref{tab:extensions}, where $\OH$ is the one-sheeted hyperboloid and $\mathrm{C}^{+}$ is the upper light cone.\footnote{As we are going to discuss in Sections~\ref{subsec:Bivector Variables and Geometric Interpretation} and~\ref{subsec:Spacetime Orientation}, the choice between upper or lower parts of $\TH_{\pm}$ and $\mathrm{C}^{\pm}$ is irrelevant for the construction of the model. We restrict our attention to timelike normal vectors lying in the upper $3$-hyperboloid and to lightlike vectors lying in the upper light cone.} These distinguished hypersurfaces in Minkowski space are depicted in Fig.~\ref{fig:hypersurfaces}.
\begin{table}[h]
\centering
\begin{tabular}{c c c}
timelike normal & $\varphi(g_v)\rightarrow \varphi(g_v;X_+)$ & $X_+\in\TH_{\pm}$\\[7pt]
lightlike normal & $\varphi(g_v)\rightarrow \varphi(g_v;X_0)$ & $X_0\in \mathrm{C}^{\pm}$\\[7pt]
spacelike normal & $\varphi(g_v)\rightarrow \varphi(g_v;X_-)$ & $\;\;\; X_-\in\OH$
\end{tabular}
\caption{Extension of the GFT field to normal vectors $X_{+},~X_{0}$ and $X_{-}$ carrying the three different possible signatures in Minkowski space.}
\label{tab:extensions}
\end{table}
As detailed in Section~\ref{subsec:Bivector Variables and Geometric Interpretation}, the normal vectors are understood as vectors in Minkowski space orthogonal to the associated tetrahedra, and they are realized as elements in quotient spaces of $\SL$ with respect to different subgroups. Let $\alpha\in\{+,0,-\}$ and denote by $\mathrm{U}^{(\alpha)}$ the $\SL$-subgroup that stabilizes the normal vectors
\begin{equation}\label{eq:reference normal vectors}
X_+ = (1,0,0,0),\quad
X_0 = \frac{1}{\sqrt{2}}(1,0,0,1),\quad
X_- = (0,0,0,1).
\end{equation}
For the timelike and spacelike case, we clearly have the isomorphisms $\mathrm{U}^{(+)}\cong \SUT$ and $\mathrm{U}^{(-)}\cong \SUO$. As shown e.g. in~\cite{Ruehl1970}, the stabilizer subgroup of $\frac{1}{\sqrt{2}}(1,0,0,1)$ is isomorphic to $\ISO$, the group of isometries acting on the Euclidean plane. 

The two conditions imposed on the GFT field, extended closure (or covariance) and simplicity, are also naturally extended to general normal vectors. They are summarized by the following two  equations
\begin{align}
\varphi(g_v;X_{\alpha}) &= \varphi(g_v h^{-1};h\cdot X_{\alpha}),\quad\forall h\in\SL,\label{eq:extended closure}\\[7pt]
\varphi(g_v;X_{\alpha}) &= \varphi(g_v u_v;X_{\alpha}),\quad \forall u_1,...,u_4\in \text{U}_{X_{\alpha}},\label{eq:simplicity}
\end{align}
where the transitive action of $\SL$ on quotient spaces $\SL/\mathrm{U}^{(\alpha)}$ as well as the stabilizer subgroups $\mathrm{U}_{X_{\alpha}}$ are defined in Appendix~\ref{subsec:Action of SL2C on Homogeneous Spaces}. Using the normal vectors, the conditions Eqs.~\eqref{eq:extended closure} and~\eqref{eq:simplicity} are imposed in a covariant and commuting fashion, and can be joined in a single \enquote{geometricity projector}. As a consequence, one has a unique definition of the model, solving ambiguities of earlier BC GFT formulations~\cite{Rovelli:2004tv,Perez:2000ec,Perez:2000ep}.

\subsection{Definition of the complete model}\label{subsec:Definition of the Complete Model}

Given the above properties of $\varphi(g_v;X_{\alpha})$, the action of the complete Barrett-Crane model is given by the sum of a kinetic and interaction part
\begin{equation}
    S[\varphi,\bar{\varphi}] = K[\varphi,\bar{\varphi}] + V[\varphi,\bar{\varphi}],
\end{equation}
which we specify in the following two subsections. We leave as understood a straightforward regularization of divergent volume factors vol$\left(\SL\right)$, vol$\left(\mathrm{U}^{(0)}\right)$ and vol$\left(\mathrm{U}^{(-)}\right)$, coming from trivial redundancies in the action, not carrying physical information, see also~\cite{Jercher:2021bie} for an exemplary discussion of this matter.

\subsubsection{Kinetic term}\label{subsubsec:Kinetic Action}

\begin{figure}[h!]
    \centering
    \includegraphics[scale = 0.35]{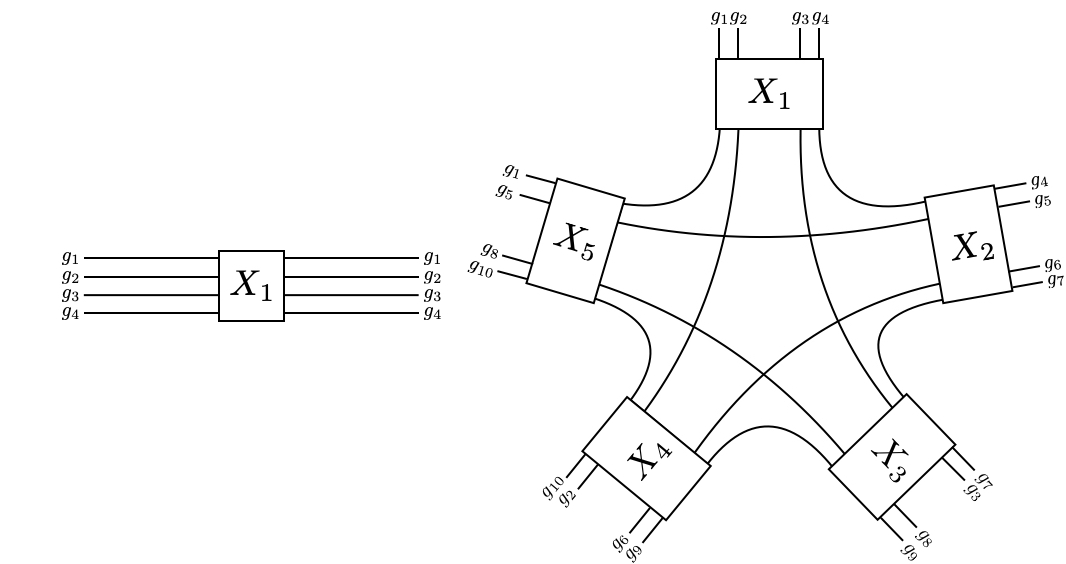}
    \caption{Left panel: A pictorial representation of the kinetic kernel, which accounts for the glueing of two tetrahedra by identifying their normal vectors, where we suppressed the signature of the normal vector. Right panel: The combinatorial structure of a $4$-simplex, formed by glueing five tetrahedra with normal vectors $X_1,...,X_5$ along ten faces. Again, we suppressed the signature of the five normal vectors.}
    \label{fig:kernels}
\end{figure}

As a straightforward generalization of the one in~\cite{Jercher:2021bie}, the kinetic term is defined as
\begin{equation}\label{eq:kinetic action}
K[\varphi,\bar{\varphi}]
=
\sum_{\alpha}\int\limits_{\SL^4}\left[\dd{g}\right]^4\int\limits_{\SL/\mathrm{U}^{(\alpha)}}\dd{X_{\alpha}}\bar{\varphi}(g_v;X_{\alpha})\varphi(g_v;X_{\alpha}),
\end{equation}
thus being a sum of contributions for each normal signature. A pictorial representation is presented in the left panel of Fig.~\ref{fig:kernels}.

Note that one could also introduce a more general kinetic kernel with the same structure as Eq.~\eqref{eq:kinetic action} but with a different mass coupling for each signature. Suggested by renormalization group analyses of related TGFT models, another possibility is that the kinetic kernel contains a differential (e.g. Laplacian) operator acting on the group domain~\cite{BenGeloun:2013mgx}\footnote{This kind of term is crucial for the renormalization flow of TGFT models and it plays also an important role to drive configurations in GFT condensate cosmology towards \enquote{dynamical isotropization}~\cite{Oriti:2016qtz,Pithis:2016wzf,Pithis:2016cxg,Gielen:2016uft,Pithis:2019tvp,Jercher:2021bie}}. Moreover, for the later construction of a CDT-like GFT model in Section~\ref{subsec:CDT and the complete BC Model}, a generalization of the kinetic kernel is desirable as it facilitates the implementation of the \enquote{dual weighting} discussed therein. In its most general form, the kinetic term is given by
\begin{equation}
K[\varphi,\bar{\varphi}]
=
\sum_{\alpha,\beta}\int\left[\dd{g}\right]^8\int\dd{X_{\alpha}}\dd{X_{\beta}}\bar{\varphi}(g_v;X_{\alpha})\mathcal{K}_{\alpha\beta}(g_v,g_w;X_{\alpha},X_{\beta})\varphi(g_w;X_{\beta}).
\end{equation}
Respecting the two conditions given in Eqs.~\eqref{eq:extended closure} and~\eqref{eq:simplicity}, the general kinetic term reduces to
\begin{equation}\label{eq:generalized kinetic action}
K[\varphi,\bar{\varphi}]
=
\sum_{\alpha}\int\left[\dd{g}\right]^4\int\dd{X_{\alpha}}\bar{\varphi}(g_v;X_{\alpha})\mathcal{K}_{\alpha}(g_v;X_{\alpha})\varphi(g_v;X_{\alpha}),
\end{equation}
where $\mathcal{K}_{\alpha}(g_v;X_{\alpha})$ can be any well-behaved function on the constrained domain.

\subsubsection{Vertex term}

For the vertex term $V[\varphi,\bar{\varphi}]$, we pose a priori no conditions on the glueing of tetrahedra to form spacetime $4$-simplices. In particular, we do not require that the signatures of normal vectors match inside a single $4$-simplex. In order to include all combinations of spacetime building blocks, we begin with the vertex term
\begin{equation}
\begin{aligned}
V[\varphi,\bar{\varphi}]
& =
\int\left[\dd{g}\right]^{10}\sum_{\alpha_1...\alpha_5}\int\dd{X_{\alpha_1}}...\int\dd{X_{\alpha_5}}\varphi_{1234}(X_{\alpha_1})\varphi_{4567}(X_{\alpha_2})\times\\[7pt]
&\times 
\varphi_{7389}(X_{\alpha_3})\varphi_{962(10)}(X_{\alpha_4})\varphi_{(10)851}(X_{\alpha_5})+\text{c.c.},
\end{aligned}
\end{equation}
summing up to $3^5 = 243$ different terms, where we introduced the short-hand notation  $\varphi_{1234}(X_{\alpha}) \equiv \varphi(g_1,g_2,g_3,g_4;X_{\alpha})$. Out of this large number of interactions, we end up with only $21$ physically distinct terms since we can reorder group labels so that only the number of timelike, lightlike and spacelike normal fields, $(n_+,n_0,n_-)$, is decisive. Consequently, we write the full vertex term as a sum
\begin{equation}\label{eq:full vertex action}
V[\varphi,\bar{\varphi}] = \sum_{\substack{ n_+,n_0,n_- \geq 0 \\ n_+ + n_0 + n_- = 5}} \frac{\lambda^{(n_+,n_0,n_-)}}{5}V(n_+,n_0,n_-),
\end{equation}
where for a specific combination of $(n_+,n_0,n_-)$, we have
\begin{equation}\label{eq:single vertex action}
\begin{aligned}
& V(n_+,n_0,n_-) =\\[7pt]
=& \int\left[\dd{g}\right]^{10}\int\left[\dd{X_{\alpha}}\right]^5\varphi_{1234}(X_{\alpha_1})\varphi_{4567}(X_{\alpha_2})\varphi_{7389}(X_{\alpha_3})\varphi_{962(10)}(X_{\alpha_4})\varphi_{(10)851}(X_{\alpha_5})+\\[7pt]
+& \text{c.c.}
\end{aligned}
\end{equation}
A pictorial representation of a the vertex term is given in the right panel of Fig.~\ref{fig:kernels}, which shows the combinatorial pattern encoding the glueing of five tetrahedra to form a spacetime $4$-simplex. Two remarks are in order: First, each of the $21$ distinct terms is a priori introduced by an independent coupling constant $\lambda^{(n_+,n_0,n_-)}$. Second, we observe that the normal vectors are integrated over separately, since we do not expect them to carry geometric information but merely serve as auxiliary variables which allow for a covariant and commutative imposition of constraints.

From the action, defined by Eqs.~\eqref{eq:kinetic action} and~\eqref{eq:full vertex action}, the equations of motion can be derived via a variation with respect to the fields $\bar{\varphi}(g_1,g_2,g_3,g_4;X_{\alpha})$\footnote{Although considered as auxiliary variables, the normal vectors enter the GFT fields as arguments and are therefore present in the equations of motion. If needed, the dependence on the normal vector can be eliminated after variation by (gauge-)fixing it to a certain value $X_0$ (using the closure/covariance condition) or by integrating over it, similar to what is done in the interaction term in Eq.~\eqref{eq:single vertex action}.} by applying
\begin{equation}
\fdv{\bar{\varphi}(g_1,g_2,g_3,g_4;X_{\alpha})}{\bar{\varphi}(h_1,h_2,h_3,h_4;Y_{\beta})}
=
\prod_{i=1}^4\delta(g_ih_i^{-1})\delta_{\alpha\beta}\delta(X_{\alpha},Y_{\beta}),
\end{equation}
which clarifies the variational principle. Notice that varying with respect to a GFT field with given normal vector picks out a particular signature, encoded in the Kronecker delta, $\delta_{\alpha\beta}$. If the signatures match, i.e. $\alpha = \beta$, then $\delta(X_{\alpha},Y_{\alpha})$ denotes the $\delta$-function on the space $\SL/\mathrm{U}^{(\alpha)}$ which are defined in Appendix~\ref{sec:Integral Geometry} for all values of $\alpha$.

Summarizing, the full action of the complete Barrett-Crane model is given by the sum of Eqs.~\eqref{eq:kinetic action} and~\eqref{eq:full vertex action}, taking into account every possible combination of five tetrahedra forming a spacetime $4$-simplex. A clear geometric understanding of the variables that enter the above defined theory, together with conclusions that can already be drawn on the classical level, are given in the following.

\subsection{Bivector variables and geometric interpretation}\label{subsec:Bivector Variables and Geometric Interpretation}

In this subsection, we perform a change of GFT variables to bivectors (i.e. Lie algebra variables), where the geometric interpretation is more transparent. Thereupon, we present and interpret classical conditions on these bivectors, arising from the simplicity constraint with respect to normal vectors of different signatures. These conditions will later serve as a comparison for restrictions on the quantum level which are derived by representation theoretic computations in Section~\ref{sec:Explicit Expressions for the Vertex Amplitudes}. 

Utilizing the non-commutative Fourier transform on Lie groups~\cite{Baratin:2010wi,Guedes:2013vi}, which has been explicitly formulated for $\SL$ in Ref.~\cite{Oriti:2018bwr}, we change from group variables $g_i\in \SL$ to Lie algebra variables $B_i \in\spl$
\begin{equation}
\Tilde{\varphi}_{X_{\alpha}}(B_1, B_2, B_3, B_4) = \int\left[\dd{g}\right]^4 \varphi(g_1,g_2,g_3,g_4;X_{\alpha})\prod_{i=1}^4 e_{g_i}(B_i),
\end{equation}
where $e_{g_i}(B_i)$ are the non-commutative generalization of plane waves. Due to the vector space-isomorphism $\spl \cong_{\mathrm{vs}}\R^{1,3}\wedge \R^{1,3}$, realized by relating the generators of $\SL$, denoted as $L^a$ and $K^a$, with bivector components $B^{AB}$ 
\begin{equation}\label{eq:SL2C generators}
L^a \defeq \frac{1}{2}\tensor{\varepsilon}{^a_{bc}}B^{bc},\quad K^a\defeq B^{0a},
\end{equation}
the Lie algebra variables are interpreted as bivectors, associated to triangles of a tetrahedron. The areas of triangles are determined as
\begin{equation}
A = \sqrt{\abs{B\cdot B}} = \sqrt{\frac{1}{2}\abs{B^{AB}B_{AB}}}.
\end{equation}
In the following, we discuss the the translation of the two constraints in Eqs.~\eqref{eq:extended closure} and~\eqref{eq:simplicity} to constraints of the bivector variables, with particular attention paid towards the simplicity constraint which encodes the embedding of tetrahedra into the plane orthogonal to $X_{\alpha}$, and ensures the geometric interpretation of the same variables. 

Upon integration over the normal vector $X_{\alpha}$, the extended closure/covariance condition Eq.~\eqref{eq:extended closure} corresponds to closing bivectors,
\begin{equation}\label{eq:closure of bivectors}
\sum_{i=1}^4 B_i = 0,
\end{equation}
which implies the closing of the associated triangles to form a tetrahedron. This condition does not depend on the signature of $X_{\alpha}$, as it can be seen from Eq.~\eqref{eq:extended closure}.

Simplicity with respect to $X_{\alpha}$, as defined by Eq.~\eqref{eq:simplicity}, imposes the linear simplicity constraint on bivectors
\begin{equation}\label{eq:linear simplicity constraint}
X_{\alpha}^A(*B)_{AB} = 0,
\end{equation}
as shown in~\cite{Baratin:2011tx} for Euclidean signature.\footnote{$*$ denotes the Hodge star operator, acting on the space of Minkowskian bivectors.} The proof can be straightforwardly extended to the Lorentzian case by utilizing the properties of the non-commutative Fourier transform on $\SL$. 

Independent of the signature $\alpha$, solutions of Eq.~\eqref{eq:linear simplicity constraint} are given by simple bivectors, i.e. by bivectors of the form $B = E_1\wedge E_2$, for some $E_1,E_2\in\R^{1,3}$. In contrast, the signature of bivectors, and thus signature of the associated triangles, is sensitive to the signature of the normal vector. We call a triangle timelike, lightlike or spacelike if $B\cdot B$ is positive, zero or negative, respectively. Table~\ref{tab:bivectors} shows the resulting form of the bivectors and their signature if simplicity is imposed with respect to the normal vectors of Eq.~\eqref{eq:reference normal vectors}, a derivation of which is presented in Appendix~\ref{sec:Classical Computations of Simplicity Constraints and Bivector Signatures}. We extract from Table~\ref{tab:bivectors} that triangles forming a spacelike tetrahedron are exclusively spacelike. We also observe that lightlike tetrahedra contain triangles which are either lightlike or spacelike. As pointed out in~\cite{Speziale:2013ifa}, the case of a lightlike bivector with a lightlike normal is degenerate, since the associated edge vectors would be parallel and therefore cannot span a triangle. Also, for a spacelike normal vector, the signature of bivectors is not constrained and hence, the associated triangles can either be timelike, spacelike or lightlike, where the latter case is not degenerate in the above sense. Finally, we note that the simplicity constraints are by construction insensitive to the orientation of the normal vector, meaning that the choice of the upper or lower parts of the $3$-hyperboloid and the light cone is irrelevant, a point we will discuss further in Section~\ref{subsec:Spacetime Orientation}. 

\begin{table}[h]
\centering
\begin{tabular}{c | c c c }
  $X$  & $(\pm 1,0,0,0)$ & $\frac{1}{\sqrt{2}}(\pm 1,0,0,1)$ & $(0,0,0,1)$  \\[7pt]
  \hline\\
  $\vb*{L}$ & $(0,0,0)$ & $(\mp B^{02},\pm B^{01},0)$ & $(B^{23},B^{31},0)$ \\[7pt]
  $\vb*{K}$ & $(B^{01},B^{02},B^{03})$ & $(B^{01},B^{02},B^{03})$ & $(0,0,B^{03})$\\[7pt]
  $B\cdot B$ & $< 0$ & $\leq 0$ & $\gtrless 0$.
\end{tabular}
\caption{An overview of the bivector components and their signature after imposing simplicity with respect to different normal vectors.}
\label{tab:bivectors}
\end{table}

Notice that for lightlike and timelike tetrahedra the signature of faces does not uniquely determine the signature of the contained edges. A mismatch of edge signatures for glued triangles however is excluded by the fact that bivectors of triangles are identified via (non-commutative) $\delta$-functions in the Lie algebra representation of the action in Eqs.~\eqref{eq:kinetic action} and~\eqref{eq:full vertex action}. 

In the following Sections~\ref{subsec:Spin Representation of the Group Field and its Action} and~\ref{sec:Explicit Expressions for the Vertex Amplitudes}, we derive the spin representation of the GFT field with all three different signatures, expanding it in terms of $\SL$-representations $(\rho,\nu)\in\R\times\mathbb{Z}/2$. This allows for a comparison with the classical bivector considerations that are summarized in Table~\ref{tab:bivectors}. In Section~\ref{sec:Explicit Expressions for the Vertex Amplitudes} we will comment on the case of bivectors satisfying $B\cdot B = 0$ in the quantum theory.

\subsection{Spin representation of the GFT field and action}\label{subsec:Spin Representation of the Group Field and its Action}

To derive the spin representation of the GFT action, we first expand the GFT fields in terms of representation labels, taking into account closure and simplicity conditions~\eqref{eq:extended closure} and~\eqref{eq:simplicity}. That is, we generalize the derivation in the appendix of~\cite{Jercher:2021bie} to the case of the normal vector being either timelike, lightlike or spacelike. 

Let us define a (pseudo-)projector $P_{\alpha}^{(\rho,\nu)}$ from the $\SL$-representation space $\mathcal{D}^{(\rho,\nu)}$ onto the $\mathrm{U}^{(\alpha)}$-invariant subspace as
\begin{equation}
P_{\alpha}^{(\rho,\nu)} \defeq \int\limits_{\mathrm{U}^{(\alpha)}}\dd{u}\vb*{D}^{(\rho,\nu)}(u),
\end{equation}
where $\vb*{D}^{(\rho,\nu)}(g)$ is an $\SL$-Wigner matrix.  Defining $\vert\mathcal{I}^{(\rho,\nu),\alpha}\rangle$ as an $\mathrm{U}^{(\alpha)}$-invariant vector in $\mathcal{D}^{(\rho,\nu)}$,
\begin{equation}
\vb*{D}^{(\rho,\nu)}(u)\vert\mathcal{I}^{(\rho,\nu),\alpha}\rangle = \vert\mathcal{I}^{(\rho,\nu),\alpha}\rangle,\quad \forall u\in \mathrm{U}^{(\alpha)}, 
\end{equation}
the projector is conveniently rewritten as
\begin{equation}
P^{(\rho,\nu)}_{\alpha} = \vert\mathcal{I}^{(\rho,\nu),\alpha}\rangle\langle\mathcal{I}^{(\rho,\nu),\alpha}\vert,
\end{equation}
with matrix coefficients in the canonical basis given by\footnote{The symbols $\mathcal{I}^{(\rho,\nu),\alpha}_{jm}$ are equal to the $\mathcal{W}$-symbols of Ref.~\cite{Perez:2000ep} for the case of $\alpha = +,-$ and they represent an extension of the work done in~\cite{Perez:2000ep} since also lightlike normal vectors are taken into account for $\alpha = 0$.}\textsuperscript{,}\footnote{In this article, we work in the canonical basis of $\SL$-representations which are denoted in bra-ket notation as $\ket{(\rho,\nu);jm}$. The Wigner matrices are best under control in this case, e.g. with respect to the orthogonality relation \eqref{eq:orthogonality relation of SL2C wigner matrices} or the behaviour under complex conjugation in Eq.~\eqref{eq:complex conjugate of Wigner matrix}. Furthermore, the interaction term of mixed type will contain the convolution of Wigner matrices arising from different normal vectors. In different bases, for instance in the pseudo-basis for functions with $\alpha = -$, the coefficients for basis change would be required explicitly, and they are not available. The price we pay for using the canonical basis is that the evaluation of $\SUO$- and $\ISO$-elements on the Wigner matrices does not yield an immediate simplification as it is for the $\SUT$-case, where $D^{(\rho,\nu)}_{jmln}(u) = \delta^{jl}D^j_{mn}(u)$ holds.}
\begin{equation}
\bra{(\rho,\nu);jm}P^{(\rho,\nu)}_{\alpha}\ket{(\rho,\nu);ln} 
=
P_{jmln}^{(\rho,\nu),\alpha}
=
\bar{\mathcal{I}}^{(\rho,\nu),\alpha}_{jm}\mathcal{I}^{(\rho,\nu),\alpha}_{ln}.
\end{equation}

In order to get a first intuition about these symbols, we anticipate some of the results of Section~\ref{subsec:Kernels of Non-Mixed Type} here, which show that indeed, the invariant coefficients project onto simple representations. Without further specification at this point, we write
\begin{align}
    P_{jmln}^{(\rho,\nu),+} &=  
    \delta_{\nu,0}P^{\rho,+}_{jmln},\label{eq:I+}\\[7pt]
    P^{(\rho,\nu),0}_{jmln} &= \delta_{\nu,0}P^{\rho,0}_{jmln},\label{eq:I0}\\[7pt]
    P^{(\rho,\nu),-}_{jmln} &= \delta_{\nu,0}P^{\rho,-}_{jmln}+\delta(\rho)\delta_{\nu\in 2\mathbb{N}^+}P^{\nu,-}_{jmln},\label{eq:I-}
\end{align}
where we refer to Section~\ref{subsec:Kernels of Non-Mixed Type} for a derivation and to Appendix~\ref{sec:Integral Geometry} for the mathematical basis given by integral geometry. As we detail in Section~\ref{sec:Explicit Expressions for the Vertex Amplitudes}, Eqs.~\eqref{eq:I+} and~\eqref{eq:I0} imply that spacelike and lightlike tetrahedra only contain spacelike faces labelled by $\rho\in\R$. Timelike tetrahedra contain a mixture of spacelike and timelike faces, as Eq.~\eqref{eq:I-} shows, where the latter are labelled by $\nu\in 2\mathbb{N}^+$. Notice that, since $\SUO$ and $\ISO$ are non-compact, the self-contractions of $\mathcal{I}^{(\rho,\nu),0}$ and $\mathcal{I}^{(\rho,\nu),-}$ lead to divergent volume factors which we regularize if appearing in our computations.

Following the definition of the invariant coefficients $\mathcal{I}^{(\rho,\nu),\alpha}_{jm}$, together with the results in~\cite{Jercher:2021bie}, we find a basis for functions on $\SL^4\times\SL/\mathrm{U}^{(\alpha)}$ which satisfy extended closure and simplicity, given by
\begin{equation}\label{eq:basis functions}
\Psi^{\rho_1\nu_1\rho_2\nu_2\rho_3\nu_3\rho_4\nu_4,\alpha}_{j_1 m_1 j_2 m_2 j_3 m_3 j_4 m_4}(g_1,g_2,g_3,g_4;X_{\alpha})
=
\prod_{i=1}^4 \sum_{l_i n_i} D^{(\rho_i,\nu_i)}_{j_i m_i l_i n_i}(g_i g_{X_{\alpha}})\bar{\mathcal{I}}^{(\rho_i,\nu_i),\alpha}_{l_i n_i},
\end{equation}
where $g_{X_{\alpha}}\in\SL$ is a representative of the equivalence class of $X_{\alpha} = [g_X]_{\alpha}\in\SL/\mathrm{U}^{(\alpha)}$, making use of the quotient structure. It is easily checked that the proposed basis functions satisfy the three properties that we seek for, i.e. closure, simplicity and invariance under change of representative $g_{X_{\alpha}}\rightarrow g_{X_{\alpha}}u$. Consequently, a function $\varphi\in L^2\left(\SL^4\times\SL/\mathrm{U}^{(\alpha)}\middle/ \sim\right)$, where $\sim$ encodes the quotient structure due to geometricity, is expanded in terms of $\SL$-representation labels as
\begin{equation}\label{eq:spin representation of group field}
\varphi(g_v;X_{\alpha})
=
\left[\prod_{i=1}^4 \sum_{\nu_i}\int\dd{\rho_i}4\left(\rho_i^2+\nu_i^2\right)\sum_{j_i m_i l_i n_i}\right]\varphi^{\rho_i\nu_i,\alpha}_{j_i m_i}\prod_i D^{(\rho_i,\nu_i)}_{j_i m_i l_i n_i}(g_i g_{X_{\alpha}})\bar{\mathcal{I}}^{(\rho_i,\nu_i),\alpha}_{l_i n_i},
\end{equation}
where the factor $4\left(\rho_i^2+\nu_i^2\right)$ stems from the Plancherel measure of functions on $\SL$. If in addition the normal vector is integrated over, as in the case of the vertex term~\eqref{eq:single vertex action}, the GFT field expansion is given by 
\begin{equation}\label{eq:spin representation of group field after normal integration}
\begin{aligned}
& \int\dd{X_{\alpha}}\varphi(g_v;X_{\alpha}) = \\[7pt]
= &
\left[\prod_{i=1}^4 \sum_{\nu_i}\int\dd{\rho_i}4\left(\rho_i^2+\nu_i^2\right)\sum_{j_i m_i l_i n_i}\right]\varphi^{\rho_i\nu_i,\alpha}_{j_i m_i}B^{\rho_i\nu_i,\alpha}_{l_i n_i}\prod_i D^{(\rho_i,\nu_i)}_{j_i m_i l_i n_i}(g_i),
\end{aligned}
\end{equation}
where we defined the generalized Barrett-Crane intertwiners via
\begin{equation}\label{eq:definition of generalized BC intertwiners}
B^{\rho_i\nu_i,\alpha}_{j_i m_i} \defeq\int\limits_{\SL/\mathrm{U}^{(\alpha)}}\dd{X}_{\alpha}\prod_{i=1}^4\sum_{l_i n_i}D^{(\rho_i,\nu_i)}_{j_i m_i l_i n_i}(g_{X_\alpha})\bar{\mathcal{I}}^{(\rho_i,\nu_i),\alpha}_{l_i n_i}.
\end{equation}

\paragraph{Spin representation of the kinetic term}

The spin representation of the action, given in Section~\ref{subsec:Definition of the Complete Model}, is derived by inserting the GFT field expansion~\eqref{eq:spin representation of group field}, using the orthogonality relation of $\SL$-Wigner matrices in the canonical basis which are given in Eq.~\eqref{eq:orthogonality relation of SL2C wigner matrices}. We obtain
\begin{equation}\label{eq:kinetic action in spin rep}
K[\varphi,\bar{\varphi}]
=
\sum_{\alpha}\left[\prod_{i=1}^4\sum_{\nu_i}\int\dd{\rho_i}4\left(\rho_i^2+\nu_i^2\right)\sum_{j_i m_i}\right]\bar{\varphi}^{\rho_i\nu_i,\alpha}_{j_i m_i}\varphi^{\rho_i\nu_i,\alpha}_{j_i m_i}.
\end{equation}

If we choose to work with a more general kinetic kernel as discussed in Section~\ref{subsec:Definition of the Complete Model}, the spin representation of the kinetic term is given by
\begin{equation}
K[\varphi,\bar{\varphi}]
=
\sum_{\alpha}\left[\prod_{i=1}^4\sum_{\nu_i}\int\dd{\rho_i}4\left(\rho_i^2+\nu_i^2\right)\sum_{j_i m_i}\right]\bar{\varphi}^{\rho_i\nu_i,\alpha}_{j_i m_i}\mathcal{K}^{\rho_i\nu_i,\alpha}_{j_i m_i}\varphi^{\rho_i\nu_i,\alpha}_{j_i m_i},
\end{equation}
thus leading to different edge amplitudes depending on the precise choice of kinetic term. A graphical representation of the kinetic kernel is given in the left panel of Fig.~\ref{fig:kernels}.

\paragraph{Spin representation of the vertex term}

Since the normal vectors entering the vertex term~\eqref{eq:single vertex action} are integrated over separately, we apply the expansion of GFT fields after normal integration, given in Eq.~\eqref{eq:spin representation of group field after normal integration} which makes use of the generalized BC intertwiners defined in Eq.~\eqref{eq:definition of generalized BC intertwiners}. We obtain
\begin{equation}\label{eq:single vertex action in spin representation}
\begin{aligned}
& V(n_+,n_0,n_-) = \\[7pt]
=& \left[\prod_{a=1}^{10}\sum_{\nu_a}\int\dd{\rho_a}4\left(\rho_a^2+\nu_a^2\right)\sum_{j_a m_a l_a n_a}(-1)^{-j_a-m_a}\right]\varphi^{\rho_1\nu_1 \rho_2\nu_2 \rho_3\nu_3 \rho_4\nu_4}_{j_1 m_1 j_2 m_2 j_3 m_3 j_4 m_4}\times\\[7pt]
\times &
\varphi^{\rho_4\nu_4 \rho_5\nu_5 \rho_6\nu_6 \rho_7\nu_7}_{j_4 -m_4 j_5 m_5 j_6 m_6 j_7 m_7}\varphi^{\rho_7\nu_7 \rho_3\nu_3 \rho_8\nu_8 \rho_9\nu_9}_{j_7 -m_7 j_3 -m_3 j_8 m_8 j_9 m_9}\varphi^{\rho_9\nu_9 \rho_6\nu_6 \rho_2\nu_2 \rho_{10}\nu_{10}}_{j_9 -m_9 j_6 -m_6 j_2 -m_2 j_{10} m_{10}}\times\\[7pt]
\times &
\varphi^{\rho_{10}\nu_{10} \rho_8\nu_8 \rho_5\nu_5 \rho_1\nu_1}_{j_{10} -m_{10} j_8 -m_8 j_5 -m_5 j_1 -m_1}\{10(\rho,\nu)\}_{(\alpha_1,...,\alpha_5)},
\end{aligned}
\end{equation}
where $\{10(\rho,\nu)\}_{(\alpha_1,...,\alpha_5)}$ is the generalized Barrett-Crane $\{10\rho\}$-symbol, defined in terms of contractions of generalized BC intertwiners by
\begin{equation}\label{eq:generalized vertex intertwiner form}
\begin{aligned}
& \{10(\rho,\nu)\}_{(\alpha_1,...,\alpha_5)}
\defeq
\left(\prod_{a=1}^{10}\sum_{l_a n_a}(-1)^{l_a+n_a}\right)B^{\rho_1\nu_1\rho_2\nu_2\rho_3\nu_3\rho_4\nu_4,\alpha_1}_{l_1 n_1 l_2 n_2 l_3 n_3 l_4 n_4}\times\\[7pt]
\times &
B^{\rho_4\nu_4 \rho_5\nu_5\rho_6\nu_6\rho_7\nu_7,\alpha_2}_{l_4 -n_4 l_5 n_5 l_6 n_6 l_7 n_7}B^{\rho_7\nu_7\rho_3\nu_3\rho_8\nu_8\rho_9\nu_9,\alpha_3}_{l_7 -n_7 l_3 -n_3 l_8 n_8 l_9 n_9}\times\\[7pt]
\times &
B^{\rho_9\nu_9\rho_6\nu_6\rho_2\nu_2\rho_{10}\nu_{10},\alpha_4}_{l_9 -n_9 l_6 -n_6 l_2 -n_2 l_{10} n_{10}}B^{\rho_{10}\nu_{10}\rho_8\nu_8\rho_5\nu_5\rho_1\nu_1,\alpha_5}_{l_{10} -n_{10} l_8 -n_8 l_5 -n_5 l_1 -n_1}.
\end{aligned}
\end{equation}
A pictorial interpretation of the generalized $\{10(\rho,\nu)\}$-symbol is given in the right panel of Fig.~\ref{fig:kernels}, visualizing that it describes the glueing of five tetrahedra to form a spacetime $4$-simplex.

In spite of its formal clarity, Eq.~\eqref{eq:generalized vertex intertwiner form} is computationally hard to handle. For actual computations of the vertex amplitude as well as the interpretation thereof, the kernel representation of Eq.~\eqref{eq:generalized vertex intertwiner form}, which we derive in the following, is important. To that end, we define $\SL$-Wigner matrices which are contracted from both sides with invariant vectors $\mathcal{I}^{(\rho,\nu),\alpha}$
\begin{equation}\label{eq:definition D-function}
D^{(\rho,\nu)}_{\alpha_1\alpha_2}(g_{X_1}^{-1}g_{X_2})
\defeq
\sum_{jmln}\mathcal{I}^{(\rho,\nu),\alpha_1}_{jm}D^{(\rho,\nu)}_{jmln}(g_{X_1}^{-1}g_{X_2})\bar{\mathcal{I}}^{(\rho,\nu),\alpha_2}_{ln},
\end{equation}
Due to the contraction with the $\mathcal{I}^{(\rho,\nu),\alpha}$, these functions effectively depend on normal vectors $X_i\in\SL/\mathrm{U}^{(\alpha_i)}$ only, for which the $g_{X_1}$ and $g_{X_2}$ are representatives. Physically, we interpret Eq.~\eqref{eq:definition D-function} as encoding the glueing of two tetrahedra with normal vectors $X_i \in \SL/\mathrm{U}^{(\alpha_i)}$ along a common face. The effective dependence on normal vectors motivates a notation similar to that of Refs.~\cite{Barrett:1999qw,Perez:2000ep}, defining the kernels 
\begin{equation}\label{eq:definition of K}
K_{\alpha_1\alpha_2}^{(\rho,\nu)}(X_1,X_2)\defeq D_{\alpha_1\alpha_2}^{(\rho,\nu)}(g_{X_1}^{-1}g_{X_2}).
\end{equation}
Notice that Eqs.~\eqref{eq:definition D-function} and~\eqref{eq:definition of K} constitute a natural generalization of the $D^{(\rho,0)}_{0000}(g_X^{-1}g_Y)$-characters used in the Barrett-Crane model with timelike normals, which can be equivalently represented as $\delta_{\nu,0}K^+_{\rho}(X,Y)$ in the notation of~\cite{Perez:2000ec}. 

An integral form of Equation~\eqref{eq:generalized vertex intertwiner form} can straightforwardly be obtained by writing the generalized BC intertwiners as integrals over normal vectors via Eq.~\eqref{eq:definition of generalized BC intertwiners}. Carrying out the contraction of the indices $(l_i n_i)$ and making use of Eqs.~\eqref{eq:definition D-function} and~\eqref{eq:definition of K}, we obtain
\begin{equation}\label{eq:generalized vertex integral form}
\{10(\rho,\nu)\}_{(\alpha_1,...,\alpha_5)}
=
\int\left[\dd{X_{\alpha}}\right]^5 \prod_{a<b}K_{\alpha_a\alpha_b}^{(\rho_{ab},\nu_{ab})}(X_a,X_b).
\end{equation}
As suggested in~\cite{Barrett:1999qw}, the regularization of the $\{10(\rho,\nu)\}$-symbol is simply achieved by fixing the vector $X_1$ and dropping its integration.

Finally, we remark that the form of the $\{10(\rho,\nu)\}$-symbol given in Eq.~\eqref{eq:generalized vertex integral form} is most convenient as it allows for the application of integral geometry methods, developed in Appendix~\ref{sec:Integral Geometry}. Since the generalized BC intertwiners still carry uncontracted magnetic indices, a clear relation to the $D$-functions of Eq.~\eqref{eq:definition D-function} is obscured. This is also the reason why the generalized BC intertwiners are defined in an implicit way here, while the $D$-functions take an explicit form, as the computations in Section~\ref{sec:Explicit Expressions for the Vertex Amplitudes} show.

\subsection{Spin foam model formulation}\label{subsec:Spin Foam Model Formulation}

With the spin representation results of the previous section, we can give the spin foam expression for the amplitudes of the complete model. These are associated to oriented $2$-complexes, dual to the Feynman diagrams of the GFT model, and denoted as $\Delta^*$, where the orientation is obtained by giving the edges $e$ a direction. $\Delta^*$ is taken to be the $2$-skeleton dual to a triangulation $\Delta$ of an oriented Lorentzian manifold. Links, triangles, tetrahedra and $4$-simplices in $\Delta$ are denoted by $l, t, \tau$ and $\sigma$, respectively, while $f,e$ and $v$ denote respectively the faces, edges and vertices of $\Delta^*$. 

Encoding the bare causal structure, tetrahedra (dual edges) carry the signature of the associated normal vector and triangles (dual faces) are either spacelike or timelike~\cite{Conrady:2010kc}. As the results of Section~\ref{sec:Explicit Expressions for the Vertex Amplitudes} and Appendix~\ref{sec:Integral Geometry} show, triangles contained in spacelike and lightlike tetrahedra are labelled by representations $(\rho,0)$, while by contrast, triangles inside timelike tetrahedra are either labelled by representations $(\rho,0)$ or $(0,\nu)$, depending on whether they are spacelike or timelike, respectively. Thus, the signatures of faces $f$ pick out either the representation $(\rho_f,0)$ or $(0,\nu_f)$.  

Finally, the spin foam amplitude for a Lorentzian $2$-complex $\Delta^*$ is defined by
\begin{equation}\label{eq:spin foam amplitude}
\mathcal{A}(\Delta^*)
=
\prod_{f}\sum_{\nu_f}\int\dd{\rho_f} \mathcal{A}_f(\rho_f,\nu_f)\prod_{e}\mathcal{A}_{e}^{\alpha_e}(\rho_{i_e},\nu_{i_e})\prod_{v}\mathcal{A}_v^{\alpha_1,...,\alpha_5}(\rho_{v_a},\nu_{v_a}),
\end{equation}
where the face amplitude $\mathcal{A}_f$ is given by the Plancherel measure
\begin{equation}
\mathcal{A}_f(\rho_f,\nu_f) = 4\left(\rho_f^2+\nu_f^2\right).
\end{equation}
Based on the kinetic and vertex actions in Eqs.~\eqref{eq:kinetic action in spin rep} and~\eqref{eq:single vertex action in spin representation}, the edge and vertex amplitudes are respectively given by
\begin{equation}
\mathcal{A}^{\alpha_e}_e(\rho_{i_e},\nu_{i_e})
=
1
\end{equation}
and
\begin{equation}
\mathcal{A}_v^{\alpha_1,...,\alpha_5}(\rho_{v_a},\nu_{v_a})
=
\{10(\rho_{v_a},\nu_{v_a})\}_{(\alpha_1,...,\alpha_5)}.
\end{equation}

The boundary states of the complete BC model are given by so-called simple spin networks, treated in~\cite{Freidel:1999jf,Barrett:1999qw,Baez:2001fh} for spacelike hypersurfaces and in~\cite{Alexandrov:2005ar} for timelike hypersurfaces. In the generalized setting we introduced here, a basis for spin networks  $s^{(\rho_e,\nu_e)}(g_e,X_v)$, where $e$ are the edges and $v$ the vertices, is obtained by a product of the kernels defined in Eq.~\eqref{eq:definition of K}~\cite{Livine:2002ak}
\begin{equation}
s^{(\rho_e,\nu_e)}(g_e,X_v) = \prod_e K^{(\rho_e,\nu_e)}_{\alpha_{s(e)}\alpha_{t(e)}}(X_{s(e)},g_e\cdot X_{t(e)}),
\end{equation}
where $s(e)$ and $t(e)$ are the starting and terminal vertices of the edge $e$, respectively.

The $2$-complex, to which the amplitudes are associated, can be introduced \enquote{by hand} in a spin foam context, and thus simply chosen to be topologically non-degenerate and oriented. However, when derived within a GFT model (which provides an unambiguous prescription for all its elements), arbitrary glueings of simplicial building blocks are generated by the GFT perturbative expansion, many of which are topologically singular. This issue has been tackled and solved in the context of generic TGFTs (see for example \cite{GurauBook}), using a colored extension of the models. To ensure that the $2$-complex is dual to a simplicial complex and orientable, we introduce a colored version of the complete BC model in the following section.

\subsection{Colored Lorentzian Barrett-Crane model}\label{subsec:colored Lorentzian BC model}

It is shown in~\cite{Gurau:2010nd} that TGFTs of rank larger than two generate Feynman diagrams which define singular topological spaces not even corresponding to triangulations of pseudo-manifolds. Coloring TGFTs resolves this issue and the generated colored Feynman diagrams are always dual to topological pseudo-manifolds.\footnote{As pointed out in~\cite{Gurau:2010ba}, coloring also allows for a $1/N$ expansion of TGFTs, in turn a key ingredient for renormalization~\cite{Gurau:2010ba,Gurau:2011aq,Gurau:2011xq,Bonzom:2012hw,Gurau:2013pca,Gurau:2012vk}.}

Our colored model is obtained by attaching labels $c \in\{0,1,2,3,4\}$ to the GFT fields
\begin{equation}
\varphi(g_v;X_{\alpha})\longrightarrow \varphi^c(g_v;X_{\alpha}).
\end{equation}
The generalized kinetic term of Eq.~\eqref{eq:generalized kinetic action} is thus modified to
\begin{equation}\label{eq:colored kinetic action}
K_{\mathrm{col}}[\varphi,\bar{\varphi}]
=
\sum_{c=0}^4\sum_{\alpha}\int\limits_{\SL^4}\left[\dd{g}\right]^4\int\limits_{\SL/\mathrm{U}^{(\alpha)}}\dd{X_{\alpha}}\bar{\varphi}^c(g_v;X_{\alpha})\mathcal{K}_{\alpha}(g_v;X_{\alpha})\varphi^c(g_v;X_{\alpha}),
\end{equation}
which implies topologically that only tetrahedra of the same color are identified. The colored vertex term is 
\begin{equation}\label{eq:colored vertex action}
\begin{aligned}
& V_{\mathrm{col}}(n_+,n_0,n_-) =\\[7pt]
=& \int\left[\dd{g}\right]^{10}\int\left[\dd{X_{\alpha}}\right]^5\sum_{\sigma}\varphi^{\sigma(0)}_{1234}(X_{\alpha_1})\varphi^{\sigma(1)}_{4567}(X_{\alpha_2})\varphi^{\sigma(2)}_{7389}(X_{\alpha_3})\varphi^{\sigma(3)}_{962(10)}(X_{\alpha_4})\varphi^{\sigma(4)}_{(10)851}(X_{\alpha_5})+\\[7pt]
+& \text{c.c.},
\end{aligned}
\end{equation}
where the sum is over all cyclic permutations of colors. We neglect factors that arise from redundancies of the sum over colors, such as a for instance a factor $5$ in front of interactions with five normals of the same type.

Notice that by using complex fields $\varphi^c:\SL^4\longrightarrow\C$, the colored Feynman diagrams are bi-partite and as a consequence, the associated $2$-complexes are orientable~\cite{Caravelli:2010nh}. The orientability of the underlying simplicial complexes is a purely topological property that does not imply any specific property at the level of quantum amplitudes. In fact, except for the models introduced in~\cite{Livine:2002rh} and~\cite{Bianchi:2021ric}, all of the known spin foam models are insensitive to the bulk time-orientation. We further elaborate on this issue in Section~\ref{subsec:Spacetime Orientation}.

The signature of tetrahedra and faces pose restrictions on possible glueings. First, the kinetic term~\eqref{eq:kinetic action} only identifies tetrahedra of the same signature, and second, faces are only glued if their signatures match. 
Notice that, since there are three types of normal vector signatures and two types of face signatures, as the results of Section~\ref{sec:Explicit Expressions for the Vertex Amplitudes} show, these causal distinctions cannot be used themselves as a form of coloring of GFT fields. Moreover, clearly the signatures do not follow the same combinatorial patterns as colors. For instance, there are interaction terms where every GFT field has a timelike normal vector, while there are no interactions where every field carries the same color.

\section{Explicit expressions for the kernels of the vertex amplitudes}\label{sec:Explicit Expressions for the Vertex Amplitudes}

As shown in Section~\ref{subsec:Spin Representation of the Group Field and its Action}, the spin representation of the vertex term leads to contractions of five generalized Barrett-Crane intertwiners, yielding the $\{10(\rho,\nu)\}_{(\alpha_1,...,\alpha_5)}$-symbol in the form of Eq.~\eqref{eq:generalized vertex intertwiner form}. One can bring this symbol to the more convenient form of Eq.~\eqref{eq:generalized vertex integral form} by making use of the kernels defined in Eq.~\eqref{eq:definition of K}. In this section we derive explicit expressions for these kernels.

First, notice that due to the inclusion of all normal signatures, there are in total six independent types of kernels, corresponding to $K_{++},K_{--},K_{00},K_{+0},K_{+-}$ and $K_{-0}$, where for instance $K_{++}$ denotes the spacelike kernel that has two timelike normal vectors as arguments. The remaining three kernels $K_{0+}, K_{-+}$ and $K_{0-}$ are obtained by the symmetry relation
\begin{equation}\label{eq:symmetry of K}
K_{\alpha_1\alpha_2}(X_1,X_2) = \overline{K_{\alpha_2\alpha_1}(X_2,X_1)}.
\end{equation}
A second important property of the kernels is that they are invariant under simultaneous action of $\SL$
\begin{equation}\label{eq:SL2C-invariance of kernels}
K_{\alpha_1\alpha_2}^{(\rho,\nu)}(X_1,X_2) = K_{\alpha_1\alpha_2}^{(\rho,\nu)}(h\cdot X_1,h\cdot X_2),\quad \forall h\in\SL,
\end{equation}
which is apparent from  their definition~\eqref{eq:definition of K}. As a consequence of this invariance, the $K_{\alpha_1\alpha_2}(X_1,X_2)$ effectively depend on the Minkowski product of $X_1$ and $X_2$, for which one can therefore choose a convenient parametrization. 

Out of the six kernels, Ref.~\cite{Perez:2000ep} presented an explicit computation for $K_{++}$ and $K_{--}$, using integral geometry tools developed in~\cite{VilenkinBook}. After restating these results from~\cite{Perez:2000ep}, we extend the computations to include a purely lightlike kernel $K_{00}$ in Section~\ref{subsec:Kernels of Non-Mixed Type}, and then present all of the mixed cases in Section~\ref{subsec:Kernels of Mixed Type}.

\subsection{Kernels of non-mixed type}\label{subsec:Kernels of Non-Mixed Type}

Since all of the non-mixed type kernels are computed in a similar fashion, namely by comparing them with the $\delta$-function on the respective homogeneous space, it is instructive to consider first the representation expansion of the $\delta$-function on $\SL$
\begin{equation}\label{eq:spin rep of delta on SL2C}
\delta(g_1^{-1}g_2)
=
\sum_{\nu}\int\dd{\rho}4\left(\rho^2+\nu^2\right)\sum_{jm}\overline{D^{(\rho,\nu)}_{jmjm}(g_1^{-1}g_2)},
\end{equation}
where we emphasize that the Wigner matrix appears as complex conjugated so that the $\delta$-function acts correctly on functions on $\SL$. Imposing simplicity with respect to the group $\mathrm{U}^{(\alpha)}$ on the $\delta$-function effectively yields the $\delta$-function on the homogeneous space $\SL/\mathrm{U}^{(\alpha)}\ni X,Y$
\begin{equation}\label{eq:spin rep of delta on general homogeneous space}
\delta(X,Y)
=
\sum_{\nu}\int\dd{\rho}4\left(\rho^2+\nu^2\right)\overline{D_{\alpha\alpha}^{(\rho,\nu)}(g_X^{-1}g_Y}) = \sum_{\nu}\int\dd{\rho}4\left(\rho^2+\nu^2\right)\overline{K_{\alpha\alpha}^{(\rho,\nu)}(X,Y)},
\end{equation}
where $D_{\alpha\alpha}$ and $K_{\alpha\alpha}$ are defined in Eqs.~\eqref{eq:definition D-function} and~\eqref{eq:definition of K}, respectively. With the expansions of the $\delta$-functions on $\SL/\mathrm{U}^{(\alpha)}$ given in Appendix~\ref{sec:Integral Geometry} and derived in~\cite{VilenkinBook}, we can relate the kernels $K_{\alpha\alpha}$ with integral geometric expressions that can be explicitly computed.

\paragraph{Timelike kernel} 

Choosing $\alpha = +$ in Eq.~\eqref{eq:spin rep of delta on general homogeneous space} and comparing with Eq.~\eqref{eq:Gel'fand expansion of lightlike delta}, we obtain the expression 
\begin{equation}\label{eq:D_++ before evaluation}
K_{++}^{(\rho,\nu)}(X,Y)
=
\frac{\delta_{\nu,0}}{2}\int\dd{\Omega}\left(X^\mu\xi_{\mu}\right)^{i\rho-1}\left(Y^{\mu}\xi_{\mu}\right)^{-i\rho-1}.
\end{equation}
The factor of $\frac{1}{2}$ appears as a consequence of change of integration range $\rho\in(-\infty,\infty)\rightarrow \rho\in[0,\infty)$, using the unitary equivalence of representations. As a consequence of Eq.~\eqref{eq:SL2C-invariance of kernels} and the fact that $X\cdot Y$ only depends on the hyperbolic distance $\eta$ between $X,Y\in\TH_+$, we can choose the parametrization 
\begin{equation}
X = (1,0,0,0),\qquad Y = (\cosh(\eta),0,0,\sinh(\eta)),\qquad \eta\in\R,
\end{equation}
with which Eq.~\eqref{eq:D_++ before evaluation} is evaluated to
\begin{equation}
K_{++}^{\rho}(\eta)
=
\int\dd{\Omega}\left(\cosh(\eta)-\sinh(\eta)\cos(\theta)\right)^{-i\rho-1},
\end{equation}
where we have defined $K^{(\rho,\nu)}_{++} \equiv \frac{\delta_{\nu,0}}{2}K^\rho_{++}$. Carrying out the integration explicitly, we obtain~\cite{Perez:2000ep}
\begin{equation}\label{eq:D++}
K_{++}^{\rho}(\eta) = \frac{\sin(\rho\eta)}{\rho\sinh(\eta)},
\end{equation}
agreeing with the results obtained in~\cite{Barrett:1999qw,Perez:2000ec}. Notice that Eq.~\eqref{eq:D++} defines a regular function in $\rho$ and $\eta$.

We interpret the result of $\nu = 0$ as reflecting the simplicity condition, which imposes a vanishing of the second $\SL$-Casimir, defined in Eq.~\eqref{eq:cas2}. At the same time, simplicity affects the area operator of triangles which is given by the first $\SL$-Casimir operator defined in Eq.~\eqref{eq:cas1} and has eigenvalues given by
\begin{equation}\label{eq:area scaling}
A^2\sim -\rho^2+\nu^2-1.
\end{equation}
On the quantum level, triangles are considered as timelike, lightlike or spacelike if $A^2$ is positive, zero or negative, respectively. By setting $\nu = 0$, the triangles of a spacelike tetrahedron clearly have a strictly negative area spectrum $-\rho^2-1 < 0$, and are therefore spacelike. This is in agreement with the classical results of Section~\ref{subsec:Bivector Variables and Geometric Interpretation}.

\paragraph{Spacelike kernel} 

If both of the normal vectors entering the kernels are spacelike, we set $\alpha = -$ in Eq.~\eqref{eq:spin rep of delta on general homogeneous space}. Comparing this expression with Eq.~\eqref{eq:delta on H21} yields
\begin{equation}\label{eq:D--}
\begin{aligned}
K_{--}^{(\rho,\nu)}(X,Y)
& =
\frac{\delta_{\nu,0}}{2}\int\dd{\Omega}\abs{X^{\mu}\xi_{\mu}}^{i\rho-1}\abs{Y^{\mu}\xi_{\mu}}^{-i\rho-1}+\\[7pt]
& +
\delta(\rho)\delta_{\nu\in 2\mathbb{N}^+}\frac{128\pi}{\abs{\nu}}\int\dd{\Omega}e^{i2\nu\Theta(X,Y)}\delta(Y^{\mu}\xi_{\mu})\delta(X^{\mu}\xi_{\mu})\\[7pt]
&\equiv
\frac{\delta_{\nu,0}}{2}K^{\rho}_{--}(X,Y) + \delta(\rho)\delta_{\nu\in 2\mathbb{N}^+}K^{\nu}_{--}(X,Y),
\end{aligned}
\end{equation}
where $\cos(\Theta) = \abs{X\cdot Y}$. Notice that the formula for $K_{--}^{\nu}$ contains the absolute value of $\nu$, guaranteeing unitary equivalence of the kernel. In the following, we treat the kernels $K^{\rho}_{--}$ and $K^{\nu}_{--}$ separately and refer to them as continuous and discrete parts, respectively.  

Focusing first on the continuous part, we choose a convenient parametrization for $X$ and $Y$, 
\begin{equation}\label{eq:parametrization of spacelike X,Y}
X = (0,0,0,1),\qquad Y = (\sinh(\eta),\cosh(\eta)\hat{\vb*{r}}),\qquad \eta\in\R,~\hat{\vb*{r}}\in S^2, 
\end{equation}
permitted by the symmetry of Eq.~\eqref{eq:SL2C-invariance of kernels}, in which  $K^{\rho}_{--}$ takes the integral form~\cite{Perez:2000ep}
\begin{equation}\label{eq:contiunuous D--}
\begin{aligned}
& K^{\rho}_{--}(\eta,\hat{r}_z) =\\[7pt]
=&
\frac{1}{4\pi}\int\dd{\phi}\int\limits_{-1}^{1}\dd{t}\abs{\sinh(\eta)-\cosh(\eta)\left(\sqrt{1-t^2}\sqrt{1-\hat{r}_z^2}\sin(\phi)+t\hat{r}_z\right)}^{-i\rho-1}\abs{t}^{i\rho-1}.
\end{aligned}
\end{equation}
In order to gain some intuition for Eq.~\eqref{eq:contiunuous D--}, consider the restriction given by $\hat{\vb*{r}} = \pm\hat{\vb*{e}}_z$, for which the integral above readily simplifies. In these particular restricted cases, Eq.~\eqref{eq:contiunuous D--} simplifies to~\cite{Perez:2000ep}
\begin{equation}\label{eq:simplified K--}
K^{\rho}_{--}(\eta,\pm 1) = \frac{\sin(\rho\eta)}{\rho\sinh(\eta)},
\end{equation}
therefore agreeing with the timelike case given in Eq.~\eqref{eq:D++}. 

In the same parametrization as Eq.~\eqref{eq:parametrization of spacelike X,Y}, the discrete part of Eq.~\eqref{eq:D--} evaluates to~\cite{Perez:2000ep}
\begin{equation}
\begin{aligned}
K^{\nu}_{--}(\eta,\hat{r}_z) = \frac{32e^{2i\nu\Theta(\eta,\hat{r}_z)}}{\abs{\nu}}\int\dd{\phi}\delta\left(\sinh(\eta)-\cosh(\eta)\sqrt{1-r_z^2}\sin(\phi)\right),
\end{aligned}
\end{equation}
which can be further simplified to~\cite{Perez:2000ep}
\begin{equation}\label{eq:discrete D--}
K^{\nu}_{--}(\eta,\hat{r}_z)
=\begin{cases}
\frac{32e^{i2\nu\Theta(\eta,\hat{r}_z)}}{\abs{\nu}\sin(\Theta)},\quad \text{for} &0\leq\Theta\leq \frac{\pi}{2},\\[7pt]
0 &\text{else},
\end{cases}
\end{equation}
where $\cos(\Theta(\eta,r_z)) = \abs{\cosh(\eta)r_z}$. As noted in~\cite{Perez:2000ep}, the real part of $K^{\nu}_{--}$ diverges for $\eta = 0$ and $\hat{r}_z = \pm 1$, which corresponds to the special case where $X$ and $Y$ are equal, and is regular otherwise.

Equation~\eqref{eq:D--} shows that for a spacelike normal, the kernels split into a sum of a continuous and a discrete part, representing a linear combination of solutions of the simplicity constraint, being $\rho = 0$ and $\nu = 0$. While the continuous part $K_{--}^\rho$ is associated to spacelike triangles, the spectrum of the area operator in Eq.~\eqref{eq:area scaling} shows that the discrete kernel $K_{--}^\nu$ dictates a discrete scaling and, in addition, restricts the values of $\nu$ to be positive even integers, $\nu\in 2\mathbb{N}^+$. As a result, the corresponding triangles have a discrete and strictly positive area spectrum $\sim \nu^2-1 > 0$, and are therefore interpreted as timelike.\footnote{The interpretation of $(\rho,\nu) = (\rho, 0)$ and $(\rho,\nu) = (0,\nu), \nu\in 2 \mathbb{N^+}$, corresponding to spacelike, respectively timelike faces stems from the way the $L^2$-spaces on the spaces $\SL/\mathrm{U}^{(\alpha)}$ decompose into simple irreducible $\SL$-representations. For a more intuitive explanation relating bivectors of different signatures to different kinds of planes and geodesics in Minkowski space, we refer to the detailed discussion in~\cite{Barrett:1999qw}.
} Consequently, we obtain a qualitatively different scaling of areas, continuous versus discrete. Moreover, faces which are labelled by $(\rho,\nu) = (0,\pm 1)$ are excluded. We further comment on this matter and give a tentative interpretation of these configurations down below.

\paragraph{Lightlike kernel} 

The computation of a lightlike kernel $K_{00}$ is novel, but we follow exactly the same lines as for the purely timelike and spacelike case above. 

Since the $\delta$-function on the light cone, introduced in detail in Appendix~\ref{subsec:Functions on the Light Cone}, is most conveniently written in a linear parametrization, we write lightlike vectors $X\in\mathrm{C}^+$ as
\begin{equation}
X = \lambda\xi,\qquad \xi = (1,\cos(\phi)\sin(\theta),\sin(\phi)\sin(\theta),\cos(\theta)),\qquad \theta\in [0,\pi),\phi\in [0,2\pi).
\end{equation}
Topologically, the light cone is given as $S^2\times [0,\infty)$, with the sphere at the origin $S^2\times \{0\}$ identified to a point. On this space, $\lambda\in\R^+$ linearly parametrizes the non-compact direction of the cone while the spatial part of $\xi$ parametrizes the sphere at a given $\lambda$.

For the purely lightlike kernel, we again consider Eq.~\eqref{eq:spin rep of delta on general homogeneous space} for which we set $\alpha = 0$, and compare the expression with Eq.~\eqref{eq:delta function on cone}, yielding
\begin{equation}\label{eq:D00}
K_{00}^{(\rho,\nu)}(\lambda'\xi',\lambda\xi)
=
\delta_{\nu,0}
\frac{\delta(\theta'-\theta)\delta(\phi'-\phi)}{\sin(\theta)}(\lambda')^{-i\rho-1}\lambda^{i\rho-1}.
\end{equation}
The term $\frac{\delta(\theta-\theta')\delta(\phi-\phi')}{\sin(\theta)}$ is interpreted as a $\delta$-function on the two-sphere, which acts regularly upon integration . In particular, $\sin(\theta)$ in the denominator is canceled when the measure $\dd{\Omega}$ on $S^2$ is considered. Since the Minkowski product of two lightlike vectors is parametrized by $\lambda\in\R^+$ and $\theta \in [0,2\pi)$, we fix the first argument of Eq.~\eqref{eq:D00} as
\begin{equation}
\lambda'\xi' = (1,0,0,1)
\end{equation}
so that the lightlike kernel is simplified to
\begin{equation}
K_{00}^{\rho}(\lambda,\theta) = \frac{\delta(\theta)}{\sin(\theta)}\lambda^{i\rho-1},
\end{equation}
where we introduced $K_{00}^{(\rho,\nu)}\equiv \delta_{\nu,0}K_{00}^\rho$. Although appearing to be irregular at $\theta = 0$, the kernel acts regularly upon integration on the sphere, as noted above.

The computations of this paragraph, based on the tools developed in Section~\ref{subsec:Functions on the Light Cone}, reveal that the discrete $\SL$-representation label vanishes if simplicity is imposed with respect to a lightlike normal vector. Consequently, triangles inside lightlike tetrahedra have a continuous and strictly negative area spectrum, and are therefore spacelike. 

\paragraph{Remark on lightlike faces}

For the three kernels of non-mixed type, $K^\rho_{++}, K^\rho_{00}$ and $K^\rho_{--}$, associated to spacelike faces, we remind that the spectrum of the area operator $\sim -\rho^2-1$ is non-zero, since $\rho\in\R$ is part of the principal series. Likewise, the computation of the discrete kernel $K^\nu_{--}$ revealed that the area of timelike faces scales like $\nu^2 - 1$, which is non-zero since $\nu\in 2\mathbb{N}^+$. Taking the perspective that lightlike faces correspond to a vanishing first Casimir operator, which upon simplicity occurs for $(\rho,\nu) = (\pm i, 0)$ or $(\rho,\nu) = (0, \pm 1)$, these results can be tentatively interpreted as to exclude such faces. In this picture, the particular case of lightlike faces inside lightlike tetrahedra is ruled out which, as mentioned above, is degenerate~\cite{Speziale:2013ifa}. Remarkably, this restriction is a purely integral geometric result which is absent at the classical level, presented in Table~\ref{tab:bivectors}. However, it is yet to be clarified if faces labelled by $(\pm i,0)$ and $(0,\pm 1)$ (and which project onto unity) are indeed lightlike or should rather be interpreted as degenerate. Importantly, work on the phase structure of models only built with spacelike tetrahedra via Landau-Ginzburg mean-field theory~\cite{Marchetti:2022igl} suggests that representations with $(\pm i,0)$ defined in the sense of hyperfunctions~\cite{Ruehl1970} play a crucial role to understand their critical behavior. It can be expected that these results will be mirrored by models built from timelike tetrahedra and mixed ones. Moreover, we conjecture that such configurations should play a quintessential role to better understand the propagation of lightlike excitations, e.g. photons and gravitons, on the lattice generated by the model. Interestingly, null triangles, reached as limits of spacelike and timelike ones, have recently also attracted some attention in the context of effective spin foam models for Lorentzian quantum gravity, see~\cite{Asante:2021zzh}. Given these motivations, we leave it to future investigations to better understand the physical nature of configurations labelled by the representations (limitting on) $(\pm i,0)$ and $(0,\pm 1)$.

\subsection{Kernels of mixed type}\label{subsec:Kernels of Mixed Type}

If at least two types of normal vectors enter the vertex amplitude in Eq.~\eqref{eq:generalized vertex integral form}, the integral will necessarily contain kernels of mixed type, $K_{\alpha_1\alpha_2}$, with $\alpha_1\neq\alpha_2$. In these cases, the methods of Section~\ref{subsec:Kernels of Non-Mixed Type} are not applicable, since the kernels do not define $\delta$-functions on the homogeneous spaces $\SL/\mathrm{U}^{(\alpha)}$. However, a detailed observation shows that mixed kernels act in a similar fashion as $\delta$-functions. To see that, consider $g_1,g_2\in\SL$ being representatives of normal vectors $X_{\alpha_i}\in\SL/\mathrm{U}^{(\alpha_i)}$. Then, the following relation holds
\begin{equation}
f(g_1) = \sum_{\nu}\int\dd{\rho}4(\rho^2+\nu^2)\int\dd{g_2}\overline{D_{\alpha_1\alpha_2}^{(\rho,\nu)}(g_1^{-1}g_2)}f(g_2).
\end{equation}
Hence, the general strategy to obtain the kernels $K_{\alpha_1\alpha_2}$ is to start with a function on the homogeneous space $\SL/\mathrm{U}^{(\alpha_1)}$ and expand it in terms of the Gel'fand transform $F$, which are given in Appendix~\ref{sec:Integral Geometry} for all signatures. As a next step one inserts for the function $F$ the inverse expression on $\SL/\mathrm{U}^{(\alpha_2)}$. The integrand of the resulting $\SL/\mathrm{U}^{(\alpha_2)}$-integration is then identified with the complex conjugate of the kernel $K_{\alpha_1\alpha_2}^{(\rho,\nu)}(X_{\alpha_1},X_{\alpha_2})$.

\paragraph{Timelike-spacelike kernel} 

Denoting $X_+\in\TH_+$ and $Y_-\in\OH$, we apply the Gel'fand expansion given in Eq.~\eqref{eq:inverse Gel'fand transform on 3-hyperboloid} to a function $f$ on $\TH_+$
\begin{equation}
f(X_+) = \int\dd{\rho}4\rho^2\int\dd{\Omega}F(\xi;\rho)(X^{\mu}_+\xi_{\mu})^{-i\rho-1}.
\end{equation}
Next, we want to insert for the Gel'fand transform $F(\xi;\rho)$ the one defined on $\OH$, which has however two components given by Eqs.~\eqref{eq:Gel'fand transform on H21 with rho} and~\eqref{eq:Gel'fand transform on H21 with nu}, corresponding to the spacelike and to the timelike part, respectively. Guided by the physical intuition that only faces of the same signature can be identified, we neglect the discrete part and only insert Eq.~\eqref{eq:Gel'fand transform on H21 with rho} into the above to obtain
\begin{equation}
f(X_+)
=
\int\dd{\rho}4\rho^2\int\dd{Y_-}\int\dd{\Omega}(X_+^{\mu}\xi_{\mu})^{-i\rho-1}\abs{Y^{\nu}_{-}\xi_{\nu}}^{i\rho-1}f(Y_-).
\end{equation}
From that, we extract the mixed kernel for a timelike and spacelike normal vector
\begin{equation}\label{eq:D+-}
K_{+-}^{(\rho,\nu)}(X_+,Y_-)
=
\frac{\delta_{\nu,0}}{2}\int\dd{\Omega}(X^{\mu}_+\xi_{\mu})^{i\rho-1}\abs{Y^{\mu}_-\xi_{\mu}}^{-i\rho-1}.
\end{equation}
Choosing an explicit parametrization
\begin{equation}
X_+ = (1,0,0,0),\qquad Y_- = (\sinh(\eta),\cosh(\eta)\hat{\vb{r}}),\qquad \eta\in\R,\hat{\vb*{r}}\in S^2,
\end{equation}
the $K_{+-}$-kernel takes a more explicit form
\begin{equation}
K_{+-}^\rho(\eta,\hat{r}_z)
=
\frac{1}{4\pi}\int\dd{t}\dd{\phi'}\abs{\sinh(\eta)-\cosh(\eta)\left(\sqrt{1-t^2}\sqrt{1-r_z^2}\sin(\phi')+tr_z\right)}^{-i\rho-1}.
\end{equation}
where we have introduced the notation $K_{+-}^{(\rho,\nu)} \equiv \frac{\delta_{\nu,0}}{2}K_{+-}^\rho$.

Similar to the case of $K_{--}^\rho$ , we can gain some further intuition of the integral above by considering $r_z = \pm 1$, which is of course a restriction of the general case where $r_z\in [-1,1]$, and then yield
\begin{equation}\label{eq:simplified K+-}
K_{+-}^{\rho}(\eta,\pm 1) = \frac{i\cos(\rho\eta)}{\rho\cosh(\eta)}.
\end{equation}
This expression appears to be of a similar structure as $K^{\rho}_{++}(\eta)$ and $K^{\rho}_{--}(\eta,\pm 1)$. Notice, that although $K_{+-}$ appears to be irregular as a function at $\rho = 0$, it is regular if considered under an integral of $\rho$, due to a factor of $4\rho^2$ arising from the Plancherel measure.

Importantly, projecting out the timelike part represented by Eq.~\eqref{eq:Gel'fand transform on H21 with nu} reflects a glueing condition of triangles on the level of quantum amplitudes. As we have shown in Sections~\ref{subsec:Bivector Variables and Geometric Interpretation} and~\ref{subsec:Kernels of Non-Mixed Type}, the faces of a spacelike tetrahedron are spacelike. If we now attach a timelike tetrahedron to that, then the connecting face is necessarily spacelike. On the quantum geometric level, this condition picks out the continuous part of the Gel'fand transform on $\OH$ labelled by $\rho$, setting $\nu = 0$.

\paragraph{Timelike-lightlike kernel} 

Proceeding similarly for the mixed case of a timelike and a lightlike normal vector, $X\in\TH_+,Y_0\in\mathrm{C}^+$, we write down the inverse Gel'fand transform of a function on $\TH_+$ according to Eq.~\eqref{eq:inverse Gel'fand transform on H21} and insert for $F(\xi;\rho)$ the Gel'fand transform for functions on the cone $\mathrm{C}^+$, given in Eq.~\eqref{eq:Gel'fand transform on cone}, leading to
\begin{equation}
f(X_+) = \int\dd{\rho}4\rho^2\int\dd{\Omega}(X_+^{\mu}\xi_{\mu})^{-i\rho-1}\int\dd{\lambda}\lambda^{-i\rho}f(\lambda\xi),
\end{equation}
where we used a linear parametrization of $Y_0 = \lambda\xi$. With the induced measure on the light cone, defined in Eq.~\eqref{eq:induced measure on C+}, we can rewrite the above to the form
\begin{equation}
f(X_+)
=
\int\dd{\rho}4\rho^2\int\limits_{\mathrm{C}^+}\dd{Y_0}\left(X^{\mu}_+Y_{0{\mu}}\right)^{-i\rho-1},
\end{equation}
from which we extract the kernel $K_{+0}$
\begin{equation}\label{eq:D+0}
K_{+0}^{(\rho,\nu)}(X_+,Y_0) = \delta_{\nu,0}(X^{\mu}_+Y_{0\mu})^{i\rho-1}.
\end{equation}

Equation~\eqref{eq:D+0} can be given an even more explicit form by choosing $X_+ = (1,0,0,0)$ and using $Y_0 = \lambda\xi$, leading to
\begin{equation}
K_{+0}^{\rho}(\lambda) = \lambda^{i\rho-1},
\end{equation}
where we used the notation $K^{(\rho,\nu)}_{+0} \equiv \delta_{\nu,0}K^\rho_{+0}$. Clearly, this kernel is regular for all values of $\lambda\in\R^+$ and $\rho\in\R$.

In comparison to the mixed $+-$ case above, tetrahedra with timelike and lightlike normal vectors both allow for spacelike faces only, and so there is no discrete part which is projected out.

\paragraph{Spacelike-lightlike kernel} 

In analogy to the timelike-spacelike kernel $K_{+-}$, we start the derivation of the mixed kernel $K_{0-}$ by considering the Gel'fand expansion of a function $f$ on the light cone, given in Eq.~\eqref{eq:inverse Gel'fand transform on cone}. For the Gel'fand transform $F(\xi;\rho)$ entering this expansion, we insert the continuous component of the Gel'fand transform on $\OH$, defined in Eq.~\eqref{eq:Gel'fand transform on H21 with rho}, therefore projecting out the discrete part depending on $\nu$. For $X_0 = \lambda\xi\in\mathrm{C}^+$ and $Y_-\in\OH$, this procedure yields
\begin{equation}
f(\lambda\xi)
=
\int\dd{\rho}4\rho^2\int\dd{Y_-}\abs{Y_-^{\mu}\xi_{\mu}}^{-i\rho-1}\lambda^{-i\rho-1}f(Y_-),
\end{equation}
from which we extract the mixed kernel for a spacelike and a lightlike normal vector
\begin{equation}\label{eq:D-0}
K_{0-}^{\rho}(X_0,Y_-) = \abs{X^{\mu}Y_{\mu}}^{i\rho-1},
\end{equation}
where we used again the simplified notation $K_{0-}^{(\rho,\nu)}\equiv \delta_{\nu,0}K^{\rho}_{0-}$. 

Choosing the parametrization
\begin{equation}
X_0 = \lambda\xi,\qquad Y_- = (0,0,0,1),
\end{equation}
Equation~\eqref{eq:D-0} further simplifies to
\begin{equation}
K^{\rho}_{0-}(\lambda,\theta) = \abs{\lambda\cos(\theta)}^{i\rho-1},
\end{equation}
which is a regular function for all $\lambda\in\R^+$, $\theta\in[0,2\pi)$ and $\rho\in\R$.

Similar to the $+-$-case, the discrete part of the Gel'fand transform on $\OH$ is projected out, leaving only terms with $\nu = 0$. This again reflects on the level of quantum amplitudes the condition that only faces of the same signature can be identified, since the triangles of lightlike tetrahedra need to be spacelike. Remarkably, in the light of all six kernels computed above, we observe that most timelike information is projected out. We suspect that the surplus of spacelike information at the microscopic level percolates to the asymmetry between spacelike and timelike directions at large scales and leave it to future research to investigate this point. 

As a concluding remark concerning all kernels, we observe the connection between the reality of the kernels, the invariance under the exchange of arguments and unitary equivalence of $\SL$-representations $(\rho,\nu)\equiv (-\rho,-\nu)$. Following Eq.~\eqref{eq:symmetry of K}, exchanging the arguments of any kernel leads to complex conjugation. In addition, we observe that complex conjugation of the kernels yields the kernel evaluated on the negative representation labels, i.e.
\begin{equation}
\overline{K^{(\rho,\nu)}_{\alpha_1\alpha_2}(X,Y)} = K_{\alpha_1\alpha_2}^{(-\rho,-\nu)}(X,Y).
\end{equation}
Hence, under integration over the representation labels, we can perform a change of variables $(-\rho,-\nu)\longrightarrow (\rho,\nu)$, effectively turning the kernels real and thus symmetric under exchange of arguments.

With the computations of all the kernels $K_{\alpha_1\alpha_2}$ achieved, each of the $21$ possible vertex amplitudes can be computed as a convolution of kernels according to Equation~\eqref{eq:generalized vertex integral form}. The next task, tackled in the following section, is to discuss the notion of spacetime-orientation, first in general terms and then for the complete model we introduced.

\subsection{Spacetime orientation}\label{subsec:Spacetime Orientation}

The work presented so far focused on the completion of the Barrett-Crane model by including all bare causal configurations. Following the discussions of the introduction in Section~\ref{sec:Introduction} and that of Refs.~\cite{Livine:2002rh,Bianchi:2021ric}, bare causality does not cover all aspects of causality but needs to be supplemented with a notion of time-orientation, which implies an ordering between causally connected events.\footnote{Simply put, bare causality corresponds to the information that two events in spacetime have a timelike, lightlike or spacelike separation, while time-orientation introduces a distinction of future and past for causally separated events in spacetime, and which is of course only Lorentz invariant for timelike separated events.} Since, in contrast to~\cite{Livine:2002rh,Bianchi:2021ric}, we work with normal vectors of all signatures, we extend the discussion to space and spacetime-orientation.

At the level of transition amplitudes between physical states, formally realized as the continuum quantum gravity path integral in the Hamiltonian formulation, we observe the absence of time-orientation caused by an integration of the lapse function $N$ over both positive and negative values~\cite{Livine:2002rh}. In the asymptotic analysis of spin foams (see~\cite{Barrett:2002ur,Freidel:2002mj} for the BC model and~\cite{Barrett:2009gg,Barrett:2009mw,Dona:2019dkf} for the EPRL model), this symmetrization leads to a vertex amplitude of the form $\mathcal{A}_v\sim e^{iS_{\mathrm{R}}}+e^{-iS_{\mathrm{R}}}\sim\cos(S_{\mathrm{R}})$, commonly referred to as the \enquote{cosine problem} (see~\cite{Rovelli:2004tv} for a discussion), where $S_{\mathrm{R}}$ is the Regge action. As clarified in~\cite{Livine:2002rh}, this is not a problem but simply a feature of the inner product of physical states, (tentatively) realized via a simplicial path integral (or, equivalently, a spin foam) construction. In the continuum, a straightforward way to break this symmetrization over time-orientations is to restrict the lapse integration to either $\R^+$ or $\R^-$, which entails a causal ordering of the boundary states~\cite{Livine:2002rh, Halliwell:1990qr,Teitelboim:1981ua}. As a consequence of the restricted lapse integration, the causal transition amplitude is no longer a solution of the Hamiltonian constraint, but a Green's function for it (analogously to the Feynman propagator for a relativistic particle). Adapting these ideas to the discrete setting, we first recall the construction of~\cite{Livine:2002rh} for the BC model with only timelike normals in the following. Thereafter, we carry the analysis over to the complete BC model.

The kernels $K_{++}^{\rho}$ given in Eq.~\eqref{eq:D++} are a-causal, which can be seen by the invariance of Eq.~\eqref{eq:D++} under the transformation $\eta\rightarrow -\eta$. This can be rephrased as a time reversal symmetry\footnote{Since the time reversal operation maps $\TH_+$ to $\TH_-$ and vice versa, one needs to apply the Gel'fand transform on the upper or lower sheet accordingly, discussed in Appendix~\ref{subsec:Functions on the Upper Sheet of the 3-Hyperboloid}. The final result will not depend on which sheet is chosen. This applies as well to the kernels with lightlike normal vectors. For details, we refer to Appendix~\ref{subsec:Functions on the Light Cone}.} of the two arguments
\begin{equation}
K_{++}^{\rho}(T\cdot X,T\cdot Y) = K_{++}^{\rho}(X,Y), 
\end{equation}
where $T$ is defined as 
\begin{equation}
T \defeq \mathrm{diag}(-1,1,1,1).
\end{equation}
Making the lack of time-orientation even more apparent, one can expand the kernel $K_{++}^{\rho}$~\cite{Livine:2002rh}
\begin{equation}\label{eq:decomposition of K++}
K_{++}^{\rho}(X,Y) = K_{++}^\rho(\eta(X,Y)) = \frac{e^{i\rho\eta}}{2i\rho\sinh(\eta)}+ \frac{e^{-i\rho\eta}}{2i\rho\sinh(-\eta)} = \sum_{\epsilon = \pm}\frac{e^{i\rho\epsilon\eta}}{2i\sinh(\epsilon\eta)},
\end{equation}
where $\eta(X,Y)\defeq \cosh^{-1}(X\cdot Y)$ is the hyperbolic distance between $X$ and $Y$ and the parameter $\epsilon=\pm$ is interpreted as the orientation of the corresponding triangle. In this form, $K_{++}^{\rho}$ is clearly an average over orientations and as such, it is insensitive to the orientation of the underlying $2$-complex. 

The arguments of~\cite{Livine:2002rh} can be generalized in two ways. First, one can consider in addition the actions of space reversal $S$ (parity transformation), defined as
\begin{equation}
S \defeq \mathrm{diag}(1,-1,-1,-1),
\end{equation}
and spacetime reversal, defined as
\begin{equation}
ST\defeq S\circ T = \mathrm{diag}(-1,-1,-1,-1).
\end{equation}
Second, one can analyze all the other kernels $K_{\alpha_1\alpha_2}$ with regard to their transformation behavior under $S, T$ and $ST$. Performing both steps simultaneously, a detailed look at Eq.~\eqref{eq:SL2C-invariance of kernels} reveals that the kernels exhibit a larger symmetry group, given by that of the whole Lorentz group $\text{O}(1,3)$ understood as the semi-direct product of the proper orthochronous Lorentz group $\text{SO}(1,3)^+$ and the Klein group $\{\one,S,T,ST\}$~\cite{BERG_2001}, or one of its double covers $\text{Pin}(1,3)$\footnote{Unlike the proper orthochronous Lorentz group, $\text{O}(1,3)$ does not have a unique universal cover. It is shown in~\cite{Janssens:2017fgb} that out of the eight distinct double covers, only two are physically admissible.}\textsuperscript{,}\footnote{Notice that the action of $\text{Pin}(1,3)$ on normal vectors is defined as a straightforward generalization of the action of $\SL$ on homogeneous spaces $\SL/\mathrm{U}^{(\alpha)}$, presented in Appendix~\ref{subsec:Action of SL2C on Homogeneous Spaces}.}
\begin{equation}\label{eq:O-invariance of kernels}
    K_{\alpha_1\alpha_2}^{(\rho,\nu)}(X_1,X_2) = K_{\alpha_1\alpha_2}^{(\rho,\nu)}(h\cdot X_1, h\cdot X_2),\qquad \forall h\in \mathrm{Pin}(1,3). 
\end{equation}
It is important not to confuse the extended symmetry under $\text{Pin}(1,3)$ with the fact that we are using a double cover instead of the Lorentz group $\text{O}(1,3)$, since the latter is simply a consequence of the symmetry $g\rightarrow -g$ of the $D$-functions defined in Eq.~\eqref{eq:definition D-function}. 

In summary, the complete model defined by the kernels $K_{\alpha_1\alpha_2}$ does neither incorporate causality in the sense of time-orientability nor it is sensitive to space and spacetime-orientation.

Taking another perspective, namely that of the BC model seen as a constrained BF-theory, the absence of orientability and in particular of causality is not surprising but rather already implied by BF-theory\footnote{If a simplicial complex with boundary is considered, it is shown in~\cite{Bianchi:2021ric} that BF-theory is in fact sensitive only to the orientation of the boundary.} and the linear simplicity constraint in Eq~\eqref{eq:linear simplicity constraint}. More precisely, the amplitudes of BF-theory do not depend on the $2$-complex $\Delta^*$, therefore being in particular blind to the orientation of $\Delta^*$~\cite{Livine:2002rh,Bianchi:2021ric} (see also \cite{Freidel:1998ua, Oriti:2006wq} for earlier and later related results in the 3d case). In addition, the linear simplicity constraint in Eq.~\eqref{eq:linear simplicity constraint} is formulated as a Lorentz vector equation and therefore transforms covariantly under the whole of $\text{O}(1,3)$. Consequently, the intrinsic geometry of tetrahedra is independent of the spacetime-orientation which is also reflected by the results of Table~\ref{tab:bivectors}. 

Modifying the amplitudes to enforce some time-orientation has been done for the spin foam formulation of the timelike normal vector BC model and the EPRL model in~\cite{Livine:2002rh} and~\cite{Bianchi:2021ric}, respectively. The same general kind of construction, but differing in details, has been studied in \cite{Engle:2011un,Engle:2015mra}. The basic idea is to explicitly break the $\mathbb{Z}_2$-symmetry of the amplitudes generated by $T$. for the timelike kernel $K_{++}$, decomposed in Eq.~\eqref{eq:decomposition of K++}, amounts to choosing a particular value of $\epsilon$ for every triangle rather than summing over both orientations. Although solving the problem of a-causality at the level of spin foam amplitudes, this approach has the drawback that the restrictions on the amplitudes is introduced by hand. We thus conclude that developing a GFT for Lorentzian quantum gravity based on representations of the group $\text{Pin}(1,3)$, chosen in such a way as to produce automatically orientation-dependent amplitudes, constitutes a compelling direction of future research (as suggested already in \cite{Oriti:2003wf}). We comment again about this issue in the conclusions in Section~\ref{sec:Discussion and Conclusion}.\footnote{For a discussion on these discrete symmetries in the context of the EPRL spin foam model, we refer to Refs.~\cite{Neiman:2011gf,Rovelli:2012yy} in particular with regard to the aspect of parity violation therein as inherited from the Palatini-Holst action~\cite{Ashtekar:1988sw,Freidel:2005sn,Contaldi:2008yz}.}

\section{Putting the complete BC Model into context of other models}\label{sec:Putting the Complete BC Model into Context of Other Models}

Sections~\ref{sec:Extended Barrett-Crane Group Field Theory Model} and~\ref{sec:Explicit Expressions for the Vertex Amplitudes} contained a detailed construction of a Lorentzian GFT model which encodes bare causality at a discrete and quantum geometric level. In order to properly put this model into context with existing models and theories, we first give a comparison to other established GFTs in Section~\ref{subsec:Relation to other GFT Models}, followed by a tentative approach to mimic CDT in GFT in Section~\ref{subsec:CDT and the complete BC Model} which turns out to be equal to a causal tensor model.

\subsection{Relation to other GFT and spin foam models}\label{subsec:Relation to other GFT Models}

We constructed the complete BC model by extending existing formulations of the BC model: One which includes timelike normal vectors only~\cite{Barrett:1999qw,Perez:2000ec} and the other one which exclusively considers spacelike normal vectors~\cite{Perez:2000ep}. We first clarify how these two models arise as a restriction of the complete model. Thereafter, we briefly compare the complete BC model to the Conrady-Hnybida extension of the EPRL model, which includes spacelike and timelike tetrahedra.

\subsubsection{Barrett-Crane model with timelike normal vectors}\label{subsubsec:Relation to the Barrett-Crane Model with Timelike Normal Vectors}

In order to retrieve the Barrett-Crane model with timelike normal vectors, we consider the restriction of the action in Eqs.~\eqref{eq:kinetic action} and~\eqref{eq:full vertex action} to the case where $\alpha = +$ everywhere. As a consequence, there is only one possible vertex term which describes the glueing of five spacelike tetrahedra, defined through a single type of kernel $K^\rho_{++}$ given in Eq.~\eqref{eq:D++}. In this restricted setting, the only relevant invariant coefficients are $\mathcal{I}^{(\rho,\nu),+}_{jm}$ which can be evaluated explicitly in the canonical basis
\begin{equation}
\mathcal{I}^{(\rho,\nu),+}_{jm} = \delta_{\nu,0}\delta_{j,0}\delta_{m,0}.
\end{equation}
This restricted model, in the specific formulation in~\cite{Perez:2000ec}, has been shown in~\cite{Crane:2001qk} to be perturbatively finite: All of the amplitudes are finite for well-defined simplicial complexes. Since the geometricity constraints are not imposed by a projector in the original formulation, the model is not uniquely defined as reflected in the ambiguity of edge amplitudes, somewhat limiting the significance of the results obtained in~\cite{Crane:2001qk}. Revisiting these studies in the less ambiguous formulation presented here (using normal vectors) would place those results on a more solid ground.

\subsubsection{Barrett-Crane model with spacelike normal vectors}\label{subsubsec:Relation to Barrett-Crane with Spacelike Normal Vectors}

The BC GFT model based on spacelike normal vectors proposed in~\cite{Perez:2000ep} is realized in the complete model by restricting $\alpha = -$ in the kinetic and vertex terms ~\eqref{eq:kinetic action} and~\eqref{eq:full vertex action}, respectively. This yields a single vertex term, which is determined by the symbol $\{10(\rho,\nu)\}_{(-,-,-,-,-)}$. Although we did not attempt to evaluate the projector coefficients $P^{(\rho,\nu),-}_{jmln}$ explicitly in the canonical basis, the results of Section~\ref{subsec:Kernels of Non-Mixed Type} show that, formally, they can be expanded as
\begin{equation}
P^{(\rho,\nu),-}_{jmln} = \delta_{\nu,0}P^{\rho,-}_{jmln} + \delta(\rho)\delta_{\nu\in2\mathbb{N}^+}P^{\nu,-}_{jmln}.
\end{equation}
As a consequence, the GFT field $\varphi(g_v;X_-)$ with spacelike normal vector $X_-\in\OH$ decomposes in the spin representation into five components, which we denote by
\begin{equation}\label{eq:spacelike group field expansion}
\varphi^-_{\rho_1\rho_2\rho_3\rho_4}+\varphi^-_{\rho_1\rho_2\rho_3\nu_4}+\varphi^-_{\rho_1\rho_2\nu_3\nu_4}+\varphi^-_{\rho_1\nu_2\nu_3\nu_4}+\varphi^-_{\nu_1\nu_2\nu_3\nu_4},
\end{equation}
where $\rho_i$ and $\nu_i$ denote the representation labels of spacelike and timelike faces, respectively, and where we suppressed the four pairs of magnetic indices $(j_im_i)$ for notational clarity. While the vertex term is defined through a single term, the decomposition of the GFT field according to the equation above, induces a large number of interactions with definite face signatures. To be more precise, there are thirty distinct possibilities to glue five fields together so that the signature of identified faces matches.

\subsubsection{Relation to the Conrady-Hnybida extension of the EPRL model}\label{subsubsec:Relation to Conrady-Hnybida Extension of the EPRL Model}

Among the viable GFT and spin foam models, the EPRL model~\cite{Engle:2007wy} (see ~\cite{BenGeloun:2010qkf,Baratin:2011hp,Oriti:2016ueo, Oriti:2014uga} for its GFT formulation) is especially interesting for its closer relation to canonical LQG, in its formulation with spacelike tetrahedra. Conrady and Hnybida (CH) formulated an extension in~\cite{Conrady:2010kc,Conrady:2010vx} that includes tetrahedra with spacelike normal vectors, therefore introducing also timelike faces.\footnote{For an asymptotic analysis of the EPRL-CH model, we refer the interested reader to Refs.~\cite{Liu:2018gfc,Simao:2021qno,Han:2021bln}.}  The EPRL-CH model differs from the complete Barrett-Crane in several aspects, the most important ones of which we briefly discuss hereafter.

Most evidently, the two models are quantizations of different classical theories, independent of which causal building blocks are taken into account. While the Barrett-Crane model constitutes a constrained BF-quantization of first-order Palatini gravity~\cite{DePietri:1998hnx,Oriti:2000hh}, the EPRL model is based on the first-order Palatini-Holst formulation of gravity~\cite{Rovelli:2004tv}, where the Holst term introduces a coupling $\gamma$, which is known as \enquote{Barbero-Immirzi parameter}. This is required to make contact with canonical LQG, based on the same classical formulation of gravity. As a consequence, the simplicity constraints are imposed differently. More precisely, in contrast to Eq.~\eqref{eq:linear simplicity constraint}, the EPRL-CH simplicity constraint is given by
\begin{equation}\label{eq:EPRL-CH simplicity}
X^A\left((*B)_{AB}+\frac{1}{\gamma}B_{AB}\right) = 0,
\end{equation}
where $X$ can either be timelike, i.e. $X\in\TH_+$, or spacelike, that is $X\in\OH$.

In the presence of $\gamma$, Eq.~\eqref{eq:EPRL-CH simplicity} is a second class constraint that has to be imposed weakly in the EPRL-CH spin foam model~\cite{Engle:2007wy}. As a result of simplicity, $\SUT$ and $\SUO$ are embedded into $\SL$ but one does not have a straightforward projection onto simple $\SL$-representations; rather one can decompose states associated to spacelike hypersurfaces into $\SUT$-based ones, as in canonical LQG. See \cite{Finocchiaro:2020xwr} for details on this decomposition. In the end, for a timelike normal, faces are spacelike and labelled by the discrete $\SUT$-label $j\in\mathbb{N}/2$, matching the kinematical boundary states of LQG. Given a spacelike normal vector, faces are either spacelike or timelike and labelled by the discrete and continuous series of $\SUO$-representations, respectively. Consequently, the EPRL-CH and the complete BC model make exactly the opposite prediction for the continuity of the area spectrum of spacelike and timelike faces. Notice that the EPRL-CH model does not have a unique formulation, due to ambiguous edge amplitudes, but, as shown in~\cite{Baratin:2011hp}, this issue cannot be resolved by going to an extended formalism with normal vectors, in contrast with the BC model, because even the covariant geometricity operator imposing both simplicity and closure fails to be a projector, due to the nature of the simplicity imposition. As a last remark, we observe that the EPRL-CH model does not treat configurations with lightlike normal vectors, since $X$ in Eq.~\eqref{eq:EPRL-CH simplicity} is either timelike or spacelike.

\subsection{CDT and the complete BC Model as a causal tensor model}\label{subsec:CDT and the complete BC Model}

The variety of causal configurations entering the complete Barrett-Crane model opens up the possibility of comparing it with other models of quantum gravity which emphasize the causal aspects of spacetime. In particular, we can give a more detailed comparison to the CDT framework~\cite{Ambjorn:2012jv}, the building blocks of which contain a mixture of spacelike and timelike faces.

To make the connection explicit, we provide a tentative formulation of a CDT-like GFT with an analogue of the foliation constraint implemented. Upon transfer to the spin representation this results in a \enquote{causal} tensor model, whose perturbative expansion generates triangulations that form a subset of those appearing in CDT. The basic idea is that the introduction of additional tensors facilitates the differentiation between timelike and spacelike edges in the dual triangulation and thus allows to map a part of the bare causal structure at the microscopic level. That multi-field models can encode efficiently causality in lower dimensions is already well-known from multi-matrix models~\cite{Ambjorn:2001br,Benedetti:2008hc,Eichhorn:2020sla}. It is here realized by using GFT fields which are extended by normal vectors of distinct signatures. Of course, to advance the discussion we need to disregard some conceptual differences between the approaches of CDT and GFT. We give a brief overview of some of those differences here, before continuing with the development of the CDT-like GFT model thereafter.

First, CDT is regarded as a quantization of Einstein-Hilbert gravity in second-order formalism, using metric variables, while the (complete) Barrett-Crane model is a quantization of first-order Palatini gravity, based on tetrads and the connection. Second, the dynamical variable in CDT is the combinatorics of the simplicial complexes, while the lengths of timelike $(+)$ and spacelike $(-)$ edges are kept fixed
\begin{equation}\label{eq:CDT length relation}
l_+^2 = \alpha a^2,\qquad l_-^2 = -a^2, 
\end{equation}
where $\alpha > 0$ is an interpolation parameter\footnote{A crucial step in the CDT analysis is to perform an analytic continuation to negative values of $\alpha$, which yields Euclideanized CDT~\cite{Ambjorn:2012jv}.} and $a\in\R$~\cite{Ambjorn:2012jv}. The continuum limit of the theory is then assumed to be obtained through a second-order phase transition in the limit $a\rightarrow 0$~\cite{Loll:2019rdj}, reflecting a conceptually different view on the lattice compared to GFT, as we are going to discuss again below. In contrast to CDT, the Barrett-Crane GFT model treats both geometric data (area labels, holonomies or fluxes) and the combinatorics of the simplicial complexes as dynamical variables. Consequently, the GFT partition function contains a sum over all possible glueings of simplices as well as a sum over geometries for given glueing, expressed e.g. as a sum over representation labels. As we are going to discuss below, this has important consequences for the construction of the CDT-like GFT model. Another difference is given by the interpretation of the discretization. In CDT, and also dynamical triangulations and (conventional) tensor models, the lattice is considered as a regulator which ultimately has to be removed without leaving any remnant signature at the continuum level. In GFT on the other hand, the elements of the kinematical Hilbert space as well as their histories are usually endowed with the physical interpretation of being fundamental building blocks of a quantum spacetime (this is more in line with the interpretation of quantum states and histories in loop quantum gravity and spin foam models). From this viewpoint, continuum physics is obtained by the collective quantum dynamics of these quantum discrete degrees of freedom, but there is no requirement that they leave no signature of their existence in the continuum dynamics. On the contrary, such signature would be interpreted, at least tentatively, as indicating potential physical quantum gravity effects. As a last important difference, we note that from the very outset, topological singularities are excluded in CDT since the rigid condition of the the foliation constraint is imposed on the partition function. This not only restricts the types of building blocks entering the theory, but also prevents the appearance of so-called branching points~\cite{Loll:2000my}, which represent causal irregularities. Different to that, GFT does not contain such a strict condition, at the outset. Although the GFT Feynman expansion produces also foliable simplicial complexes, these do not single out any preferred foliation, and in general there are many other terms being generated, even in the colored case. Moreover, the CDT building blocks are endowed with a time direction and causal gluing conditions are imposed between them to enforce causal ordering. 
To obtain similar causal amplitudes in the context of the complete BC model, we would then need a time-oriented formulation of our model.

\begin{figure}[ht]
\includegraphics[width=6cm]{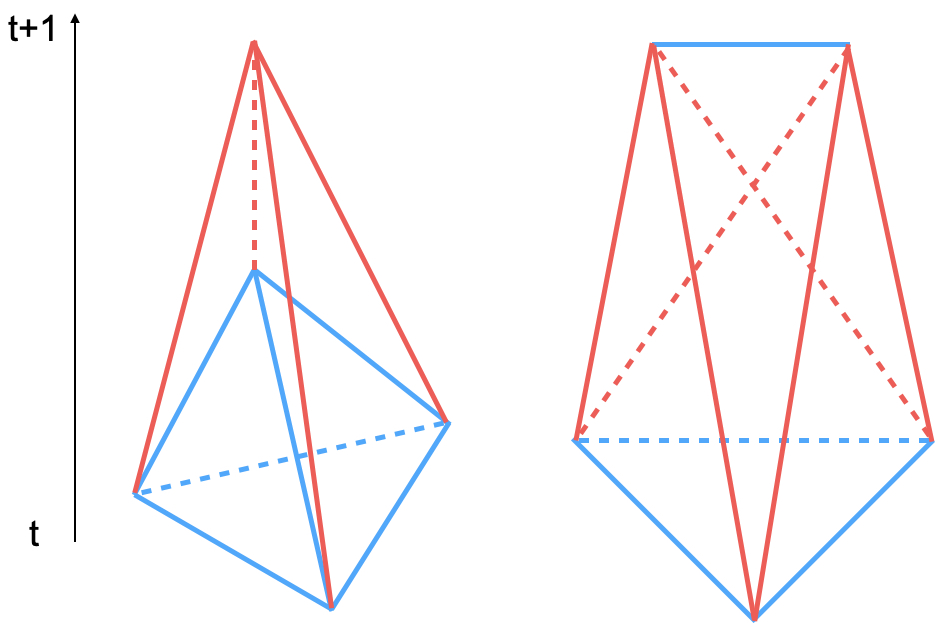}
\centering
\caption{In CDT a triangulation is composed of $4d$ triangulated layers which are built from two fundamental building blocks, i.e. $(4,1)$-simplices (left) and $(3,2)$-simplices (right). These interpolate between two consecutive spatial hypersurfaces at integer times $t$ and $t+1$. The numbers in brackets indicate the number of vertices in the triangulations of the respective constant-time slices. In this sense, the tetrahedron which forms the base of the $(4,1)$-simplex lies entirely in the $t$-hypersurface and thus is spacelike, indicated by blue edges. Timelike edges are colored in red. Consequently, all the other tetrahedra in both $4d$ building blocks are timelike. See also~\cite{Ambjorn:2012jv,Loll:2019rdj} for further details.}\label{fig:CDTsimplices}
\end{figure}

The following construction will be guided in particular by the restricted set of building blocks in CDT, being the $(4,1)$- and the $(3,2)$-simplices depicted in Fig.~\ref{fig:CDTsimplices}. Reconstructing these in GFT, we observe that the $(4,1)$-simplex contains one spacelike and four timelike tetrahedra, while the $(3,2)$-simplex is made up out of five timelike tetrahedra. Consequently, we need to employ the vertex amplitudes $\{10(\rho,\nu)\}_{(+,-,-,-,-)}$ and $\{10(\rho,\nu)\}_{(-,-,-,-,-)}$, which are generally defined in Eq.~\eqref{eq:generalized vertex integral form}. In addition, all of the triangles inside the $4$-simplices have a definite spacetime signature. For triangles inside timelike tetrahedra, this amounts to a choice between the continuous label $\rho\in\R$ or the discrete label $\nu$, corresponding to spacelike and timelike triangles, respectively. Let us firstly inspect the $(4,1)$-simplex. While the four triangles of the spacelike tetrahedron, say $0$, are labelled by $\rho_{0b}$, $b\in\{1,2,3,4\}$, one quickly observes that the remaining triangles are timelike and therefore labelled by $\nu_{ab}$, $0< a < b$. For the $(3,2)$-simplex, the triangle $(12)$ is spacelike, and hence labelled by $\rho_{01}$, while all the other ones are timelike and labelled by $\nu_{ab}$, $(ab)\neq (01)$. Consequently, there are three types of tetrahedra, a spacelike one represented by $\varphi^{\rho_1\rho_2\rho_3\rho_4,c,+}$ and two timelike ones, represented by $\varphi^{\rho_1\nu_2\nu_3\nu_4,c,-}$ and $\varphi^{\nu_1\nu_2\nu_3\nu_4,c,-}$, where we suppressed magnetic indices $(jm)$. Following these choices, and employing a coloring of the model so that only non-singular simplicial complexes are generated, the corresponding action in spin representation is given by
\begin{equation}\label{eq:CDT-like GFT action in spin rep}
\begin{aligned}
& S[\varphi,\bar{\varphi}] 
= 
\sum_{c = 0}^4 \Bigg{\{ }\Bigg{[}\prod_{i=1}^4\int\dd{\rho_i}4\rho_i^2\sum_{j_i m_i l_i n_i}\Bigg{]}\bar{\varphi}^{\rho_1\rho_2\rho_3\rho_4,c,+}_{j_i m_i}\left(C^{-1}\right)^{\rho_1\rho_2\rho_3\rho_4}_{j_i m_i l_i n_i}\varphi^{\rho_1\rho_2\rho_3\rho_4,c,+}_{l_i n_i}+\\[7pt]
+&
\Bigg{[}\int\dd{\rho_1}4\rho_1^2\sum_{j_1 m_1}\prod_{i=2}^4\sum_{\nu_i}4\nu_i^2\sum_{j_i m_i}\Bigg{]}\bar{\varphi}^{\rho_1\nu_2\nu_3\nu_4,c,-}_{j_i m_i}\varphi^{\rho_1\nu_2\nu_3\nu_4,c,-}_{j_i m_i}+\\[7pt]
+&
\Bigg{[}\prod_{i=1}^4\sum_{\nu_i}4\nu_i^2\sum_{j_i m_i}\Bigg{]}\bar{\varphi}^{\nu_1\nu_2\nu_3\nu_4,c,-}_{j_i m_i}\varphi^{\nu_1\nu_2\nu_3\nu_4,c,-}_{j_i m_i}\Bigg{\}}+\\[7pt]
+&
\sum_\sigma\Bigg{\{ }\Bigg{[}\prod_{b=1}^4\int\dd{\rho_{0b}}4\rho_{0b}^2\prod_{0<a<b}\sum_{\nu_{ab}}4\nu_{ab}^2\Bigg{]}\mathcal{A}^{(4,1)}(\rho_{0b},\nu_{ab})\varphi^{\rho_{01}\rho_{02}\rho_{03}\rho_{04},\sigma(0),+}\cdot\\[7pt]
\cdot& 
\varphi^{\rho_{04}\nu_{14}\nu_{24}\nu_{34},\sigma(4),-}\cdot\varphi^{\nu_{34}\rho_{03}\nu_{13}\nu_{23},\sigma(3),-}\cdot\varphi^{\nu_{23}\nu_{24}\rho_{02}\nu_{12},\sigma(2),-}\cdot\varphi^{\nu_{12}\nu_{13}\nu_{14}\rho_{01},\sigma(1),-}+\\[7pt]
+&
\Bigg{[}\int\dd{\rho_{01}}4\rho_{01}^2\prod_{(ab)\neq (01)}\sum_{\nu_{ab}}4\nu_{ab}^2\Bigg{]}\mathcal{A}^{(3,2)}(\rho_{01},\nu_{ab})\varphi^{\rho_{01}\nu_{02}\nu_{03}\nu_{04},\sigma(0),-}\cdot\\[7pt]
\cdot&
\varphi^{\nu_{04}\nu_{14}\nu_{24}\nu_{34},\sigma(4),-}\cdot\varphi^{\nu_{34}\nu_{03}\nu_{13}\nu_{23},\sigma(3),-}\cdot\varphi^{\nu_{23}\nu_{24}\nu_{02}\nu_{12},\sigma(2),-}\cdot\varphi^{\nu_{12}\nu_{13}\nu_{14}\rho_{01},\sigma(1),-}+\text{c.c.}\Bigg{\}},
\end{aligned}
\end{equation} 
where vertex kernels $\mathcal{A}^{(4,1)}$ and $\mathcal{A}^{(3,2)}$ are defined below in Eqs.~\eqref{eq:(4,1)-vertex} and~\eqref{eq:(3,2)-vertex}. The details of this action are explained subsequently.

To gain a better understanding of the above formula, consider the following remarks. First, for better readability of the action, we have suppressed the $\SL$-magnetic indices $(jm)$ and according factors of $(-1)^{-j-m}$ in the interaction, where the explicit contraction pattern is encoded in the \enquote{$\cdot$} which can be extracted from Eq.~\eqref{eq:single vertex action in spin representation}. Like in the general formula for the colored vertex action in Eq.~\eqref{eq:colored vertex action}, the sum over $\sigma$ is understood as a sum over all cyclic permutations, assuring that the colors are distributed evenly among the fields. Another point to remark is that the kinetic kernel for $\varphi^+$ is contaminated by the operator $C$, which, as we discuss in more detail below, will be used to realize a dual weighting. In summary, Eq.~\eqref{eq:CDT-like GFT action in spin rep} is a restriction of the full colored action given in Eqs.~\eqref{eq:colored kinetic action} and~\eqref{eq:colored vertex action} in a two-fold way. First, only spacelike and timelike tetrahedra and only two interactions among them are included in the theory. Second, two out of five possible timelike tetrahedra are picked out, which have either one or no spacelike face. Further explanations follow momentarily.

The last ingredients of the action in Eq.~\eqref{eq:CDT-like GFT action in spin rep} that need further explanations are the vertex amplitudes $\mathcal{A}^{(4,1)}$ and $\mathcal{A}^{(3,2)}$. Using the integral form of Eq.~\eqref{eq:generalized vertex integral form}, these are given by
\begin{equation}\label{eq:(4,1)-vertex}
\mathcal{A}^{(4,1)}
=
\int\left[\dd{X}\right]^5\prod_{b=1}^4\delta_{\nu_{0b},0}K_{+-}^{\rho_{0b}}(X_0^+,X_b^-)\prod_{0<a<b}\delta(\rho_{ab})\delta_{\nu_{ab}\in 2 \mathbb{N}^+}K_{--}^{\nu_{ab}}(X_a^-,X_b^-).
\end{equation}
and
\begin{equation}\label{eq:(3,2)-vertex}
\begin{aligned}
\mathcal{A}^{(3,2)}
=
\int\left[\dd{X}\right]^5\delta_{\nu_{01},0}K_{+-}^{\rho_{01}}(X_0^+,X_1^-)\prod_{0 < a < b}\delta(\rho_{ab})\delta_{\nu_{ab}\in 2\mathbb{N}^+}K_{--}^{\nu_{ab}}(X_a^-,X_b^-)
\end{aligned}
\end{equation}
understood as restrictions of the two symbols $\{10(\rho,\nu)\}_{(+,-,-,-,-)}$ and $\{10(\rho,\nu)\}_{(-,-,-,-,-)}$, respectively. 

As a next step, we impose yet another restriction on the model, which aims at mimicking the fixed lengths of timelike and spacelike edges in CDT, summarized in Eq.~\eqref{eq:CDT length relation}. Since we do not control the edge lengths in GFT but rather the areas in terms of bivectors, we relate the areas of timelike $(+)$ and spacelike $(-)$ triangles by
\begin{equation}\label{eq:CDT area relation}
\abs{A_+} = \frac{\sqrt{1+4\alpha}}{4}a^2,\qquad \abs{A_-} = \frac{\sqrt{3}}{4}a^2,
\end{equation}
similar to the Eq.~\eqref{eq:CDT length relation}. Translating this condition to the spectra of area operators, Eq.~\eqref{eq:CDT area relation} induces a connection between discrete and continuous representation labels
\begin{equation}\label{eq:rho nu relation}
\rho^2 = 3\frac{1-\nu^2}{1+4\alpha}-1,
\end{equation}
depending on $\alpha$.\footnote{In the context of GFT condensate cosmology~\cite{Oriti:2016qtz,Marchetti:2020umh,Pithis:2019tvp,Jercher:2021bie}, restricting to fixed representation labels reflects the isotropy of spacelike tetrahedra at the quantum level. It has been shown that in this context, equal representation labels arise dynamically during relational evolution~\cite{Gielen:2016uft,Pithis:2016cxg,Jercher:2021bie}.} In addition to relating representation labels, Eq.~\eqref{eq:CDT area relation} implies a of fixing areas which, at the level of $\SL$-representations, is realized by
\begin{equation}
(\rho_{ab},\nu_{ab}) = (\rho^*,\nu^*),\qquad \forall\; 0\leq a < b\leq 4.
\end{equation}

\begin{figure}[ht]
\includegraphics[width=15cm]{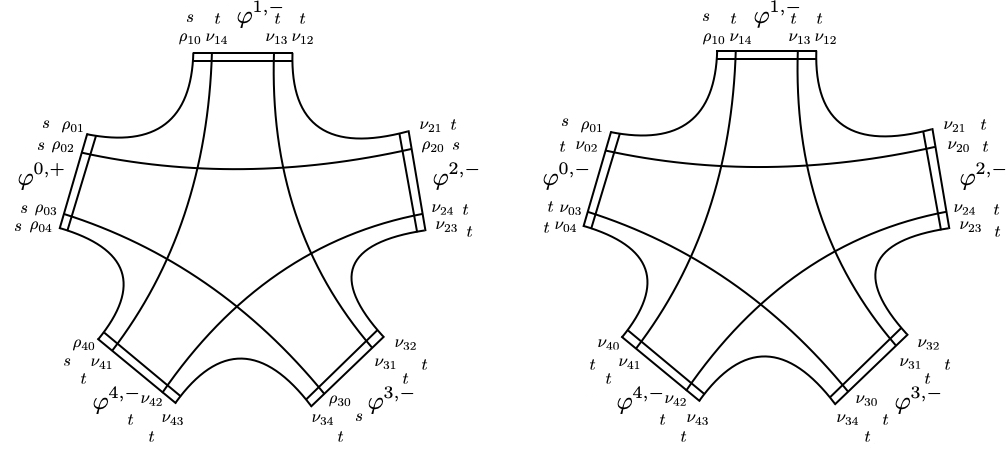}
\centering
\caption{Combinatorics of the two simplicial interaction terms via stranded diagrams: The left panel shows the interaction term corresponding to the $(4,1)$-simplex of CDT while the right panel depicts the respective $(3,2)$-simplex. Each diagram corresponds to one representative from the sum over cyclic permutations over the color degrees of freedom. Faces in the corresponding tetrahedra are labelled with their respective representation labels $\rho$ or $\nu$ fixed to $\rho^{*}$ and $\nu^{*}$ and corresponding signature $s$ for spacelike or $t$ for timelike. Since the model is colored, strands are bi-colored as indicated by their respective representation labels which come with a double index with the property that $\rho_{ij}=\rho_{ji}$ and $\nu_{ij}=\nu_{ji}$. Analogue explanations apply to the complex conjugated versions of these vertices.}\label{fig:combinatoricsCDTCTMGFT}
\end{figure}

As a result of all these restrictions, the action in Eq.~\eqref{eq:CDT-like GFT action in spin rep} effectively defines a causal tensor model in the $\SL$-magnetic indices $(jm)$ with three types of tensors
\begin{equation}
\varphi^{\rho^*\rho^*\rho^*\rho^*,c,+}_{j_1 m_1 j_2 m_3 j_3 m_3 j_4 m_4},\qquad \varphi^{\rho^*\nu^*\nu^*\nu^*,c,-}_{j_1 m_1 j_2 m_3 j_3 m_3 j_4 m_4},\qquad \varphi^{\nu^*\nu^*\nu^*\nu^*,c,-}_{j_1 m_1 j_2 m_3 j_3 m_3 j_4 m_4},
\end{equation}
and with two interactions, encoded by $\mathcal{A}^{(4,1)}(\rho^*,\nu^*)$ and $\mathcal{A}^{(3,2)}(\rho^*,\nu^*)$. A visual interpretation of the interactions is presented in Fig.~\ref{fig:combinatoricsCDTCTMGFT}. Since the unitary $\SL$-representations are infinite-dimensional, reflected by the index $j$ being unbounded from above, we introduce a cutoff $M$ in $j$ to ensure the index set to be finite to meet the precise definition of a tensor model~\cite{GurauBook}. Consequently, the trace over the identity in the two simple representations $(\rho^*,0)$ and $(0,\nu^*)$ is respectively given by
\begin{equation}\label{eq:cutoff identity}
N_{\rho^*} = \sum_{j = 0}^M (2j+1),\qquad N_{\nu^*} = \sum_{j = \abs{\nu^*}}^M (2j+1).
\end{equation}

\begin{figure}[ht]
\includegraphics[width=10cm]{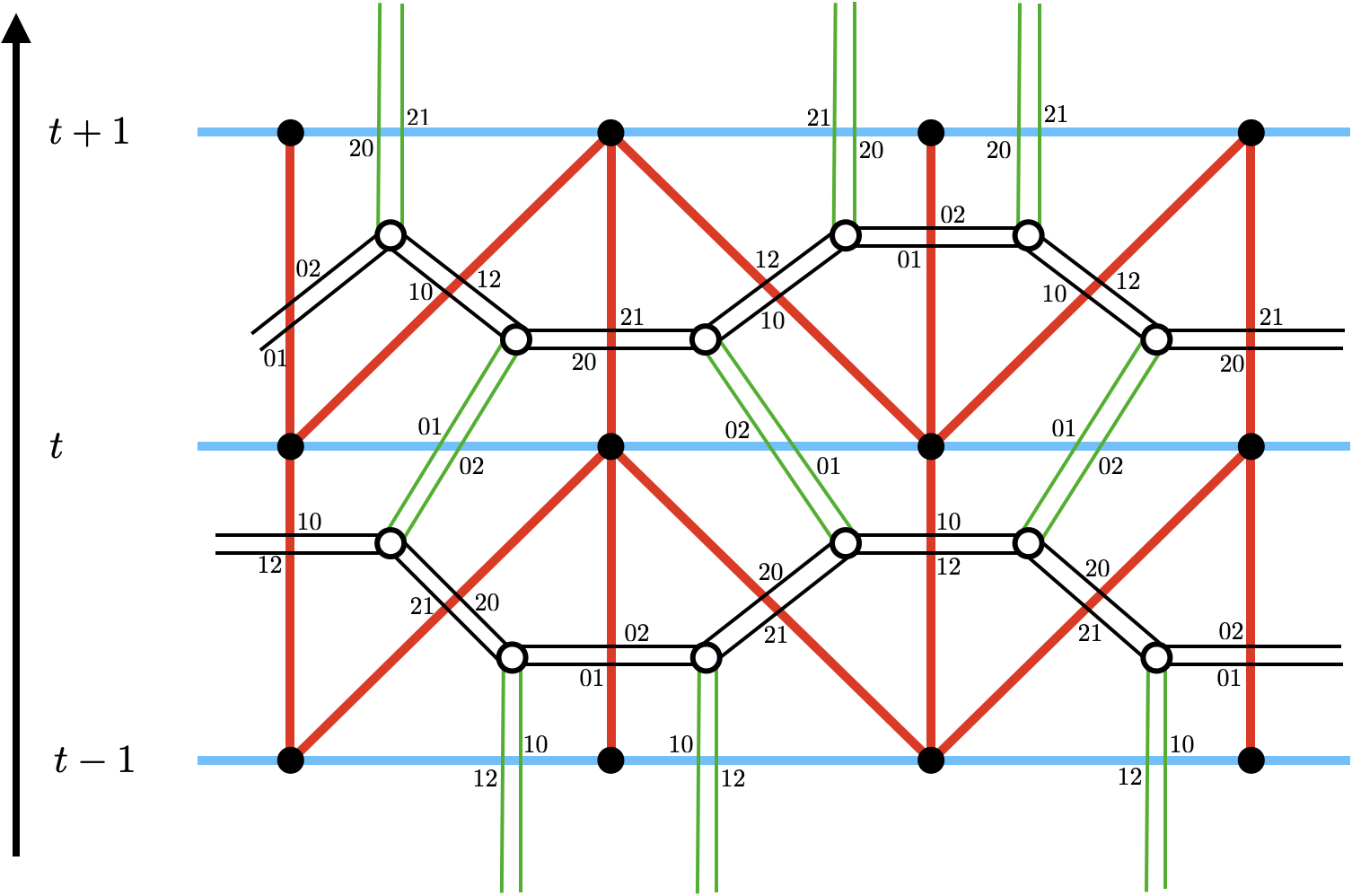}
\centering
\caption{Snippet of an exemplary triangulation in $1+1$ dimensions with time slices at integer time steps $t-1$, $t$ and $t+1$ with superimposed dual ribbon graph. In the triangulation vertices are indicated by black bullets, blue edges are spacelike while red ones are timelike. In contrast, in the dual graph vertices are indicated by circles, black strands are spacelike while green ones are timelike. For an easier comparison with the present model, the triangulation is colored so that strands of the ribbon graph are bi-colored with the property that for the double color index $ij=ji$ holds. A face in the dual graph clearly has two timelike edges and we call these facetime faces here. In $2+1$ and $3+1$ dimensions an additional type of face occurs in the dual graph which, however, only consist of spacelike edges.}\label{fig:1+1triangulation}
\end{figure}

While in CDT the type of triangulations and the topology thereof are part of the input and hence chosen to be well-behaved at the outset, the partition function of the CDT-like GFT model with a standard kinetic term would generate simplicial complexes with spatial topology change of the trouser-geometry type. In order to take care of this issue, we introduce a generalized kinetic kernel, guided by the arguments of~\cite{Benedetti:2008hc} where a dually weighted two-matrix model is presented which is capable of generating causal dynamical triangulations in $1+1$ dimensions. In order to locally prevent spatial topology change where the causal structure would degenerate, one uses the key observation in CDT in $1+1$ dimensions that any vertex in the triangulation is only coordinated by two spacelike edges, see Fig.~\ref{fig:1+1triangulation} for an exemplary triangulation. At the global level this corresponds to the assumption that there exists a global time foliation~\cite{Ambjorn:2012jv}. To map spacelike and timelike edges, one simply uses two different matrices while the foliation constraint is enforced by a dual weighting via a specific external matrix. In $2+1$ dimensions this foliation constraint simply translates to the requirement that a spacelike edge is only coordinated by two spacelike triangles while in $3+1$ dimensions one demands that any spacelike triangle is only coordinated by two spacelike tetrahedra~\cite{Ambjorn:2012jv}. Crucially, in all cases these requirements translate at the level of the dual graph to the fact that spacetime faces are allowed to have two timelike edges only, see for instance Fig.~\ref{fig:1+1triangulation} in $1+1$ dimensions. Bringing this together with the above, to generate causal dynamical triangulations in $3+1$ dimensions via a causal tensor model one needs three types of tensors, where the propagator for the spacelike tensors $\varphi^{\rho^*\rho^*\rho^*\rho^*,c,+}$ is modified by a dual weighting, specified hereafter, such that the foliation constraint of CDT is satisfied. The colorization of such a model guarantees that no singular topologies are generated upon perturbative expansion.\footnote{Notice that the dual weighting for matrix models was introduced in~\cite{Kazakov:1995ae} and transferred to colored tensor models in~\cite{Benedetti:2011nn}.}

Thus, the idea is to introduce now a dual weighting of the field $\varphi^{c,+}$, encoded by the operator $C^{-1}$ entering Eq.~\eqref{eq:CDT-like GFT action in spin rep}, which enforces spacetime faces in the dual ribbon graph to only have two timelike dual edges. In order to simplify the contraction pattern of the $C$-tensor according to a spacetime face, we consider the following product ansatz
\begin{equation}
C^{\rho^*}_{j_1m_1...j_4m_4l_1n_1...l_4n_4} = c^{\rho^*}_{j_1 m_1 l_1 n_1}...c^{\rho^*}_{j_4 m_4 l_4 n_4},
\end{equation}
where the $c$ are symmetric matrices. This ansatz allows to control properties of individual faces in the dual ribbon graph. Then the condition of spacetime faces having only two timelike dual edges is reflected in the conditions
\begin{equation}\label{eq:dual weighting}
\Tr\left(\left(c^{\rho^*}\right)^{m^{ab}_{\mathrm{st}}}\right)=N_{\rho^*}\delta_{m^{ab}_{\mathrm{st}},2},\quad m^{ab}_{\mathrm{st}}\geq 1,
\end{equation}
wherein $N_{\rho^*}$ is defined in Eq.~\eqref{eq:cutoff identity} and $m^{ab}_{\mathrm{st}}$ denotes the number of internal time-like lines of a space-time face $(ab)$ of colours $(ab)$ in the dual ribbon graph.\footnote{Note that it is in principle sufficient to work with the abstract definition of the matrix $c$ in terms of the dual weighting condition~\eqref{eq:dual weighting}. In the context of the causal matrix model for causal dynamical triangulations in $1+1$ dimensions~\cite{Benedetti:2008hc}, an explicit algorithm is developed in Ref.~\cite{Castro:2020dzt} yielding a representation of this operator which can then be used for concrete computations. These considerations could be readily applied in the present context.}

As a consequence, the amplitudes are not changed but rather a restriction on the possible glueings of the building blocks is imposed. Although we do not prove this here, we assume that the arguments of~\cite{Benedetti:2008hc} can be translated to the $3+1$-dimensional case, so that the dual weighting of the field $\varphi^+$ does in fact prevent spatial topology change. 

Finally, the partition function expanded in terms of Feynman diagrams is given by
\begin{equation}\label{eq:CDT-like partition function}
Z = \sum_{\Gamma}\frac{1}{\mathrm{sym}(\Gamma)}\left(4\rho^{*2}N_{\rho^*}\right)^{F_s}\left(4\nu^{*2} N_{\nu^*}\right)^{F_t}\left(\lambda_{(4,1)}\mathcal{A}^{(4,1)}\right)^{V_{(4,1)}}\left(\lambda_{(3,2)}\mathcal{A}^{(3,2)}\right)^{V_{(3,2)}},
\end{equation}
where $\Gamma$ are Feynman graphs dual to simplicial complexes restricted to topologically non-singular complexes without spatial topology change, ensured by coloring and the dual-weighting respectively. $V_{(4,1)}$ and $V_{(3,2)}$ are accordingly the number of $(4,1)$- and $(3,2)$-vertices, and $\lambda_{(4,1)},\lambda_{(3,2)}$ are the corresponding couplings. Arising from the $\SL$-Plancherel measure, we obtain factors of the fixed parameters $4(\rho^*)^2$ and $4(\nu^*)^2$, where $F_{t,s}$ denotes the number of timelike and spacelike faces respectively. Rewriting the above as the sum over an exponential, we can bring the GFT partition function to a form that is similar to that of CDT
\begin{equation}
Z = \sum_{\Gamma}\frac{1}{\mathrm{sym}(\Gamma)}e^{-S_{\mathrm{eff}}(\Gamma)},
\end{equation}
where the effective action $S(\Gamma)$ is given by
\begin{equation}\label{eq:effective CDT-like action}
\begin{aligned}
S_{\mathrm{eff}}(\Gamma) &= -F_s(\Gamma)\ln(4\rho^{*2}N_{\rho^*})-F_t(\Gamma)\ln(4\nu^{*2}N_{\nu^*})+\\[7pt]
-&
V_{(4,1)}(\Gamma)\ln(\lambda_{(4,1)}\mathcal{A}^{(4,1)})-V_{(3,2)}(\Gamma)\ln(\lambda_{(3,2)}\mathcal{A}^{(3,2)}).
\end{aligned}
\end{equation}
This expression can be compared with the form of the Lorentzian CDT action~\cite{Ambjorn:2012jv},
\begin{equation}\label{eq:Lorentzian CDT action}
\begin{aligned}
S_{\mathrm{CDT}}(\Gamma) &= f_1(\alpha,a,G_{\mathrm{N}},\Lambda)F_s +f_2(\alpha,a,G_{\mathrm{N}},\Lambda)F_t +\\[7pt] &+
f_3(\alpha,a,G_{\mathrm{N}},\Lambda)V_{(4,1)} + f_4(\alpha,a,G_{\mathrm{N}},\Lambda)V_{(3,2)},
\end{aligned}
\end{equation}
where the $f_i$ are functions of Newton's constant $G_{\mathrm{N}}$, the bare cosmological constant $\Lambda$, the fixed edge length $a$ and the parameter $\alpha$, explicitly given in~\cite{Ambjorn:2012jv}. In principle, comparing the  coefficients in Eqs.~\eqref{eq:effective CDT-like action} and~\eqref{eq:Lorentzian CDT action} yields a tentative interpretation of the parameters $\rho^*,\lambda_{(4,1)},\lambda_{(3,2)}, N_{\rho^*}$ and $N_{\nu^*}$ in terms of the ones entering the CDT action. 

Despite the numerous restrictions we imposed on the complete BC model, there are still various technical differences between the CDT-like GFT model and actual CDT. We notice in particular that the number of configurations which is generated by CDT is generically smaller than the one produced by our causal tensor model. This is due to the fact that the latter also creates disconnected subgraphs through multi-trace terms also dubbed as touching interactions. We refer to~\cite{Ambjorn:2001br,Eichhorn:2020sla} for an exhaustive discussion of this matter in lower dimensions in the case of matrix models where such configurations correspond to branched trees of spherical bubbles forbidden by construction by CDT~\cite{Ambjorn:2012jv}. Finally, the set of configurations in the present restricted GFT model could be extended by lightlike tetrahedra. In contrast, lightlike configurations are usually not considered in CDT.\footnote{As mentioned in~\cite{Ambjorn:2012jv}, CDT could include spacelike and null edges by setting $\alpha = 0$, which would however spoil the subsequent analysis based on a Wick rotation in $\alpha$.} 

We close this survey by suggesting a different way of implementing the ideas of CDT in the TGFT formalism in a less restrictive fashion. First notice that the introduction of a global time foliation is undesirable since it singles out a preferred Lorentzian reference frame. This issue can be alleviated by lifting the foliation constraint in the spirit of locally causal DT~\cite{Jordan:2013awa,jordan2013globally,Loll:2015yaa} where consequently many more interactions than just the $(4,1)$- and $(3,2)$-simplex are then allowed. More specifically, ignoring for the moment lightlike tetrahedra, one has a priori $49$ possibilities to glue five group fields of the types
\begin{equation}
\varphi^{c,+}_{\rho_1\rho_2\rho_3\rho_4},\; \varphi^{c,-}_{\rho_1\rho_2\rho_3\rho_4},\; \varphi^{c,-}_{\rho_1\rho_2\rho_3\nu_4},\; \varphi^{c,-}_{\rho_1\rho_2\nu_3\nu_4},\; \varphi^{c,-}_{\rho_1\nu_2\nu_3\nu_4},\; \varphi^{c,-}_{\nu_1\nu_2\nu_3\nu_4}.
\end{equation}
Similar to locally causal DT in $2+1$ dimensions~\cite{Jordan:2013awa} we notice that there are many simplices which do not have the correct signature if the areas and their signature is fixed, including for instance the simplex built out of spacelike triangles only. In addition to causality conditions at the level of single $4$-simplices, the glueing of the remaining spacetime simplices should be restricted by an analogue of the so-called \enquote{vertex causality}~\cite{Jordan:2013awa}. This condition requires that at every vertex of the simplicial complex, there is exactly one upper and one lower light cone and in this way suppresses the light cone degeneracies characteristic of topology change~\cite{Loll:2015yaa}. A possible implementation could be realized by constraints on bubbles of the colored graphs and may potentially be achieved through a different kind of dual weighting than the one given above. We also expect that the class of local causality conditions for piecewise flat manifolds discussed in the context of discrete Lorentzian Regge gravity in recent work~\cite{Asante:2021phx} will probably be very useful to further clarify and refine aspects of causality in our model going beyond the mere differentiation between spacelike, timelike and lightlike building blocks implemented so far therein.

\paragraph{Remark on closed timelike curves}

Closed timelike curves (CTCs)~\cite{ashtekar2014springer} and spatial topology change are excluded in CDT and the CDT-like GFT model through the foliation constraint or the dual weighting, respectively. In contrast, CTCs occur generically if these constraints are lifted, see for instance~\cite{Loll:2015yaa} for the case of locally causal DT. That spatial topology change, CTCs and degenerate geometries are not independent phenomena has been investigated in~\cite{Horowitz1991}. Following this work, it was shown in a theorem by Geroch~\cite{Geroch:1967fs} that independent of any field equations, a topology-changing spacetime must either have CTCs or singularities. If one embraces the idea of topology change~\cite{wheeler1957nature,Sorkin:1997gi} while excluding pathologies like CTCs, this theorem implies that configurations with degenerate metrics at finite, isolated spacetime points should be included. Such configurations are naturally included in first-order form of general relativity and it was shown in~\cite{Horowitz1991} that for this formulation exist smooth solutions allowing for topology change. If one permits degenerate metrics also in the second-order formulation then these arguments can be easily carried over. Though topology change thus seems inevitable in general relativity according to these theoretical considerations, it is unclear if quantum or other effects lead to its suppression and why it is not observable.

\section{Discussion and conclusion}\label{sec:Discussion and Conclusion}

First formulated for Euclidean~\cite{Barrett:1997gw} and then for Lorentzian signature~\cite{Barrett:1999qw}, the Barrett-Crane model is a popular spin foam and GFT model for quantum gravity based on the constrained BF-quantization scheme. Still, the usual formulations of the model only generate a special class of discretizations with exclusively spacelike or timelike tetrahedra, see~\cite{Perez:2000ec} and~\cite{Perez:2000ep}, respectively. This motivated us to develop, in this article, a completion of the BC TGFT (and spin foam) model including normal vectors of all signatures and hence spacelike, lightlike and timelike tetrahedra. To this aim, we firstly generalized the simplicity constraint in Eq.~\eqref{eq:linear simplicity constraint} to all types of normal vectors and secondly set up a TGFT action, the interaction term of which incorporates all possible glueings of tetrahedra to form $4$-simplices with an arbitrary mixture of normal signatures. The resulting model, defined through the action given in Eqs.~\eqref{eq:kinetic action} and~\eqref{eq:full vertex action}, includes all bare causal configurations and all of their possible interactions. Setting up the model, deriving the spin foam amplitudes thereof and explicitly computing the corresponding kernels, constitutes the first main result of this article. 

Studying a selected number of properties of the complete BC was the second objective of the present work. First, we provided a quantum geometric interpretation of the kernels in Section~\ref{sec:Explicit Expressions for the Vertex Amplitudes}, showing that spacelike and lightlike tetrahedra contain spacelike faces only, while timelike tetrahedra contain a mixture of spacelike and timelike faces, labelled by $\rho\in\R$ and $\nu\in 2\mathbb{N}^+$, respectively. Faces which lie in the intersection of a timelike and a spacelike or lightlike tetrahedron are spacelike, which, at the level of amplitudes, is reflected by the fact that the discrete part of the mixed kernels $K_{-+}$ and $K_{-0}$ is projected out.\footnote{Following the discussion of~\cite{Alexandrov:2005ar}, an analogy of continuous space and discrete time can be found in the 't Hooft model of a point particle in $2+1$-dimensional quantum gravity~\cite{Matschull:1997du,tHooft:1993jwb}.} Addressing the time-orientation aspect of causality, we analyzed how the kernels behave under space, time and spacetime reversal, with the result that all kernels are un-oriented and in fact possess a larger symmetry group given by the group $\text{Pin}(1,3)$. We discuss the physical implications of this symmetry, and possible developments concerning it, in more detail below. As a last result, we constructed a CDT-like TGFT model by coloring the complete BC model and restricting it to the configurations that appear also in CDT. This involved the exclusion of lightlike tetrahedra as well as only considering two possible interactions terms. Fixing areas, and therefore the representation labels, the CDT-like GFT turned into a causal tensor model for which we introduced a dual weighting to realize the foliation constraint of CDT.

In the following, we outline possible further developments from our results. Studying the perturbative properties of the complete BC model would lead to a better understanding of the amplitudes and in particular, of what kind of causal configurations would be dominant in different limits. As an example, one could extend the analysis of the asymptotic behavior and of the perturbative finiteness for the BC model with timelike normal, given respectively in Refs.~\cite{Barrett:2002ur,Baez:2002rx} and~\cite{Crane:2001as,Crane:2001qk}, to the complete model.  Another interesting point to be investigated is the large $N$ behavior of the new model. This would necessitate to study the scaling behavior of Feynman amplitudes with respect to a cut-off in the representation labels. Developed for colored tensor models and colored TGFTs on a compact domain without simplicity constraint imposed~\cite{Gurau:2010ba,Gurau:2011aq,Gurau:2011xq}, such an expansion has never been performed for viable Lorentzian quantum gravity TGFT models, where the domain is non-compact and the degrees of freedom are subject to simplicity conditions. These studies could reveal if there are preferred vertex interaction terms corresponding to specific choices of $4$-dimensional causal building blocks.

Going beyond perturbative aspects, a Landau-Ginzburg mean-field analysis along the lines of~\cite{Pithis:2018eaq,Pithis:2019mlv,Marchetti:2021fix} as well as a full-fledged FRG analysis building on results on simpler models~\cite{Benedetti:2015et,BenGeloun:2015ej,Benedetti:2016db,BenGeloun:2016kw,Carrozza:2016tih,Carrozza:2017vkz,BenGeloun:2018ekd,Pithis:2020kio,Pithis:2020sxm,Lahoche:2021nba} would shed light on the phase diagram and continuum behavior of the complete model, making explicit use of the field theoretic setting of GFTs (such full-fledged FRG analysis appears however quite challenging, at present). A crucial step in this direction is provided by the application of Landau-Ginzburg theory to models with simplicial and tensor-invariant interactions where the group fields are those used here restricted to the case of a timelike normal vector~\cite{Marchetti:2022igl}. A particularly interesting question is whether the enlarged number of configurations and interactions would lead to a different phase structure than the one obtained from the BC model with timelike normal only. Also in such continuum limit, and starting from the most democratic action which includes all possible interactions, it may turn out that only specific kinds of interactions are relevant, while others are dynamically suppressed. Such analyses should at any rate be complemented by scrutinizing the local causal structure present in the array of $4$-simplices proposed here in the spirit of e.g.~\cite{Jordan:2013awa,Asante:2021phx} which might lead to an exclusion of certain of these and thus to a refinement of the model proposed here.

As a first example of physical applications of the complete model, we suspect that in order to properly couple matter to quantum gravity, it is necessary to incorporate at least all spacelike and timelike configurations. Work addressing the coupling of matter in GFTs and spin foams is presented, for example, in~\cite{Li:2017uao} for minimally coupled scalar fields and in~\cite{Bianchi:2010bn} for fermions and Yang-Mills fields (for earlier work, see~\cite{Oriti:2006jk, Dowdall:2010ej, Oriti:2002bn, Speziale:2007mt}). In both cases, the underlying simplicial complex consists of spacelike tetrahedra only. Since the derivative entering the matter action is discretized on dual links, the restriction to spacelike tetrahedra implies that all the derivatives are in timelike direction. We expect that including spacelike and lightlike dual links, and therefore timelike and lightlike tetrahedra, alters the coupling of fermions~\cite{Bianchi:2010bn}, since the latter depends on the signature of the chosen normal vector. For minimally coupled scalar fields the full bare causal structure may have also another type of physical implication, as these fields are commonly used as physical reference frames in the form of rods and clocks~\cite{Giesel:2012rb,Marchetti:2020umh}. The causal structure may influence, in fact, how clocks are distinguished from rods, in terms of the respective gradients. A construction of proper Lorentzian material reference frames via the complete model could have far-reaching consequences on the cosmological behavior of TGFT condensates too. In particular, it is possible that the mismatch of cosmological perturbations in TGFT~\cite{Marchetti:2021gcv} with those of general relativity for intermediate and small wavelengths could be naturally resolved (if it needs to be resolved at all) along these lines.\footnote{In passing, we remark that for the coupling of lightlike excitations like photons and gravitons the incorporation of lightlike geometric configurations into the model (to identify lightlike paths) seems necessary, see also the discussion on lightlike faces in the main text.}

In light of the numerous results in TGFT condensate cosmology, such as the emergence of Friedmann-like dynamics exhibiting a quantum bounce~\cite{Oriti:2016qtz,Jercher:2021bie}, dynamical isotropization~\cite{Pithis:2016cxg,Jercher:2021bie} or early and late time phase of accelerated expansion~\cite{deCesare:2016rsf,Oriti:2021rvm}, it is interesting to ask what kind of consequences the inclusion of other than spacelike building blocks would have on the cosmological evolution. Since cosmology is characterized by the evolution of homogeneous spatial hypersurfaces, the spacelike building blocks entering the EPRL-like and BC TGFT condensate cosmology models~\cite{Oriti:2016qtz,Jercher:2021bie} appear sufficient. However, including more than only spacelike tetrahedra would still allow for computing the transition between spatial hypersurfaces, with the additional causal degrees of freedom contributing to the bulk dynamics. A priori, one would expect that the resulting dynamics will differ from what has been obtained already. A precise characterization of these differences, however, would require looking at other observables than only the spatial $3$-volume like the $4$-volume.

The complete model allows to consider lightlike as well as timelike boundaries. The most prominent example of a spacetime admitting timelike boundaries, and thus now amenable to analysis in the complete model, is Anti-de Sitter (AdS) space~\cite{kroon2016conformal}, which is of enormous theoretical and physical (if not observational) interest. Highly interesting would also be the study of lightlike boundaries, which are quintessential to describe cosmological and black hole horizons. On general grounds, a possible classical starting point for such an analysis is the development of Plebanski gravity with lightlike boundary terms and a TGFT quantization thereof. Attacking the problem from the quantum perspective, a convenient context to model quantum geometric horizons of spherical symmetry could be that of~\cite{Oriti:2018qty}, using so-called \enquote{shell condensate states} (first introduced in~\cite{Oriti:2015rwa}), which realize a foliation of spatial hypersurfaces into surfaces of spherical topology. In the enlarged setting of the complete model, it would be possible to revisit the observations of~\cite{Oriti:2018qty} in terms of lightlike hypersurfaces, foliated into spheres. Studies in this direction could strengthen the area law results of~\cite{Oriti:2018qty} and offer a way to enforce more detailed horizon conditions, with the horizon being understood as a lightlike boundary. Another possibility to model a black hole in the TGFT framework could be to consider the whole spacetime as a two-component quantum fluid of spacelike and timelike building blocks in a separated phase, where the interface represents the horizon. This would realize the idea that the role of space and time are interchanged by passing through the horizon.

Another potential research direction is the construction of a spacetime-oriented version of the complete model, in the spirit of \cite{Livine:2002rh}. Following the discussion of Section~\ref{subsec:Spacetime Orientation}, the current formulation of the complete BC model is insensitive to space, time and spacetime-orientation. We take the viewpoint, that the larger symmetry stems from the very definition of the discrete quantum gravity path integral as realizing the physical inner product of boundary states. With this interpretation, the construction of oriented amplitudes is given by explicitly breaking the invariance under the action of the Klein group $\{\one,S,T,ST\}$. This possibility has been studied in~\cite{Livine:2002rh} for the BC model with only timelike normal, at the level of spin foam amplitudes only. A causal TGFT model that exclusively generates oriented amplitudes would indeed be very interesting. We suspect that such an extension necessitates the construction of a GFT model based on the group $\text{Pin}(1,3)$, which in turn requires further attention to the representation theory of $\text{Pin}(1,3)$, only scarcely analyzed in the literature (see for instance~\cite{Naimark1964}). As one of the consequences, we expect that, in such oriented model, the unitary equivalence of representations $(\rho,\nu)\equiv (-\rho,-\nu)$ dissolves, leading to complex-valued kernels which are then no longer invariant under exchange of arguments. 

An oriented formulation of the complete model could become relevant to answer questions about the superposition of causally ordered elements, with connections to quantum information theory, e.g. quantum causal histories, and quantum causality (see for example~\cite{Brukner2014}).  From a more structural perspective, a desirable feature of an oriented model is the possibility to define the transition amplitudes between kinematical states as corresponding to a locally unitary evolution, in the sense of quantum causal histories~\cite{Markopoulou:1997wi,Markopoulou:1999cz,Markopoulou:1999ht}.\footnote{The evolution may still fail to be globally unitary, and a globally unitary evolution is tricky in a quantum gravity context, in general, due to the absence of any preferred temporal direction} In general, constructing an oriented model would facilitate a bridge between the complete BC model, and thus TGFT and spin foam models, and quantum causal sets and quantum causal histories~\cite{Markopoulou:1997wi,Markopoulou:1999cz,Markopoulou:1999ht}, an example of which is given in~\cite{Livine:2002rh}.  In addition, a fully causal (oriented) model could be important when considering $4$-simplices formed out of lightlike tetrahedra only, to be used as discrete building blocks of causal diamonds, considered in causal set theory~\cite{Khetrapal:2012ux}, holography~\cite{deBoer:2016pqk} and cosmology~\cite{Bousso:2000nf}. Finally, such a model would allow to compute retarded and advanced propagators and an important physical application could then be the characterization of quantum gravitational radiation phenomena using them. 

\subsection*{Acknowledgements}

The authors thank Luca Marchetti for helpful comments on the manuscript. The authors are also grateful to the Jena group and in particular Jos\'{e} Diogo Sim\~{a}o and Sebastian Steinhaus for insightful discussions. D. Oriti and A. Pithis acknowledge funding from DFG research grants OR432/3-1 and OR432/4-1. A. Pithis and A. Jercher kindly acknowledge the generous financial support from the MCQST via the seed funding Aost 862983-4.

\appendix

\section{Aspects of $\SL$ and its representation theory}\label{sec:Aspects of SL2C and its Representation Theory}

In this appendix, we provide a short summary of necessary formulas for computations involving $\SL$-representation theory. For detailed studies of $\SL$ and its representation theory, we refer to Refs.~\cite{Martin-Dussaud:2019ypf,Ruehl1970}.

Among the many subgroups of $\SL$, a comprehensive list of which is given in~\cite{Martin-Dussaud:2019ypf}, there are three subgroups of particular interest to us, given by $\SUT$, $\ISO$ and $\text{SU}(1,1)$ and denoted as $\mathrm{U}^{(+)}, \mathrm{U}^{(0)}$ and $\mathrm{U}^{(-)}$, respectively. Explicitly, these are defined as~\cite{Ruehl1970}
\begin{align}
\SUT &\defeq \left\{g\in\SL\;\middle\vert\; gg^{\dagger} = e\right\},\\[7pt]
\ISO &\defeq \left\{g\in\SL\;\middle\vert\; g(e+\sigma_3)g^{\dagger} = e+\sigma_3\right\},\\[7pt]
\SUO &\defeq \left\{g\in\SL\;\middle\vert\; g\sigma_3 g^{\dagger} = \sigma_3\right\}.
\end{align}
Forming the quotient space $\SL/\mathrm{U}^{(\alpha)}$ with respect to these groups yields homogeneous spaces which can be understood as embedded manifolds in $\R^{1,3}$. A summary of the quotient spaces and the stabilized normal vectors is given in Table~\ref{tab:homogeneous spaces} and a graphical representation of these distinguished hypersurfaces in Minkowski space is given in the main body of this article in Fig.~\ref{fig:hypersurfaces}.

\begin{table}[h!]
\centering
\begin{tabular}{c|c c}
$\alpha$ & $\SL/\mathrm{U}^{(\alpha)}$ & stabilized normal  \\[7pt]
\hline\\
$+$ & two-sheeted hyperboloid $\TH_{\pm}$ & $(\pm 1,0,0,0)$\\[7pt]
$0$ & upper and lower light cone $\mathrm{C}^{\pm}$ & $\frac{1}{\sqrt{2}}(\pm 1,0,0,1)$\\[7pt]
$-$ & one-sheeted hyperboloid $\OH$ & $(0,0,0,1)$
\end{tabular}
\caption{Homogeneous spaces as quotients with respect to the three-dimensional subgroups $\mathrm{U}^{(\alpha)}$.}
\label{tab:homogeneous spaces}
\end{table}

\subsection{Action of $\SL$ on homogeneous spaces}\label{subsec:Action of SL2C on Homogeneous Spaces}

The action of $\SL$ on homogeneous spaces can be defined in various ways, two of which we present now. In the main article, we choose freely between different presentations, depending on the considered problem. 

Since the homogeneous spaces $\SL/\mathrm{U}^{(\alpha)}$ arise as quotient spaces, the respective elements are given in terms of equivalence classes. So for $a\in\SL$,
\begin{equation}
[a]_{\alpha}\in\SL/\mathrm{U}^{(\alpha)}
\end{equation}
denotes an equivalence class, which satisfies
\begin{equation}
[a]_{\alpha} = [au]_{\alpha},\quad\forall u\in \mathrm{U}^{(\alpha)}.
\end{equation}
On $\SL/\mathrm{U}^{(\alpha)}$, $\SL$ acts in a canonical way, defined by
\begin{equation}\label{eq:SL2C-action on equivalence classes}
g\cdot [a]_{\alpha} \defeq [ga]_{\alpha},
\end{equation}
from which it follows clearly that the stabilizer subgroup $U_{[a]_{\alpha}}$ of $[a]_{\alpha}$ is given by
\begin{equation}\label{eq:definition of stabilizer subgroup}
U_{[a]_{\alpha}}
\defeq
\left\{aua^{-1}\;\middle\vert\; u\in \mathrm{U}^{(\alpha)}\right\} = a\mathrm{U}^{(\alpha)}a^{-1}.
\end{equation}
Since conjugation is a group isomorphism the stabilizer subgroup $U_{[a]_{\alpha}}$ is isomorphic to $\mathrm{U}^{(\alpha)}$.

A different way of defining the $\SL$-action on homogeneous spaces is by exploiting the isomorphism of Minkowski space $\R^{1,3}$ and the space of $2\times 2$ Hermitian matrices
\begin{align}
\Phi: \R^{1,3} &\overset{\cong}{\longrightarrow} \mathrm{H}_2(\C)\\[7pt]
X &\longmapsto \Phi(X),
\end{align}
given in explicit form in Ref.~\cite{Martin-Dussaud:2019ypf}. If the Minkowski inner product of $X$ is either $+1,0$ or $-1$, then $X$ can be represented as an equivalence class in the respective $\SL$-quotient space. In this representation, the equivalence of $\SL$-quotient spaces and submanifolds of Minkowski space, as shown in Table~\ref{tab:homogeneous spaces} is made transparent with the cost of the action, defined by
\begin{equation}\label{eq:SL2C action on normal using H2}
g\cdot X \defeq  \Phi^{-1}(g\Phi(X)g^{\dagger}),
\end{equation}
being less straightforward compared to Eq.~\eqref{eq:SL2C-action on equivalence classes}.

\subsection{Representation theory of $\SL$}\label{subsec:Representation Theory of SL2C}

The unitary irreducible representation spaces of $\SL$ in the principal series, denoted by $\mathcal{D}^{(\rho,\nu)}$, are labelled by pairs $(\rho,\nu)\in\R\times\mathbb{Z}/2$, realized on the space of homogeneous functions on $\C^2$ with degree $(i\rho+\nu-1,i\rho-\nu-1)$. 

In the canonical basis, the Wigner matrices for $\SL$ are denoted by $\vb*{D}^{(\rho,\nu)}$ with matrix elements $D^{(\rho,\nu)}_{jmln}$, where
\begin{equation}
j,l\in\{\abs{\nu},\abs{\nu}+1,...\}, \quad m\in\{-j,...,j\},\quad n\in\{-l,...,l\}.
\end{equation}
Most important for the computation of the spin representation of the action in Eq.~\eqref{eq:kinetic action} is the orthogonality relation~\cite{Martin-Dussaud:2019ypf}
\begin{equation}\label{eq:orthogonality relation of SL2C wigner matrices}
\int\limits_{\SL}\dd{h}\overline{D^{(\rho_1,\nu_1)}_{j_1 m_1 l_1 n_1}(h)}D^{(\rho_2,\nu_2)}_{j_2 m_2 l_2 n_2}(h)
=
\frac{\delta(\rho_1-\rho_2)\delta_{\nu_1, \nu_2}\delta_{j_1, j_2}\delta_{l_1, l_2}\delta_{m_1, m_2}\delta_{n_1, n_2}}{4\left(\rho_1^2+\nu_1^2\right)},
\end{equation}
as well as the complex conjugation property~\cite{Speziale:2016axj}
\begin{equation}\label{eq:complex conjugate of Wigner matrix}
\overline{D^{(\rho,\nu)}_{jmln}(g)}
=
(-1)^{j-l+m-n}D^{(\rho,\nu)}_{j-ml-n}(g).
\end{equation}

Appearing as the area operator for triangles and playing an important role for the simplicity constraint, the two Casimir operators of $\SL$ are given by
\begin{equation}
\begin{aligned}
\cas_1 & = \abs{\vb*{L}}^2-\abs{\vb*{K}}^2,\\[7pt]
\cas_2 & = \vb*{K}\cdot\vb*{L},
\end{aligned}
\end{equation}
in terms the $\SL$-generators $\vb*{K}$ and $\vb*{L}$, introduced in Eq.~\eqref{eq:SL2C generators}. On states in the canonical basis $\ket{(\rho,\nu);jm}\in\mathcal{D}^{(\rho,\nu)}$, the Casimir operators act by
\begin{align}
\cas_1\ket{(\rho,\nu);jm} & =  -\rho^2+\nu^2-1\ket{(\rho,\nu);jm}\label{eq:cas1}\\[7pt]
\cas_2\ket{(\rho,\nu);jm} & = \rho\nu\ket{(\rho,\nu);jm}\label{eq:cas2}.
\end{align}
Equation~\eqref{eq:cas2} clearly shows that the imposition of simplicity leads to either $\rho$ or $\nu$ vanishing. In Section~\ref{sec:Explicit Expressions for the Vertex Amplitudes}, this fact is confirmed by the integral geometric computations which show that for a timelike and a lightlike normal $\nu$ is set to zero, while a spacelike normal leads to a linear combination of $\rho = 0$ and $\nu = 0$.

\section{Integral geometry}\label{sec:Integral Geometry}

In this appendix, we introduce the key notions of integral geometry as developed by the authors of Ref.~\cite{VilenkinBook}, which turn out to be crucial for the computation of the kernels in Section~\ref{sec:Explicit Expressions for the Vertex Amplitudes}. 

\subsection{Functions on the upper sheet of the 3-hyperboloid}\label{subsec:Functions on the Upper Sheet of the 3-Hyperboloid}

The evaluation of $K_{++}$-kernels appearing in Eq.~(\ref{eq:generalized vertex integral form}) requires the decomposition of functions defined on $\TH_{\pm}$ into irreducible $\SL$-representations of the principal series. Following~\cite{VilenkinBook}, the Gel'fand transform of $f\in L^2\left(\TH_{\pm}\right)$ is given by
\begin{equation}\label{eq:Gel'fand transform on 3-hyperboloid}
F(\xi;\rho) = \int\limits_{\TH_{\pm}}\dd{X}f(X)\left(\pm X^{\mu}\xi_{\mu}\right)^{i\rho-1},
\end{equation}
with inverse defined as
\begin{equation}\label{eq:inverse Gel'fand transform on 3-hyperboloid}
f(X) = \int\limits_{0}^{\infty}\dd{\rho}4\rho^2\int\limits_{S^2}\dd{\Omega}F(\xi;\rho)\left(\pm X^{\mu}\xi_{\mu}\right)^{-i\rho-1},
\end{equation}
where we absorbed the prefactor of $(4\pi)^3$ appearing in~\cite{VilenkinBook} into the measure. Here, the null vector $\xi\in\mathrm{C}^+$ is parametrized as 
\begin{equation}\label{eq:definition if xi}
\xi = (1,\hat{\vb*{\xi}}(\phi,\theta)),
\end{equation}
where $\hat{\vb*{\xi}}(\phi,\theta)\in S^2$, $\phi$ and $\theta$ are the angles on the sphere. The normalized measure on the sphere is
\begin{equation}
\dd{\Omega} = \frac{1}{4\pi}\sin(\theta)\dd{\phi}\dd{\theta},
\end{equation}
repeatedly used in the remainder of this appendix. 

Inserting Eq.~\eqref{eq:Gel'fand transform on 3-hyperboloid} into Eq.~\eqref{eq:inverse Gel'fand transform on 3-hyperboloid} and imposing that this reproduces the original function $f(X)$, we conclude that the $\delta$-function on $\TH_{\pm}$ is written as
\begin{equation}\label{eq:Gel'fand expansion of lightlike delta}
\delta(X,Y) = \int\dd{\rho}4\rho^2\int\dd{\Omega}\left(Y^{\mu}\xi_{\mu}\right)^{i\rho-1}\left(X^{\mu}\xi_{\mu}\right)^{-i\rho-1}.
\end{equation}
Notice that this definition holds on $\TH_+$ and $\TH_-$ as the $\pm$ cancels.

\subsection{Functions on the one-sheeted hyperboloid}\label{subsec:Functions on the One-Sheeted Hyperboloid}

Similar to the previous appendix, a decomposition of functions on imaginary Lobachevskian space $\OH/\mathbb{Z}_2$ has been derived in~\cite{VilenkinBook}. Importantly, this space differs from the one-sheeted hyperboloid $\OH$ by the fact that opposite points are identified, $X = -X$. The expansion of functions on $\OH/\mathbb{Z}_2$ contains components with both, discrete and continuous labels $\rho$ and $\nu$. Explicitly, for $f\in L^2(\OH)$, it is given by~\cite{VilenkinBook,Perez:2000ep}
\begin{equation}\label{eq:inverse Gel'fand transform on H21}
\begin{aligned}
f(X) &= \int\limits_0^{\infty}\dd{\rho}4\rho^2\int\dd{\Omega}F(\xi;\rho)\abs{X^{\mu}\xi_{\mu}}^{-i\rho-1} +\\[7pt]
&+
128\pi\sum_{k = 1}^{\infty} 4k^2\int\dd{\Omega}F(\xi,X;2k)\delta(X^{\mu}\xi_{\mu}),
\end{aligned}
\end{equation}
with inverses given by~\cite{VilenkinBook}
\begin{align}
F(\xi;\rho) &= \int\limits_{\OH}\dd{X}f(X)\abs{X^{\mu}\xi_{\mu}}^{i\rho-1},\label{eq:Gel'fand transform on H21 with rho}\\[7pt]
F(\xi,X;2k) &= \frac{1}{k}\int\limits_{\OH}\dd{Y}f(Y)e^{-i2k\Theta(X,Y)}\delta(Y^{\mu}\xi_{\mu}),\label{eq:Gel'fand transform on H21 with nu}
\end{align}
where $k\in\mathbb{N}^+$ and $\cos(\Theta) \defeq \abs{X\cdot Y}$. It is important to notice that for the Gel'fand transform of Eq.~\eqref{eq:Gel'fand transform on H21 with nu}, the discrete representation parameter $\nu\in\mathbb{Z}/2$ is restricted to positive and even integers, $\nu\in 2\mathbb{N}^+$, which plays an important role for the spacelike kernel in Section~\ref{subsec:Kernels of Non-Mixed Type}.

Similar to the previous section, we insert Eqs.~\eqref{eq:Gel'fand transform on H21 with rho} and~\eqref{eq:Gel'fand transform on H21 with nu} into Eq.~\eqref{eq:inverse Gel'fand transform on H21} and impose that this gives back the original function, yielding the form of the $\delta$-function on $\OH$
\begin{equation}\label{eq:delta on H21}
\begin{aligned}
\delta(X,Y)
=&
\int\limits_{0}^{\infty}\dd{\rho}4\rho^2\int\dd{\Omega}\abs{Y^{\mu}\xi_{\mu}}^{i\rho-1}\abs{X^{\mu}\xi_{\mu}}^{-i\rho-1}+\\[7pt]
+&
128\pi\sum_{k=1}^{\infty}4k^2\int\dd{\Omega}\frac{1}{k}e^{-i2k\Theta(X,Y)}\delta(Y^{\mu}\xi_{\mu})\delta(X^{\mu}\xi_{\mu}).
\end{aligned}
\end{equation}
Notice the symmetry of the $\delta$-distribution under $(X,Y)\rightarrow (-X,-Y)$ which assures that it is effectively defined on $\OH/\mathbb{Z}_2$.

\subsection{Functions on the light cone}\label{subsec:Functions on the Light Cone}

Following~\cite{VilenkinBook}, the Gel'fand transform of $f\in L^2(\mathrm{C}^{\pm})$ and its inverse are given by~\cite{VilenkinBook}
\begin{align}\label{eq:Gel'fand transform on cone}
F(X;\rho) &= \int\limits_{0}^{\infty}\dd{t}f(tX)t^{-i\rho},\\[7pt]
f(X) &= \int\limits_{\R}\dd{\rho}4\rho^2F(X;\rho)\label{eq:inverse Gel'fand transform on cone},
\end{align}
where, in comparison to~\cite{VilenkinBook}, we absorbed a factor of $4\rho^2/4\pi$ into the measure for convenience. Physically, the fact that $\nu = 0$ in the expansion reflects the classical result shown in Appendix~\ref{sec:Classical Computations of Simplicity Constraints and Bivector Signatures}, stating that tetrahedra with a lightlike normal vector cannot have timelike faces. Remarkably, the condition of $\nu = 0$ further implies that the faces need to be spacelike, since the first Casimir operator in Eq.~\eqref{eq:cas1} is strictly negative. For a discussion on the absence of representation $(\rho,\nu) = (\pm i, 0)$, tentatively interpreted as lightlike faces, we refer to the end of Section~\ref{subsec:Kernels of Non-Mixed Type}.

It turns out useful to introduce a parametrization of vectors $X\in\mathrm{C}^{\pm}$ 
\begin{equation}\label{eq:parametrization of lightlike vectors}
X = \lambda\xi,
\end{equation}
where $\xi$ is as defined in Eq.~\eqref{eq:definition if xi}. $\lambda\in\R^{\pm}$ on the other hand parametrizes the distance along the lightlike direction and can therefore be interpreted as the radius of the sphere $S^2$ which is encoded in $\xi$. 
Following~\cite{VilenkinBook}, the measure on the cone is induced from the measure on $\R^{1,3}$
\begin{equation}
\dd{X_0}\wedge \dd{X_1}\wedge \dd{X_2}\wedge \dd{X_3}
\end{equation}
by differentiating the relation 
\begin{equation}
X_0^2-X_1^2-X_2^2-X_3^2 = 0,
\end{equation}
yielding~\cite{VilenkinBook}
\begin{equation}
\dd{X} = \frac{\dd{X_1}\wedge\dd{X_2}\wedge\dd{X_3}}{\sqrt{X_1^2+X_2^2+X_3^2}}.
\end{equation}
As a consequence, the parametrization given in Eq.~\eqref{eq:parametrization of lightlike vectors}, which is similar to a change to spherical coordinates, the induced measure is
\begin{equation}\label{eq:induced measure on C+}
\dd{X} = \lambda\dd{\lambda}\dd{\Omega}.
\end{equation}
 Next, we insert Eq.~\eqref{eq:Gel'fand transform on cone} into Eq.~\eqref{eq:inverse Gel'fand transform on cone} and impose that we get back the original function, yielding an expression for the $\delta$-function on $\mathrm{C}^{\pm}$
\begin{equation}\label{eq:delta function on cone}
\delta(\lambda\xi,\lambda'\xi') 
=
\frac{\delta(\theta-\theta')\delta(\phi-\phi')}{\sin(\theta)}\int\dd{\rho}\lambda^{i\rho-1}(\lambda')^{-i\rho-1}.
\end{equation}
where $(\theta,\phi)$ and $(\theta',\phi')$ are the angles parametrizing $\xi$ and $\xi'$, respectively. 

\section{Classical simplicity constraints and bivector signatures}\label{sec:Classical Computations of Simplicity Constraints and Bivector Signatures}

In this appendix, we derive a convenient form of the linear simplicity constraints for a general normal vector $X^A_{\alpha}$, given in Eq.~\eqref{eq:linear simplicity constraint}, in order to determine the effect of the signature of $X^A_{\alpha}$ on the signature of the respective bivectors. The conclusions we draw are summarized in Table~\ref{tab:bivectors} and allow for a comparison with the representation theoretic results that we derive in Section~\ref{sec:Explicit Expressions for the Vertex Amplitudes}. 

Starting with the $\SL$-generators $L^a$ and $K^a$, defined in Eq.~\eqref{eq:SL2C generators}, the timelike and spatial components of Eq.~\eqref{eq:linear simplicity constraint} yield a scalar and vector equation, respectively,
\begin{align}
    \vb*{X}_{\alpha}\cdot \vb*{L} &= 0,\label{eq:scalar part of ls}\\[7pt]
    \vb*{X}_{\alpha}\times\vb*{K} &= X^0_{\alpha}\vb*{L},\label{eq:vector part of ls}
\end{align}
out of which only three equations are independent. The simplest choices for a timelike, lightlike and spacelike normal vector are given in Eqs.~\eqref{eq:reference normal vectors}.

For $X_+ = (1,0,0,0)$, Eq.~\eqref{eq:vector part of ls} clearly implies that $\vb*{L} = 0$, while there is no condition imposed on $\vb*{K}$. As a consequence, the signature of the face with bivector $B$ is given by
\begin{equation}
B\cdot B \defeq \frac{1}{2}B^{AB}B_{AB} = \abs{\vb*{L}}^2 - \abs{\vb*{K}}^2 = -\abs{\vb*{K}}^2 < 0
\end{equation}
which is clearly negative for non-zero $\vb*{K}$. Notice that choosing instead for $X_+$ the vector $(-1,0,0,0)$, which lies in the lower sheet of $\TH$, the simplicity constraint does not change and the gauge-invariant data, given by the two scalars $B\cdot B$ and $B\cdot *B$, remains the same. 

Let the lightlike vector $X_0\in\mathrm{C}^+$ be given by the simplest choice $X_0 = \frac{1}{\sqrt{2}}(1,0,0,1)$. Then, Eqs.~\eqref{eq:scalar part of ls} and~\eqref{eq:vector part of ls} imply the following relations
\begin{equation}
L^3 = 0,\qquad L^1 = -K^2,\qquad L^2 = K^1,
\end{equation}
which can be rephrased in terms of the translation generators~\cite{Speziale:2013ifa}
\begin{equation}
P_{+}^1 \defeq L^1 +K^2,\qquad P_{+}^2 \defeq L^2-K^1
\end{equation}
as the conditions
\begin{equation}
L^3 = 0,\qquad P_{+}^1 = 0,\qquad P_{+}^2 = 0.
\end{equation}
As a result, for a lightlike normal vector, $B\cdot B$ is given by
\begin{equation}
B\cdot B = -(K^3)^2 \leq 0,
\end{equation}
which is either zero or negative. As discussed in Section~\ref{subsec:Bivector Variables and Geometric Interpretation}, faces of zero signature with a null normal vector are degenerate by the arguments of~\cite{Speziale:2013ifa}. For a quantum treatment of this issue in the context of the complete BC model, we refer to the end of Section~\ref{subsec:Kernels of Non-Mixed Type}.

If we choose the simplest lightlike vector in the lower light cone instead, given by $\frac{1}{\sqrt{2}}(-1,0,0,1)$, then the simplicity condition changes to
\begin{equation}
L^3 = 0,\qquad L^1 = K^2,\qquad L^2 = -K^1,
\end{equation}
which, written in terms of the translation generators
\begin{equation}
P_{-}^1 \defeq L^1 -K^2,\qquad P_{-}^2 \defeq L^2+K^1,
\end{equation}
is given by
\begin{equation}
L^3 = 0,\qquad P_{-}^1 = 0,\qquad P_{-}^2 = 0.
\end{equation}
Although the simplicity condition on the vectors $\vb*{L}$ and $\vb*{K}$ changes slightly, the gauge invariant data, given by the two Casimirs, stays the same. Hence, there is no invariant difference if a lightlike vector is chosen to lie in the lower light cone instead of the upper light cone.

Given the spacelike vector $X_-$ as above, the linear simplicity constraints yield that $K^1,K^2$ and $L^3$ vanish, while $K^3, L^1$ and $L^2$ are free variables. Hence, $B\cdot B$ is in the spacelike case given by
\begin{equation}
B\cdot B = (L^1)^2+(L^2)^2-(K^3)^2,
\end{equation}
which can either be negative, zero or positive, therefore allowing for all signatures of bivectors, which we denoted as $\gtrless 0$ in Table~\ref{tab:bivectors}.

From their very definition, the Casimir operators expressed in terms of $\SL$-generators as $B\cdot B$ and $B\cdot *B$, respectively, exhibit not only an $\SL$-invariance but in addition an invariance under space, time and spacetime reversal $\{S, T, ST\}$. As a consequence of time reversal symmetry, imposition of simplicity with respect to a timelike or lightlike normal vector does not depend on the choice of upper or lower part of the two-sheeted hyperboloid or light cone, respectively. Therefore, we choose to work with $\TH_+$ and $\mathrm{C}^+$ without restricting the geometric configurations that are generated in the TGFT partition function. Similarly, invariance under spacetime reversal $ST$ justifies to work with Lobachevskian space $\OH/\mathbb{Z}_2$ in the case of a spacelike normal vector, for which the integral geometric methods of~\cite{VilenkinBook} have been developed.

\bibliographystyle{JHEP}
\bibliography{references.bib}

\providecommand{\href}[2]{#2}\begingroup\raggedright\begin{thebibliography}{100}

\bibitem{Malament}
D.B.~Malament, \emph{The class of continuous timelike curves determines the
  topology of spacetime}, \href{https://doi.org/10.1063/1.523436}{\emph{Journal
  of Mathematical Physics} {\bfseries 18} (1977) 1399}
  [\href{https://arxiv.org/abs/https://doi.org/10.1063/1.523436}{{\ttfamily
  https://doi.org/10.1063/1.523436}}].

\bibitem{Livine:2002rh}
E.R.~Livine and D.~Oriti, \emph{{Implementing causality in the spin foam
  quantum geometry}},
  \href{https://doi.org/10.1016/S0550-3213(03)00378-X}{\emph{Nucl. Phys. B}
  {\bfseries 663} (2003) 231}
  [\href{https://arxiv.org/abs/gr-qc/0210064}{{\ttfamily gr-qc/0210064}}].

\bibitem{Bianchi:2021ric}
E.~Bianchi and P.~Martin-Dussaud, \emph{{Causal structure in spin-foams}},
  \href{https://arxiv.org/abs/2109.00986}{{\ttfamily 2109.00986}}.

\bibitem{HawkingBook}
S.~Hawking and G.~Ellis, \emph{The Large Scale Structure of Space-Time},
  Cambridge University Press (1973).

\bibitem{Loll:2019rdj}
R.~Loll, \emph{{Quantum Gravity from Causal Dynamical Triangulations: A
  Review}}, \href{https://doi.org/10.1088/1361-6382/ab57c7}{\emph{Class. Quant.
  Grav.} {\bfseries 37} (2020) 013002}
  [\href{https://arxiv.org/abs/1905.08669}{{\ttfamily 1905.08669}}].

\bibitem{Ambjorn:2012jv}
J.~Ambjorn, A.~Goerlich, J.~Jurkiewicz and R.~Loll, \emph{{Nonperturbative
  Quantum Gravity}},
  \href{https://doi.org/10.1016/j.physrep.2012.03.007}{\emph{Phys. Rept.}
  {\bfseries 519} (2012) 127}
  [\href{https://arxiv.org/abs/1203.3591}{{\ttfamily 1203.3591}}].

\bibitem{Surya:2019ndm}
S.~Surya, \emph{{The causal set approach to quantum gravity}},
  \href{https://doi.org/10.1007/s41114-019-0023-1}{\emph{Living Rev. Rel.}
  {\bfseries 22} (2019) 5} [\href{https://arxiv.org/abs/1903.11544}{{\ttfamily
  1903.11544}}].

\bibitem{Oriti:2011jm}
D.~Oriti, \emph{{The microscopic dynamics of quantum space as a group field
  theory}},  in \emph{{Foundations of Space and Time: Reflections on Quantum
  Gravity}}, pp.~257--320, 10, 2011
  [\href{https://arxiv.org/abs/1110.5606}{{\ttfamily 1110.5606}}].

\bibitem{Krajewski:2011zzu}
T.~Krajewski, \emph{{Group field theories}},
  \href{https://doi.org/10.22323/1.140.0005}{\emph{PoS} {\bfseries QGQGS2011}
  (2011) 005} [\href{https://arxiv.org/abs/1210.6257}{{\ttfamily 1210.6257}}].

\bibitem{Carrozza:2013oiy}
S.~Carrozza, \emph{{Tensorial methods and renormalization in Group Field
  Theories}}, Ph.D. thesis, Orsay, LPT, 2013.
\newblock \href{https://arxiv.org/abs/1310.3736}{{\ttfamily 1310.3736}}.
\newblock 10.1007/978-3-319-05867-2.

\bibitem{Oriti:2014uga}
D.~Oriti, \emph{{Group Field Theory and Loop Quantum Gravity}},  8, 2014
  [\href{https://arxiv.org/abs/1408.7112}{{\ttfamily 1408.7112}}].

\bibitem{Perez:2012wv}
A.~Perez, \emph{{The Spin Foam Approach to Quantum Gravity}},
  \href{https://doi.org/10.12942/lrr-2013-3}{\emph{Living Rev. Rel.} {\bfseries
  16} (2013) 3} [\href{https://arxiv.org/abs/1205.2019}{{\ttfamily
  1205.2019}}].

\bibitem{Ashtekar:2004eh}
A.~Ashtekar and J.~Lewandowski, \emph{{Background independent quantum gravity:
  A Status report}},
  \href{https://doi.org/10.1088/0264-9381/21/15/R01}{\emph{Class. Quant. Grav.}
  {\bfseries 21} (2004) R53}
  [\href{https://arxiv.org/abs/gr-qc/0404018}{{\ttfamily gr-qc/0404018}}].

\bibitem{Rovelli:2011eq}
C.~Rovelli, \emph{{Zakopane lectures on loop gravity}},
  \href{https://doi.org/10.22323/1.140.0003}{\emph{PoS} {\bfseries QGQGS2011}
  (2011) 003} [\href{https://arxiv.org/abs/1102.3660}{{\ttfamily 1102.3660}}].

\bibitem{DiFrancesco:1993cyw}
P.~Di~Francesco, P.H.~Ginsparg and J.~Zinn-Justin, \emph{{2-D Gravity and
  random matrices}},
  \href{https://doi.org/10.1016/0370-1573(94)00084-G}{\emph{Phys. Rept.}
  {\bfseries 254} (1995) 1}
  [\href{https://arxiv.org/abs/hep-th/9306153}{{\ttfamily hep-th/9306153}}].

\bibitem{GurauBook}
R.~Gurau, \emph{{Random Tensors}}, Oxford University Press (2016).

\bibitem{Reisenberger:1997sk}
M.P.~Reisenberger, \emph{{A Lattice world sheet sum for 4-d Euclidean general
  relativity}},  \href{https://arxiv.org/abs/gr-qc/9711052}{{\ttfamily
  gr-qc/9711052}}.

\bibitem{Freidel:1998pt}
L.~Freidel and K.~Krasnov, \emph{{Spin foam models and the classical action
  principle}}, \href{https://doi.org/10.4310/ATMP.1998.v2.n6.a1}{\emph{Adv.
  Theor. Math. Phys.} {\bfseries 2} (1999) 1183}
  [\href{https://arxiv.org/abs/hep-th/9807092}{{\ttfamily hep-th/9807092}}].

\bibitem{Baratin:2011hp}
A.~Baratin and D.~Oriti, \emph{{Group field theory and simplicial gravity path
  integrals: A model for Holst-Plebanski gravity}},
  \href{https://doi.org/10.1103/PhysRevD.85.044003}{\emph{Phys. Rev. D}
  {\bfseries 85} (2012) 044003}
  [\href{https://arxiv.org/abs/1111.5842}{{\ttfamily 1111.5842}}].

\bibitem{Finocchiaro:2018hks}
M.~Finocchiaro and D.~Oriti, \emph{{Spin foam models and the Duflo map}},
  \href{https://doi.org/10.1088/1361-6382/ab58da}{\emph{Class. Quant. Grav.}
  {\bfseries 37} (2020) 015010}
  [\href{https://arxiv.org/abs/1812.03550}{{\ttfamily 1812.03550}}].

\bibitem{Barrett:1999qw}
J.W.~Barrett and L.~Crane, \emph{{A Lorentzian signature model for quantum
  general relativity}},
  \href{https://doi.org/10.1088/0264-9381/17/16/302}{\emph{Class. Quant. Grav.}
  {\bfseries 17} (2000) 3101}
  [\href{https://arxiv.org/abs/gr-qc/9904025}{{\ttfamily gr-qc/9904025}}].

\bibitem{Perez:2000ec}
A.~Perez and C.~Rovelli, \emph{{Spin foam model for Lorentzian general
  relativity}}, \href{https://doi.org/10.1103/PhysRevD.63.041501}{\emph{Phys.
  Rev. D} {\bfseries 63} (2001) 041501}
  [\href{https://arxiv.org/abs/gr-qc/0009021}{{\ttfamily gr-qc/0009021}}].

\bibitem{Baratin:2011tx}
A.~Baratin and D.~Oriti, \emph{{Quantum simplicial geometry in the group field
  theory formalism: reconsidering the Barrett-Crane model}},
  \href{https://doi.org/10.1088/1367-2630/13/12/125011}{\emph{New J. Phys.}
  {\bfseries 13} (2011) 125011}
  [\href{https://arxiv.org/abs/1108.1178}{{\ttfamily 1108.1178}}].

\bibitem{Jercher:2021bie}
A.F.~Jercher, D.~Oriti and A.G.A.~Pithis, \emph{{Emergent cosmology from
  quantum gravity in the Lorentzian Barrett-Crane tensorial group field theory
  model}}, \href{https://doi.org/10.1088/1475-7516/2022/01/050}{\emph{JCAP}
  {\bfseries 01} (2022) 050}
  [\href{https://arxiv.org/abs/2112.00091}{{\ttfamily 2112.00091}}].

\bibitem{Oriti:2016qtz}
D.~Oriti, L.~Sindoni and E.~Wilson-Ewing, \emph{{Emergent Friedmann dynamics
  with a quantum bounce from quantum gravity condensates}},
  \href{https://doi.org/10.1088/0264-9381/33/22/224001}{\emph{Class. Quant.
  Grav.} {\bfseries 33} (2016) 224001}
  [\href{https://arxiv.org/abs/1602.05881}{{\ttfamily 1602.05881}}].

\bibitem{Marchetti:2020umh}
L.~Marchetti and D.~Oriti, \emph{{Effective relational cosmological dynamics
  from Quantum Gravity}},
  \href{https://doi.org/10.1007/JHEP05(2021)025}{\emph{JHEP} {\bfseries 05}
  (2021) 025} [\href{https://arxiv.org/abs/2008.02774}{{\ttfamily
  2008.02774}}].

\bibitem{Loll:1998aj}
R.~Loll, \emph{{Discrete approaches to quantum gravity in four-dimensions}},
  \href{https://doi.org/10.12942/lrr-1998-13}{\emph{Living Rev. Rel.}
  {\bfseries 1} (1998) 13}
  [\href{https://arxiv.org/abs/gr-qc/9805049}{{\ttfamily gr-qc/9805049}}].

\bibitem{Ambjorn:1991pq}
J.~Ambjorn and J.~Jurkiewicz, \emph{{Four-dimensional simplicial quantum
  gravity}}, \href{https://doi.org/10.1016/0370-2693(92)90709-D}{\emph{Phys.
  Lett. B} {\bfseries 278} (1992) 42}.

\bibitem{Ambjorn:2001br}
J.~Ambjorn, J.~Jurkiewicz, R.~Loll and G.~Vernizzi, \emph{{Lorentzian 3-D
  gravity with wormholes via matrix models}},
  \href{https://doi.org/10.1088/1126-6708/2001/09/022}{\emph{JHEP} {\bfseries
  09} (2001) 022} [\href{https://arxiv.org/abs/hep-th/0106082}{{\ttfamily
  hep-th/0106082}}].

\bibitem{Eichhorn:2020sla}
A.~Eichhorn, A.D.~Pereira and A.G.A.~Pithis, \emph{{The phase diagram of the
  multi-matrix model with ABAB-interaction from functional renormalization}},
  \href{https://doi.org/10.1007/JHEP12(2020)131}{\emph{JHEP} {\bfseries 12}
  (2020) 131} [\href{https://arxiv.org/abs/2009.05111}{{\ttfamily
  2009.05111}}].

\bibitem{Benedetti:2008hc}
D.~Benedetti and J.~Henson, \emph{{Imposing causality on a matrix model}},
  \href{https://doi.org/10.1016/j.physletb.2009.06.027}{\emph{Phys. Lett. B}
  {\bfseries 678} (2009) 222}
  [\href{https://arxiv.org/abs/0812.4261}{{\ttfamily 0812.4261}}].

\bibitem{Castro:2020dzt}
A.~Castro and T.~Koslowski, \emph{{Renormalization Group Approach to the
  Continuum Limit of Matrix Models of Quantum Gravity with Preferred
  Foliation}}, \href{https://doi.org/10.3389/fphy.2021.531766}{\emph{Front. in
  Phys.} {\bfseries 9} (2021) 114}
  [\href{https://arxiv.org/abs/2008.10090}{{\ttfamily 2008.10090}}].

\bibitem{Perez:2000ep}
A.~Perez and C.~Rovelli, \emph{{3+1 spinfoam model of quantum gravity with
  space - like and time - like components}},
  \href{https://doi.org/10.1103/PhysRevD.64.064002}{\emph{Phys. Rev. D}
  {\bfseries 64} (2001) 064002}
  [\href{https://arxiv.org/abs/gr-qc/0011037}{{\ttfamily gr-qc/0011037}}].

\bibitem{Alexandrov:2005ar}
S.~Alexandrov and Z.~Kadar, \emph{{Timelike surfaces in Lorentz covariant loop
  gravity and spin foam models}},
  \href{https://doi.org/10.1088/0264-9381/22/17/010}{\emph{Class. Quant. Grav.}
  {\bfseries 22} (2005) 3491}
  [\href{https://arxiv.org/abs/gr-qc/0501093}{{\ttfamily gr-qc/0501093}}].

\bibitem{Alexandrov:2002br}
S.~Alexandrov and E.R.~Livine, \emph{{SU(2) loop quantum gravity seen from
  covariant theory}},
  \href{https://doi.org/10.1103/PhysRevD.67.044009}{\emph{Phys. Rev. D}
  {\bfseries 67} (2003) 044009}
  [\href{https://arxiv.org/abs/gr-qc/0209105}{{\ttfamily gr-qc/0209105}}].

\bibitem{Dupuis:2010jn}
M.~Dupuis and E.R.~Livine, \emph{{Lifting SU(2) Spin Networks to Projected Spin
  Networks}}, \href{https://doi.org/10.1103/PhysRevD.82.064044}{\emph{Phys.
  Rev. D} {\bfseries 82} (2010) 064044}
  [\href{https://arxiv.org/abs/1008.4093}{{\ttfamily 1008.4093}}].

\bibitem{Speziale:2013ifa}
S.~Speziale and M.~Zhang, \emph{{Null twisted geometries}},
  \href{https://doi.org/10.1103/PhysRevD.89.084070}{\emph{Phys. Rev. D}
  {\bfseries 89} (2014) 084070}
  [\href{https://arxiv.org/abs/1311.3279}{{\ttfamily 1311.3279}}].

\bibitem{Neiman:2012fu}
Y.~Neiman, \emph{{Causal cells: spacetime polytopes with null hyperfaces}},
  \href{https://arxiv.org/abs/1212.2916}{{\ttfamily 1212.2916}}.

\bibitem{Conrady:2010kc}
F.~Conrady and J.~Hnybida, \emph{{A spin foam model for general Lorentzian
  4-geometries}},
  \href{https://doi.org/10.1088/0264-9381/27/18/185011}{\emph{Class. Quant.
  Grav.} {\bfseries 27} (2010) 185011}
  [\href{https://arxiv.org/abs/1002.1959}{{\ttfamily 1002.1959}}].

\bibitem{Conrady:2010vx}
F.~Conrady, \emph{{Spin foams with timelike surfaces}},
  \href{https://doi.org/10.1088/0264-9381/27/15/155014}{\emph{Class. Quant.
  Grav.} {\bfseries 27} (2010) 155014}
  [\href{https://arxiv.org/abs/1003.5652}{{\ttfamily 1003.5652}}].

\bibitem{Simao:2021qno}
J.D.~Sim\~ao and S.~Steinhaus, \emph{{Asymptotic analysis of spin-foams with
  time-like faces in a new parameterisation}},
  \href{https://arxiv.org/abs/2106.15635}{{\ttfamily 2106.15635}}.

\bibitem{Liu:2018gfc}
H.~Liu and M.~Han, \emph{{Asymptotic analysis of spin foam amplitude with
  timelike triangles}},
  \href{https://doi.org/10.1103/PhysRevD.99.084040}{\emph{Phys. Rev. D}
  {\bfseries 99} (2019) 084040}
  [\href{https://arxiv.org/abs/1810.09042}{{\ttfamily 1810.09042}}].

\bibitem{Han:2021bln}
M.~Han, W.~Kaminski and H.~Liu, \emph{{Finiteness of spinfoam vertex amplitude
  with timelike polyhedra and the regularization of full amplitude}},
  \href{https://doi.org/10.1103/PhysRevD.105.084034}{\emph{Phys. Rev. D}
  {\bfseries 105} (2022) 084034}
  [\href{https://arxiv.org/abs/2110.01091}{{\ttfamily 2110.01091}}].

\bibitem{Halliwell:1990qr}
J.J.~Halliwell and J.B.~Hartle, \emph{{Wave functions constructed from an
  invariant sum over histories satisfy constraints}},
  \href{https://doi.org/10.1103/PhysRevD.43.1170}{\emph{Phys. Rev. D}
  {\bfseries 43} (1991) 1170}.

\bibitem{Oriti:2004yu}
D.~Oriti, \emph{{The Feynman propagator for quantum gravity: Spin foams, proper
  time, orientation, causality and timeless-ordering}},
  \href{https://doi.org/10.1590/S0103-97332005000300019}{\emph{Braz. J. Phys.}
  {\bfseries 35} (2005) 481}
  [\href{https://arxiv.org/abs/gr-qc/0412035}{{\ttfamily gr-qc/0412035}}].

\bibitem{Teitelboim:1981ua}
C.~Teitelboim, \emph{{Quantum Mechanics of the Gravitational Field}},
  \href{https://doi.org/10.1103/PhysRevD.25.3159}{\emph{Phys. Rev. D}
  {\bfseries 25} (1982) 3159}.

\bibitem{Barrett:2002ur}
J.W.~Barrett and C.M.~Steele, \emph{{Asymptotics of relativistic spin
  networks}}, \href{https://doi.org/10.1088/0264-9381/20/7/307}{\emph{Class.
  Quant. Grav.} {\bfseries 20} (2003) 1341}
  [\href{https://arxiv.org/abs/gr-qc/0209023}{{\ttfamily gr-qc/0209023}}].

\bibitem{Freidel:2002mj}
L.~Freidel and D.~Louapre, \emph{{Asymptotics of 6j and 10j symbols}},
  \href{https://doi.org/10.1088/0264-9381/20/7/303}{\emph{Class. Quant. Grav.}
  {\bfseries 20} (2003) 1267}
  [\href{https://arxiv.org/abs/hep-th/0209134}{{\ttfamily hep-th/0209134}}].

\bibitem{Alesci:2007tx}
E.~Alesci and C.~Rovelli, \emph{{The Complete LQG propagator. I. Difficulties
  with the Barrett-Crane vertex}},
  \href{https://doi.org/10.1103/PhysRevD.76.104012}{\emph{Phys. Rev. D}
  {\bfseries 76} (2007) 104012}
  [\href{https://arxiv.org/abs/0708.0883}{{\ttfamily 0708.0883}}].

\bibitem{Baratin:2010wi}
A.~Baratin and D.~Oriti, \emph{{Group field theory with non-commutative metric
  variables}},
  \href{https://doi.org/10.1103/PhysRevLett.105.221302}{\emph{Phys. Rev. Lett.}
  {\bfseries 105} (2010) 221302}
  [\href{https://arxiv.org/abs/1002.4723}{{\ttfamily 1002.4723}}].

\bibitem{Engle:2007uq}
J.~Engle, R.~Pereira and C.~Rovelli, \emph{{The Loop-quantum-gravity
  vertex-amplitude}},
  \href{https://doi.org/10.1103/PhysRevLett.99.161301}{\emph{Phys. Rev. Lett.}
  {\bfseries 99} (2007) 161301}
  [\href{https://arxiv.org/abs/0705.2388}{{\ttfamily 0705.2388}}].

\bibitem{Dittrich:2021kzs}
B.~Dittrich, \emph{{Modified Graviton Dynamics From Spin Foams: The Area Regge
  Action}},  \href{https://arxiv.org/abs/2105.10808}{{\ttfamily 2105.10808}}.

\bibitem{Asante:2020iwm}
S.K.~Asante, B.~Dittrich and H.M.~Haggard, \emph{{Discrete gravity dynamics
  from effective spin foams}},
  \href{https://doi.org/10.1088/1361-6382/ac011b}{\emph{Class. Quant. Grav.}
  {\bfseries 38} (2021) 145023}
  [\href{https://arxiv.org/abs/2011.14468}{{\ttfamily 2011.14468}}].

\bibitem{Asante:2020qpa}
S.K.~Asante, B.~Dittrich and H.M.~Haggard, \emph{{Effective Spin Foam Models
  for Four-Dimensional Quantum Gravity}},
  \href{https://doi.org/10.1103/PhysRevLett.125.231301}{\emph{Phys. Rev. Lett.}
  {\bfseries 125} (2020) 231301}
  [\href{https://arxiv.org/abs/2004.07013}{{\ttfamily 2004.07013}}].

\bibitem{Asante:2021zzh}
S.K.~Asante, B.~Dittrich and J.~Padua-Arguelles, \emph{{Effective spin foam
  models for Lorentzian quantum gravity}},
  \href{https://doi.org/10.1088/1361-6382/ac1b44}{\emph{Class. Quant. Grav.}
  {\bfseries 38} (2021) 195002}
  [\href{https://arxiv.org/abs/2104.00485}{{\ttfamily 2104.00485}}].

\bibitem{Gielen:2013kla}
S.~Gielen, D.~Oriti and L.~Sindoni, \emph{{Cosmology from Group Field Theory
  Formalism for Quantum Gravity}},
  \href{https://doi.org/10.1103/PhysRevLett.111.031301}{\emph{Phys. Rev. Lett.}
  {\bfseries 111} (2013) 031301}
  [\href{https://arxiv.org/abs/1303.3576}{{\ttfamily 1303.3576}}].

\bibitem{Gielen:2013naa}
S.~Gielen, D.~Oriti and L.~Sindoni, \emph{{Homogeneous cosmologies as group
  field theory condensates}},
  \href{https://doi.org/10.1007/JHEP06(2014)013}{\emph{JHEP} {\bfseries 06}
  (2014) 013} [\href{https://arxiv.org/abs/1311.1238}{{\ttfamily 1311.1238}}].

\bibitem{Gielen:2016dss}
S.~Gielen and L.~Sindoni, \emph{{Quantum Cosmology from Group Field Theory
  Condensates: a Review}},
  \href{https://doi.org/10.3842/SIGMA.2016.082}{\emph{SIGMA} {\bfseries 12}
  (2016) 082} [\href{https://arxiv.org/abs/1602.08104}{{\ttfamily
  1602.08104}}].

\bibitem{Oriti:2016acw}
D.~Oriti, \emph{{The universe as a quantum gravity condensate}},
  \href{https://doi.org/10.1016/j.crhy.2017.02.003}{\emph{Comptes Rendus
  Physique} {\bfseries 18} (2017) 235}
  [\href{https://arxiv.org/abs/1612.09521}{{\ttfamily 1612.09521}}].

\bibitem{Pithis:2019tvp}
A.G.A.~Pithis and M.~Sakellariadou, \emph{{Group field theory condensate
  cosmology: An appetizer}},
  \href{https://doi.org/10.3390/universe5060147}{\emph{Universe} {\bfseries 5}
  (2019) 147} [\href{https://arxiv.org/abs/1904.00598}{{\ttfamily
  1904.00598}}].

\bibitem{VilenkinBook}
I.~Gel'fand, M.I.~Graev and N.Y.~Vilenkin, \emph{{Generalized Functions: Volume
  5, Integral Geometry and Representation Theory}}, Academic Press New York
  (1966).

\bibitem{Ruehl1970}
W.~Ruehl, \emph{Lorentz group and harmonic analysis}, W A Benjamin, Inc, United
  States (1970).

\bibitem{Rovelli:2004tv}
C.~Rovelli, \emph{{Quantum gravity}}, Cambridge Monographs on Mathematical
  Physics, Univ. Pr., Cambridge, UK (2004),
  \href{https://doi.org/10.1017/CBO9780511755804}{10.1017/CBO9780511755804}.

\bibitem{BenGeloun:2013mgx}
J.~Ben~Geloun, \emph{{On the finite amplitudes for open graphs in Abelian
  dynamical colored Boulatov-Ooguri models}},
  \href{https://doi.org/10.1088/1751-8113/46/40/402002}{\emph{J. Phys. A}
  {\bfseries 46} (2013) 402002}
  [\href{https://arxiv.org/abs/1307.8299}{{\ttfamily 1307.8299}}].

\bibitem{Pithis:2016wzf}
A.G.A.~Pithis, M.~Sakellariadou and P.~Tomov, \emph{{Impact of nonlinear
  effective interactions on group field theory quantum gravity condensates}},
  \href{https://doi.org/10.1103/PhysRevD.94.064056}{\emph{Phys. Rev. D}
  {\bfseries 94} (2016) 064056}
  [\href{https://arxiv.org/abs/1607.06662}{{\ttfamily 1607.06662}}].

\bibitem{Pithis:2016cxg}
A.G.A.~Pithis and M.~Sakellariadou, \emph{{Relational evolution of effectively
  interacting group field theory quantum gravity condensates}},
  \href{https://doi.org/10.1103/PhysRevD.95.064004}{\emph{Phys. Rev. D}
  {\bfseries 95} (2017) 064004}
  [\href{https://arxiv.org/abs/1612.02456}{{\ttfamily 1612.02456}}].

\bibitem{Gielen:2016uft}
S.~Gielen, \emph{{Emergence of a low spin phase in group field theory
  condensates}},
  \href{https://doi.org/10.1088/0264-9381/33/22/224002}{\emph{Class. Quant.
  Grav.} {\bfseries 33} (2016) 224002}
  [\href{https://arxiv.org/abs/1604.06023}{{\ttfamily 1604.06023}}].

\bibitem{Guedes:2013vi}
C.~Guedes, D.~Oriti and M.~Raasakka, \emph{{Quantization maps, algebra
  representation and non-commutative Fourier transform for Lie groups}},
  \href{https://doi.org/10.1063/1.4818638}{\emph{J. Math. Phys.} {\bfseries 54}
  (2013) 083508} [\href{https://arxiv.org/abs/1301.7750}{{\ttfamily
  1301.7750}}].

\bibitem{Oriti:2018bwr}
D.~Oriti and G.~Rosati, \emph{{Noncommutative Fourier transform for the Lorentz
  group via the Duflo map}},
  \href{https://doi.org/10.1103/PhysRevD.99.106005}{\emph{Phys. Rev. D}
  {\bfseries 99} (2019) 106005}
  [\href{https://arxiv.org/abs/1812.08616}{{\ttfamily 1812.08616}}].

\bibitem{Freidel:1999jf}
L.~Freidel and K.~Krasnov, \emph{{Simple spin networks as Feynman graphs}},
  \href{https://doi.org/10.1063/1.533203}{\emph{J. Math. Phys.} {\bfseries 41}
  (2000) 1681} [\href{https://arxiv.org/abs/hep-th/9903192}{{\ttfamily
  hep-th/9903192}}].

\bibitem{Baez:2001fh}
J.C.~Baez and J.W.~Barrett, \emph{{Integrability for relativistic spin
  networks}}, \href{https://doi.org/10.1088/0264-9381/18/21/316}{\emph{Class.
  Quant. Grav.} {\bfseries 18} (2001) 4683}
  [\href{https://arxiv.org/abs/gr-qc/0101107}{{\ttfamily gr-qc/0101107}}].

\bibitem{Livine:2002ak}
E.R.~Livine, \emph{{Projected spin networks for Lorentz connection: Linking
  spin foams and loop gravity}},
  \href{https://doi.org/10.1088/0264-9381/19/21/316}{\emph{Class. Quant. Grav.}
  {\bfseries 19} (2002) 5525}
  [\href{https://arxiv.org/abs/gr-qc/0207084}{{\ttfamily gr-qc/0207084}}].

\bibitem{Gurau:2010nd}
R.~Gurau, \emph{{Lost in Translation: Topological Singularities in Group Field
  Theory}}, \href{https://doi.org/10.1088/0264-9381/27/23/235023}{\emph{Class.
  Quant. Grav.} {\bfseries 27} (2010) 235023}
  [\href{https://arxiv.org/abs/1006.0714}{{\ttfamily 1006.0714}}].

\bibitem{Gurau:2010ba}
R.~Gurau, \emph{{The 1/N expansion of colored tensor models}},
  \href{https://doi.org/10.1007/s00023-011-0101-8}{\emph{Annales Henri
  Poincare} {\bfseries 12} (2011) 829}
  [\href{https://arxiv.org/abs/1011.2726}{{\ttfamily 1011.2726}}].

\bibitem{Gurau:2011aq}
R.~Gurau and V.~Rivasseau, \emph{{The 1/N expansion of colored tensor models in
  arbitrary dimension}},
  \href{https://doi.org/10.1209/0295-5075/95/50004}{\emph{EPL} {\bfseries 95}
  (2011) 50004} [\href{https://arxiv.org/abs/1101.4182}{{\ttfamily
  1101.4182}}].

\bibitem{Gurau:2011xq}
R.~Gurau, \emph{{The complete 1/N expansion of colored tensor models in
  arbitrary dimension}},
  \href{https://doi.org/10.1007/s00023-011-0118-z}{\emph{Annales Henri
  Poincare} {\bfseries 13} (2012) 399}
  [\href{https://arxiv.org/abs/1102.5759}{{\ttfamily 1102.5759}}].

\bibitem{Bonzom:2012hw}
V.~Bonzom, R.~Gurau and V.~Rivasseau, \emph{{Random tensor models in the large
  N limit: Uncoloring the colored tensor models}},
  \href{https://doi.org/10.1103/PhysRevD.85.084037}{\emph{Phys. Rev. D}
  {\bfseries 85} (2012) 084037}
  [\href{https://arxiv.org/abs/1202.3637}{{\ttfamily 1202.3637}}].

\bibitem{Gurau:2013pca}
R.~Gurau, \emph{{The 1/N Expansion of Tensor Models Beyond Perturbation
  Theory}}, \href{https://doi.org/10.1007/s00220-014-1907-2}{\emph{Commun.
  Math. Phys.} {\bfseries 330} (2014) 973}
  [\href{https://arxiv.org/abs/1304.2666}{{\ttfamily 1304.2666}}].

\bibitem{Gurau:2012vk}
R.~Gurau, \emph{{A review of the 1/N expansion in random tensor models}},  in
  \emph{{17th International Congress on Mathematical Physics}}, 9, 2012
  [\href{https://arxiv.org/abs/1209.3252}{{\ttfamily 1209.3252}}].

\bibitem{Caravelli:2010nh}
F.~Caravelli, \emph{{A Simple Proof of Orientability in Colored Group Field
  Theory}}, \href{https://doi.org/10.1186/2193-1801-1-6}{\emph{SpringerPlus}
  {\bfseries 1} (2012) 6} [\href{https://arxiv.org/abs/1012.4087}{{\ttfamily
  1012.4087}}].

\bibitem{Marchetti:2022igl}
L.~Marchetti, D.~Oriti, A.G.A.~Pithis and J.~Th\"urigen, \emph{{Phase
  transitions in TGFT: a Landau-Ginzburg analysis of Lorentzian quantum
  geometric models}},  \href{https://arxiv.org/abs/2209.04297}{{\ttfamily
  2209.04297}}.

\bibitem{Barrett:2009gg}
J.W.~Barrett, R.J.~Dowdall, W.J.~Fairbairn, H.~Gomes and F.~Hellmann,
  \emph{{Asymptotic analysis of the EPRL four-simplex amplitude}},
  \href{https://doi.org/10.1063/1.3244218}{\emph{J. Math. Phys.} {\bfseries 50}
  (2009) 112504} [\href{https://arxiv.org/abs/0902.1170}{{\ttfamily
  0902.1170}}].

\bibitem{Barrett:2009mw}
J.W.~Barrett, R.J.~Dowdall, W.J.~Fairbairn, F.~Hellmann and R.~Pereira,
  \emph{{Lorentzian spin foam amplitudes: Graphical calculus and asymptotics}},
  \href{https://doi.org/10.1088/0264-9381/27/16/165009}{\emph{Class. Quant.
  Grav.} {\bfseries 27} (2010) 165009}
  [\href{https://arxiv.org/abs/0907.2440}{{\ttfamily 0907.2440}}].

\bibitem{Dona:2019dkf}
P.~Don\`a, M.~Fanizza, G.~Sarno and S.~Speziale, \emph{{Numerical study of the
  Lorentzian Engle-Pereira-Rovelli-Livine spin foam amplitude}},
  \href{https://doi.org/10.1103/PhysRevD.100.106003}{\emph{Phys. Rev. D}
  {\bfseries 100} (2019) 106003}
  [\href{https://arxiv.org/abs/1903.12624}{{\ttfamily 1903.12624}}].

\bibitem{BERG_2001}
M.~BERG, C.~DeWITT-MORETTE, S.~GWO and E.~KRAMER, \emph{{THE} {PIN} {GROUPS}
  {IN} {PHYSICS}: C, p {AND} t},
  \href{https://doi.org/10.1142/s0129055x01000922}{\emph{Reviews in
  Mathematical Physics} {\bfseries 13} (2001) 953}.

\bibitem{Janssens:2017fgb}
B.~Janssens, \emph{{Pin Groups in General Relativity}},
  \href{https://doi.org/10.1103/PhysRevD.101.021702}{\emph{Phys. Rev. D}
  {\bfseries 101} (2020) 021702}
  [\href{https://arxiv.org/abs/1709.02742}{{\ttfamily 1709.02742}}].

\bibitem{Freidel:1998ua}
L.~Freidel and K.~Krasnov, \emph{{Discrete space-time volume for
  three-dimensional BF theory and quantum gravity}},
  \href{https://doi.org/10.1088/0264-9381/16/2/003}{\emph{Class. Quant. Grav.}
  {\bfseries 16} (1999) 351}
  [\href{https://arxiv.org/abs/hep-th/9804185}{{\ttfamily hep-th/9804185}}].

\bibitem{Oriti:2006wq}
D.~Oriti and T.~Tlas, \emph{{Causality and matter propagation in 3-D spin foam
  quantum gravity}},
  \href{https://doi.org/10.1103/PhysRevD.74.104021}{\emph{Phys. Rev. D}
  {\bfseries 74} (2006) 104021}
  [\href{https://arxiv.org/abs/gr-qc/0608116}{{\ttfamily gr-qc/0608116}}].

\bibitem{Engle:2011un}
J.~Engle, \emph{{Proposed proper Engle-Pereira-Rovelli-Livine vertex
  amplitude}}, \href{https://doi.org/10.1103/PhysRevD.87.084048}{\emph{Phys.
  Rev. D} {\bfseries 87} (2013) 084048}
  [\href{https://arxiv.org/abs/1111.2865}{{\ttfamily 1111.2865}}].

\bibitem{Engle:2015mra}
J.~Engle and A.~Zipfel, \emph{{Lorentzian proper vertex amplitude: Classical
  analysis and quantum derivation}},
  \href{https://doi.org/10.1103/PhysRevD.94.064024}{\emph{Phys. Rev. D}
  {\bfseries 94} (2016) 064024}
  [\href{https://arxiv.org/abs/1502.04640}{{\ttfamily 1502.04640}}].

\bibitem{Oriti:2003wf}
D.~Oriti, \emph{{Spin foam models of quantum space-time}},  other thesis, 11,
  2003, [\href{https://arxiv.org/abs/gr-qc/0311066}{{\ttfamily
  gr-qc/0311066}}].

\bibitem{Neiman:2011gf}
Y.~Neiman, \emph{{Parity and reality properties of the EPRL spinfoam}},
  \href{https://doi.org/10.1088/0264-9381/29/6/065008}{\emph{Class. Quant.
  Grav.} {\bfseries 29} (2012) 065008}
  [\href{https://arxiv.org/abs/1109.3946}{{\ttfamily 1109.3946}}].

\bibitem{Rovelli:2012yy}
C.~Rovelli and E.~Wilson-Ewing, \emph{{Discrete Symmetries in Covariant LQG}},
  \href{https://doi.org/10.1103/PhysRevD.86.064002}{\emph{Phys. Rev. D}
  {\bfseries 86} (2012) 064002}
  [\href{https://arxiv.org/abs/1205.0733}{{\ttfamily 1205.0733}}].

\bibitem{Ashtekar:1988sw}
A.~Ashtekar, A.P.~Balachandran and S.~Jo, \emph{{The {CP} Problem in Quantum
  Gravity}}, \href{https://doi.org/10.1142/S0217751X89000649}{\emph{Int. J.
  Mod. Phys. A} {\bfseries 4} (1989) 1493}.

\bibitem{Freidel:2005sn}
L.~Freidel, D.~Minic and T.~Takeuchi, \emph{{Quantum gravity, torsion, parity
  violation and all that}},
  \href{https://doi.org/10.1103/PhysRevD.72.104002}{\emph{Phys. Rev. D}
  {\bfseries 72} (2005) 104002}
  [\href{https://arxiv.org/abs/hep-th/0507253}{{\ttfamily hep-th/0507253}}].

\bibitem{Contaldi:2008yz}
C.R.~Contaldi, J.~Magueijo and L.~Smolin, \emph{{Anomalous CMB polarization and
  gravitational chirality}},
  \href{https://doi.org/10.1103/PhysRevLett.101.141101}{\emph{Phys. Rev. Lett.}
  {\bfseries 101} (2008) 141101}
  [\href{https://arxiv.org/abs/0806.3082}{{\ttfamily 0806.3082}}].

\bibitem{Crane:2001qk}
L.~Crane, A.~Perez and C.~Rovelli, \emph{{A Finiteness proof for the Lorentzian
  state sum spin foam model for quantum general relativity}},
  \href{https://arxiv.org/abs/gr-qc/0104057}{{\ttfamily gr-qc/0104057}}.

\bibitem{Engle:2007wy}
J.~Engle, E.~Livine, R.~Pereira and C.~Rovelli, \emph{{LQG vertex with finite
  Immirzi parameter}},
  \href{https://doi.org/10.1016/j.nuclphysb.2008.02.018}{\emph{Nucl. Phys. B}
  {\bfseries 799} (2008) 136}
  [\href{https://arxiv.org/abs/0711.0146}{{\ttfamily 0711.0146}}].

\bibitem{BenGeloun:2010qkf}
J.~Ben~Geloun, R.~Gurau and V.~Rivasseau, \emph{{EPRL/FK Group Field Theory}},
  \href{https://doi.org/10.1209/0295-5075/92/60008}{\emph{EPL} {\bfseries 92}
  (2010) 60008} [\href{https://arxiv.org/abs/1008.0354}{{\ttfamily
  1008.0354}}].

\bibitem{Oriti:2016ueo}
D.~Oriti, L.~Sindoni and E.~Wilson-Ewing, \emph{{Bouncing cosmologies from
  quantum gravity condensates}},
  \href{https://doi.org/10.1088/1361-6382/aa549a}{\emph{Class. Quant. Grav.}
  {\bfseries 34} (2017) 04LT01}
  [\href{https://arxiv.org/abs/1602.08271}{{\ttfamily 1602.08271}}].

\bibitem{DePietri:1998hnx}
R.~De~Pietri and L.~Freidel, \emph{{so(4) Plebanski action and relativistic
  spin foam model}},
  \href{https://doi.org/10.1088/0264-9381/16/7/303}{\emph{Class. Quant. Grav.}
  {\bfseries 16} (1999) 2187}
  [\href{https://arxiv.org/abs/gr-qc/9804071}{{\ttfamily gr-qc/9804071}}].

\bibitem{Oriti:2000hh}
D.~Oriti and R.M.~Williams, \emph{{Gluing 4 simplices: A Derivation of the
  Barrett-Crane spin foam model for Euclidean quantum gravity}},
  \href{https://doi.org/10.1103/PhysRevD.63.024022}{\emph{Phys. Rev. D}
  {\bfseries 63} (2001) 024022}
  [\href{https://arxiv.org/abs/gr-qc/0010031}{{\ttfamily gr-qc/0010031}}].

\bibitem{Finocchiaro:2020xwr}
M.~Finocchiaro, Y.~Jeong and D.~Oriti, \emph{{Quantum geometric maps and their
  properties}},  \href{https://arxiv.org/abs/2012.11536}{{\ttfamily
  2012.11536}}.

\bibitem{Loll:2000my}
R.~Loll, \emph{{Discrete Lorentzian quantum gravity}},
  \href{https://doi.org/10.1016/S0920-5632(01)00957-4}{\emph{Nucl. Phys. B
  Proc. Suppl.} {\bfseries 94} (2001) 96}
  [\href{https://arxiv.org/abs/hep-th/0011194}{{\ttfamily hep-th/0011194}}].

\bibitem{Kazakov:1995ae}
V.A.~Kazakov, M.~Staudacher and T.~Wynter, \emph{{Character expansion methods
  for matrix models of dually weighted graphs}},
  \href{https://doi.org/10.1007/BF02101902}{\emph{Commun. Math. Phys.}
  {\bfseries 177} (1996) 451}
  [\href{https://arxiv.org/abs/hep-th/9502132}{{\ttfamily hep-th/9502132}}].

\bibitem{Benedetti:2011nn}
D.~Benedetti and R.~Gurau, \emph{{Phase Transition in Dually Weighted Colored
  Tensor Models}},
  \href{https://doi.org/10.1016/j.nuclphysb.2011.10.015}{\emph{Nucl. Phys. B}
  {\bfseries 855} (2012) 420}
  [\href{https://arxiv.org/abs/1108.5389}{{\ttfamily 1108.5389}}].

\bibitem{Jordan:2013awa}
S.~Jordan and R.~Loll, \emph{{Causal Dynamical Triangulations without Preferred
  Foliation}},
  \href{https://doi.org/10.1016/j.physletb.2013.06.007}{\emph{Phys. Lett. B}
  {\bfseries 724} (2013) 155}
  [\href{https://arxiv.org/abs/1305.4582}{{\ttfamily 1305.4582}}].

\bibitem{jordan2013globally}
S.~Jordan, \emph{Globally and locally causal dynamical triangulations}, [Sl:
  sn] (2013).

\bibitem{Loll:2015yaa}
R.~Loll and B.~Ruijl, \emph{{Locally Causal Dynamical Triangulations in Two
  Dimensions}}, \href{https://doi.org/10.1103/PhysRevD.92.084002}{\emph{Phys.
  Rev. D} {\bfseries 92} (2015) 084002}
  [\href{https://arxiv.org/abs/1507.04566}{{\ttfamily 1507.04566}}].

\bibitem{Asante:2021phx}
S.K.~Asante, B.~Dittrich and J.~Padua-Arg\"uelles, \emph{{Complex actions and
  causality violations: Applications to Lorentzian quantum cosmology}},
  \href{https://arxiv.org/abs/2112.15387}{{\ttfamily 2112.15387}}.

\bibitem{ashtekar2014springer}
A.~Ashtekar and V.~Petkov, \emph{Springer handbook of spacetime}, Springer
  (2014).

\bibitem{Horowitz1991}
G.T.~Horowitz, \emph{Topology change in classical and quantum gravity},
  \href{https://doi.org/10.1088/0264-9381/8/4/007}{\emph{Classical and Quantum
  Gravity} {\bfseries 8} (1991) 587}.

\bibitem{Geroch:1967fs}
R.P.~Geroch, \emph{{Topology in general relativity}},
  \href{https://doi.org/10.1063/1.1705276}{\emph{J. Math. Phys.} {\bfseries 8}
  (1967) 782}.

\bibitem{wheeler1957nature}
J.A.~Wheeler, \emph{On the nature of quantum geometrodynamics}, {\emph{Annals
  of Physics} {\bfseries 2} (1957) 604}.

\bibitem{Sorkin:1997gi}
R.D.~Sorkin, \emph{{Forks in the road, on the way to quantum gravity}},
  \href{https://doi.org/10.1007/BF02435709}{\emph{Int. J. Theor. Phys.}
  {\bfseries 36} (1997) 2759}
  [\href{https://arxiv.org/abs/gr-qc/9706002}{{\ttfamily gr-qc/9706002}}].

\bibitem{Barrett:1997gw}
J.W.~Barrett and L.~Crane, \emph{{Relativistic spin networks and quantum
  gravity}}, \href{https://doi.org/10.1063/1.532254}{\emph{J. Math. Phys.}
  {\bfseries 39} (1998) 3296}
  [\href{https://arxiv.org/abs/gr-qc/9709028}{{\ttfamily gr-qc/9709028}}].

\bibitem{Matschull:1997du}
H.-J.~Matschull and M.~Welling, \emph{{Quantum mechanics of a point particle in
  (2+1)-dimensional gravity}},
  \href{https://doi.org/10.1088/0264-9381/15/10/008}{\emph{Class. Quant. Grav.}
  {\bfseries 15} (1998) 2981}
  [\href{https://arxiv.org/abs/gr-qc/9708054}{{\ttfamily gr-qc/9708054}}].

\bibitem{tHooft:1993jwb}
G.~'t~Hooft, \emph{{Canonical quantization of gravitating point particles in
  (2+1)-dimensions}},
  \href{https://doi.org/10.1088/0264-9381/10/8/022}{\emph{Class. Quant. Grav.}
  {\bfseries 10} (1993) 1653}
  [\href{https://arxiv.org/abs/gr-qc/9305008}{{\ttfamily gr-qc/9305008}}].

\bibitem{Baez:2002rx}
J.C.~Baez, J.D.~Christensen and G.~Egan, \emph{{Asymptotics of 10j symbols}},
  \href{https://doi.org/10.1088/0264-9381/19/24/315}{\emph{Class. Quant. Grav.}
  {\bfseries 19} (2002) 6489}
  [\href{https://arxiv.org/abs/gr-qc/0208010}{{\ttfamily gr-qc/0208010}}].

\bibitem{Crane:2001as}
L.~Crane, A.~Perez and C.~Rovelli, \emph{{Perturbative finiteness in spin-foam
  quantum gravity}},
  \href{https://doi.org/10.1103/PhysRevLett.87.181301}{\emph{Phys. Rev. Lett.}
  {\bfseries 87} (2001) 181301}.

\bibitem{Pithis:2018eaq}
A.G.A.~Pithis and J.~Th\"urigen, \emph{{Phase transitions in group field
  theory: The Landau perspective}},
  \href{https://doi.org/10.1103/PhysRevD.98.126006}{\emph{Phys. Rev. D}
  {\bfseries 98} (2018) 126006}
  [\href{https://arxiv.org/abs/1808.09765}{{\ttfamily 1808.09765}}].

\bibitem{Pithis:2019mlv}
A.G.A.~Pithis, \emph{{Aspects of quantum gravity}}, Ph.D. thesis, King's Coll.
  London, 2019.
\newblock \href{https://arxiv.org/abs/1903.07735}{{\ttfamily 1903.07735}}.

\bibitem{Marchetti:2021fix}
L.~Marchetti, D.~Oriti, A.G.A.~Pithis and J.~Th\"urigen, \emph{{Phase
  transitions in tensorial group field theories: Landau-Ginzburg analysis of
  models with both local and non-local degrees of freedom}},
  \href{https://arxiv.org/abs/2110.15336}{{\ttfamily 2110.15336}}.

\bibitem{Benedetti:2015et}
D.~Benedetti, J.~Ben~Geloun and D.~Oriti, \emph{{Functional Renormalisation
  Group Approach for Tensorial Group Field Theory: a Rank-3 Model}},
  {\emph{JHEP} {\bfseries 03} (2015) 084}
  [\href{https://arxiv.org/abs/1411.3180}{{\ttfamily 1411.3180}}].

\bibitem{BenGeloun:2015ej}
J.~Ben~Geloun, R.~Martini and D.~Oriti, \emph{{Functional Renormalization Group
  analysis of a Tensorial Group Field Theory on ${R}^3$}}, {\emph{EPL}
  {\bfseries 112} (2015) 31001}
  [\href{https://arxiv.org/abs/1508.01855}{{\ttfamily 1508.01855}}].

\bibitem{Benedetti:2016db}
D.~Benedetti and V.~Lahoche, \emph{{Functional renormalization group approach
  for tensorial group field theory: a rank-6 model with closure constraint}},
  {\emph{Classical And Quantum Gravity} {\bfseries 33} (2016) }
  [\href{https://arxiv.org/abs/1508.06384}{{\ttfamily 1508.06384}}].

\bibitem{BenGeloun:2016kw}
J.~Ben~Geloun, R.~Martini and D.~Oriti, \emph{{Functional renormalization group
  analysis of tensorial group field theories on $R^d$}}, {\emph{Phys. Rev. D}
  {\bfseries 94} (2016) 024017}
  [\href{https://arxiv.org/abs/1601.08211}{{\ttfamily 1601.08211}}].

\bibitem{Carrozza:2016tih}
S.~Carrozza and V.~Lahoche, \emph{{Asymptotic safety in three-dimensional SU(2)
  Group Field Theory: evidence in the local potential approximation}},
  \href{https://doi.org/10.1088/1361-6382/aa6d90}{\emph{Class. Quant. Grav.}
  {\bfseries 34} (2017) 115004}
  [\href{https://arxiv.org/abs/1612.02452}{{\ttfamily 1612.02452}}].

\bibitem{Carrozza:2017vkz}
S.~Carrozza, V.~Lahoche and D.~Oriti, \emph{{Renormalizable Group Field Theory
  beyond melonic diagrams: an example in rank four}},
  \href{https://doi.org/10.1103/PhysRevD.96.066007}{\emph{Phys. Rev. D}
  {\bfseries 96} (2017) 066007}
  [\href{https://arxiv.org/abs/1703.06729}{{\ttfamily 1703.06729}}].

\bibitem{BenGeloun:2018ekd}
J.~Ben~Geloun, T.A.~Koslowski, D.~Oriti and A.D.~Pereira, \emph{{Functional
  Renormalization Group analysis of rank 3 tensorial group field theory: The
  full quartic invariant truncation}},
  \href{https://doi.org/10.1103/PhysRevD.97.126018}{\emph{Phys. Rev. D}
  {\bfseries 97} (2018) 126018}
  [\href{https://arxiv.org/abs/1805.01619}{{\ttfamily 1805.01619}}].

\bibitem{Pithis:2020kio}
A.G.A.~Pithis and J.~Th\"urigen, \emph{{Phase transitions in TGFT: functional
  renormalization group in the cyclic-melonic potential approximation and
  equivalence to O$(N)$ models}},
  \href{https://doi.org/10.1007/JHEP12(2020)159}{\emph{JHEP} {\bfseries 12}
  (2020) 159} [\href{https://arxiv.org/abs/2009.13588}{{\ttfamily
  2009.13588}}].

\bibitem{Pithis:2020sxm}
A.G.A.~Pithis and J.~Th\"urigen, \emph{{(No) phase transition in tensorial
  group field theory}},
  \href{https://doi.org/10.1016/j.physletb.2021.136215}{\emph{Phys. Lett. B}
  {\bfseries 816} (2021) 136215}
  [\href{https://arxiv.org/abs/2007.08982}{{\ttfamily 2007.08982}}].

\bibitem{Lahoche:2021nba}
V.~Lahoche, B.-B.B.~Natta and D.O.~Samary, \emph{{No Ward-Takahashi identity
  violation for an Abelian tensorial group field theories with closure
  constraint}},  \href{https://arxiv.org/abs/2108.10979}{{\ttfamily
  2108.10979}}.

\bibitem{Li:2017uao}
Y.~Li, D.~Oriti and M.~Zhang, \emph{{Group field theory for quantum gravity
  minimally coupled to a scalar field}},
  \href{https://doi.org/10.1088/1361-6382/aa85d2}{\emph{Class. Quant. Grav.}
  {\bfseries 34} (2017) 195001}
  [\href{https://arxiv.org/abs/1701.08719}{{\ttfamily 1701.08719}}].

\bibitem{Bianchi:2010bn}
E.~Bianchi, M.~Han, C.~Rovelli, W.~Wieland, E.~Magliaro and C.~Perini,
  \emph{{Spinfoam fermions}},
  \href{https://doi.org/10.1088/0264-9381/30/23/235023}{\emph{Class. Quant.
  Grav.} {\bfseries 30} (2013) 235023}
  [\href{https://arxiv.org/abs/1012.4719}{{\ttfamily 1012.4719}}].

\bibitem{Oriti:2006jk}
D.~Oriti and J.~Ryan, \emph{{Group field theory formulation of 3-D quantum
  gravity coupled to matter fields}},
  \href{https://doi.org/10.1088/0264-9381/23/22/027}{\emph{Class. Quant. Grav.}
  {\bfseries 23} (2006) 6543}
  [\href{https://arxiv.org/abs/gr-qc/0602010}{{\ttfamily gr-qc/0602010}}].

\bibitem{Dowdall:2010ej}
R.J.~Dowdall and W.J.~Fairbairn, \emph{{Observables in 3d spinfoam quantum
  gravity with fermions}},
  \href{https://doi.org/10.1007/s10714-010-1107-0}{\emph{Gen. Rel. Grav.}
  {\bfseries 43} (2011) 1263}
  [\href{https://arxiv.org/abs/1003.1847}{{\ttfamily 1003.1847}}].

\bibitem{Oriti:2002bn}
D.~Oriti and H.~Pfeiffer, \emph{{A Spin foam model for pure gauge theory
  coupled to quantum gravity}},
  \href{https://doi.org/10.1103/PhysRevD.66.124010}{\emph{Phys. Rev. D}
  {\bfseries 66} (2002) 124010}
  [\href{https://arxiv.org/abs/gr-qc/0207041}{{\ttfamily gr-qc/0207041}}].

\bibitem{Speziale:2007mt}
S.~Speziale, \emph{{Coupling gauge theory to spinfoam 3d quantum gravity}},
  \href{https://doi.org/10.1088/0264-9381/24/20/014}{\emph{Class. Quant. Grav.}
  {\bfseries 24} (2007) 5139}
  [\href{https://arxiv.org/abs/0706.1534}{{\ttfamily 0706.1534}}].

\bibitem{Giesel:2012rb}
K.~Giesel and T.~Thiemann, \emph{{Scalar Material Reference Systems and Loop
  Quantum Gravity}},
  \href{https://doi.org/10.1088/0264-9381/32/13/135015}{\emph{Class. Quant.
  Grav.} {\bfseries 32} (2015) 135015}
  [\href{https://arxiv.org/abs/1206.3807}{{\ttfamily 1206.3807}}].

\bibitem{Marchetti:2021gcv}
L.~Marchetti and D.~Oriti, \emph{{Effective dynamics of scalar cosmological
  perturbations from quantum gravity}},
  \href{https://arxiv.org/abs/2112.12677}{{\ttfamily 2112.12677}}.

\bibitem{deCesare:2016rsf}
M.~de~Cesare, A.G.A.~Pithis and M.~Sakellariadou, \emph{{Cosmological
  implications of interacting Group Field Theory models: cyclic Universe and
  accelerated expansion}},
  \href{https://doi.org/10.1103/PhysRevD.94.064051}{\emph{Phys. Rev. D}
  {\bfseries 94} (2016) 064051}
  [\href{https://arxiv.org/abs/1606.00352}{{\ttfamily 1606.00352}}].

\bibitem{Oriti:2021rvm}
D.~Oriti and X.~Pang, \emph{{Phantom-like dark energy from quantum gravity}},
  \href{https://arxiv.org/abs/2105.03751}{{\ttfamily 2105.03751}}.

\bibitem{kroon2016conformal}
J.A.V.~Kroon, \emph{Conformal methods in general relativity}, Cambridge
  University Press (2016).

\bibitem{Oriti:2018qty}
D.~Oriti, D.~Pranzetti and L.~Sindoni, \emph{{Black Holes as Quantum Gravity
  Condensates}}, \href{https://doi.org/10.1103/PhysRevD.97.066017}{\emph{Phys.
  Rev. D} {\bfseries 97} (2018) 066017}
  [\href{https://arxiv.org/abs/1801.01479}{{\ttfamily 1801.01479}}].

\bibitem{Oriti:2015rwa}
D.~Oriti, D.~Pranzetti and L.~Sindoni, \emph{{Horizon entropy from quantum
  gravity condensates}},
  \href{https://doi.org/10.1103/PhysRevLett.116.211301}{\emph{Phys. Rev. Lett.}
  {\bfseries 116} (2016) 211301}
  [\href{https://arxiv.org/abs/1510.06991}{{\ttfamily 1510.06991}}].

\bibitem{Naimark1964}
M.~Naimark, \emph{Linear representations of the lorentz group},
  {\emph{Elsevier} {\bfseries 61} (1964) 464}.

\bibitem{Brukner2014}
{\v{C}}.~Brukner, \emph{Quantum causality},
  \href{https://doi.org/10.1038/nphys2930}{\emph{Nature Physics} {\bfseries 10}
  (2014) 259}.

\bibitem{Markopoulou:1997wi}
F.~Markopoulou and L.~Smolin, \emph{{Causal evolution of spin networks}},
  \href{https://doi.org/10.1016/S0550-3213(97)00488-4}{\emph{Nucl. Phys. B}
  {\bfseries 508} (1997) 409}
  [\href{https://arxiv.org/abs/gr-qc/9702025}{{\ttfamily gr-qc/9702025}}].

\bibitem{Markopoulou:1999cz}
F.~Markopoulou, \emph{{Quantum causal histories}},
  \href{https://doi.org/10.1088/0264-9381/17/10/302}{\emph{Class. Quant. Grav.}
  {\bfseries 17} (2000) 2059}
  [\href{https://arxiv.org/abs/hep-th/9904009}{{\ttfamily hep-th/9904009}}].

\bibitem{Markopoulou:1999ht}
F.~Markopoulou, \emph{{An Insider's guide to quantum causal histories}},
  \href{https://doi.org/10.1016/S0920-5632(00)00791-X}{\emph{Nucl. Phys. B
  Proc. Suppl.} {\bfseries 88} (2000) 308}
  [\href{https://arxiv.org/abs/hep-th/9912137}{{\ttfamily hep-th/9912137}}].

\bibitem{Khetrapal:2012ux}
S.~Khetrapal and S.~Surya, \emph{{Boundary Term Contribution to the Volume of a
  Small Causal Diamond}},
  \href{https://doi.org/10.1088/0264-9381/30/6/065005}{\emph{Class. Quant.
  Grav.} {\bfseries 30} (2013) 065005}
  [\href{https://arxiv.org/abs/1212.0629}{{\ttfamily 1212.0629}}].

\bibitem{deBoer:2016pqk}
J.~de~Boer, F.M.~Haehl, M.P.~Heller and R.C.~Myers, \emph{{Entanglement,
  holography and causal diamonds}},
  \href{https://doi.org/10.1007/JHEP08(2016)162}{\emph{JHEP} {\bfseries 08}
  (2016) 162} [\href{https://arxiv.org/abs/1606.03307}{{\ttfamily
  1606.03307}}].

\bibitem{Bousso:2000nf}
R.~Bousso, \emph{{Positive vacuum energy and the N bound}},
  \href{https://doi.org/10.1088/1126-6708/2000/11/038}{\emph{JHEP} {\bfseries
  11} (2000) 038} [\href{https://arxiv.org/abs/hep-th/0010252}{{\ttfamily
  hep-th/0010252}}].

\bibitem{Martin-Dussaud:2019ypf}
P.~Martin-Dussaud, \emph{{A Primer of Group Theory for Loop Quantum Gravity and
  Spin-foams}}, \href{https://doi.org/10.1007/s10714-019-2583-5}{\emph{Gen.
  Rel. Grav.} {\bfseries 51} (2019) 110}
  [\href{https://arxiv.org/abs/1902.08439}{{\ttfamily 1902.08439}}].

\bibitem{Speziale:2016axj}
S.~Speziale, \emph{{Boosting Wigner\textquoteright{}s nj-symbols}},
  \href{https://doi.org/10.1063/1.4977752}{\emph{J. Math. Phys.} {\bfseries 58}
  (2017) 032501} [\href{https://arxiv.org/abs/1609.01632}{{\ttfamily
  1609.01632}}].

\end{thebibliography}\endgroup

\end{document}